\documentclass[11pt]{article}

\usepackage[a4paper, total={5.8in, 9in}]{geometry}
\usepackage[indent,skip=4pt]{parskip}
\usepackage{amsmath,amssymb,amsfonts,amsthm,mathrsfs,mathtools,cases}
\usepackage{dsfont}
\usepackage{authblk}
\usepackage[colorlinks=true,linkcolor=blue, citecolor=blue, bookmarks]{hyperref}
\usepackage{cleveref}
\usepackage{microtype}
\usepackage{comment}
\usepackage{graphicx,adjustbox} 
\usepackage{subfig}
\usepackage{wrapfig}
\usepackage[font=small,labelfont=bf]{caption}
\usepackage[usenames,dvipsnames]{xcolor}  
\usepackage{booktabs}
\usepackage{braket}
\usepackage[nottoc]{tocbibind}
\usepackage{nicematrix}
\usepackage{tikz}
\usepackage{pgfplots}
\usepackage{cite}
\pgfplotsset{compat=1.18}
\usepackage{enumitem}

\usetikzlibrary{matrix}
\usetikzlibrary{decorations.markings}
\usetikzlibrary{patterns}
\usetikzlibrary{arrows.meta}
\usepackage{color,ifthen,amsthm,amsmath,amsxtra,amsfonts,dsfont,graphicx,bm,tikz,scalerel,wasysym,bbm,graphicx,amsthm,braket,physics,empheq,CircuitTikz,float}

\numberwithin{equation}{section}

\crefname{figure}{Fig.}{Figs.}
\crefname{equation}{Eq.}{Eqs.}
\crefname{section}{Sec.}{Secs.}
\crefname{appendix}{Appendix}{Appendices}
  
\colorlet{darkerblue}{blue!80!black}
\colorlet{lightblue}{blue!70!white}
\usepackage{braket}
\hypersetup{
    colorlinks,
    linkcolor={darkerblue},
    citecolor={darkerblue},
    urlcolor={darkerblue},
    backref=true
}

\usepackage{CircuitTikz,braket}
\newcommand{\be}{\begin{equation}}
\newcommand{\ee}{\end{equation}}
\newcommand{\bea}{\begin{eqnarray}}
\newcommand{\eea}{\end{eqnarray}}
\newcommand{\bse}{\begin{subequations}}
\newcommand{\ese}{\end{subequations}}
\newcommand{\1}{\mathbbm{1}}

\usepackage{authblk}

\title{\bf Entanglement and information dynamics in charged dual-unitary circuits with impurities}
\author[1,2]{Shachar Fraenkel}
\author[3]{Colin Rylands}

\affil[1]{Technical University of Munich, TUM School of Natural Sciences, Physics Department, 85748 Garching, Germany}
\affil[2]{Munich Center for Quantum Science and Technology (MCQST), Schellingstr.~4, 80799 M\"unchen, Germany}
\affil[3]{Centre for Fluid and Complex Systems, Coventry University, Coventry, CV1 2TT, United Kingdom}

\date{}

\begin{document}
\maketitle

\begin{abstract}
Local parts of a closed, non-equilibrium quantum many-body system can reach stationarity through the buildup of entanglement between different regions. In this paper we investigate how this process is affected by the presence of a quantum impurity that breaks a global symmetry of the bulk. In particular, we study a charged dual-unitary quantum circuit perturbed by a symmetry-breaking, non-dual-unitary gate acting on a spatially fixed pair of sites. In this setup we calculate the entanglement entropy for a subsystem that does not contain the impurity, as well as the mutual information and logarithmic negativity for a bipartite subsystem that straddles the impurity.  We obtain exact results in the early time regime for these quantities from the dual unitarity of the bulk system. Beyond this regime, we compute the predictions of the quasiparticle and entanglement membrane pictures, supporting them with exact calculations obtained by taking the corresponding limits of the bulk entangling power. We find that the presence of a symmetry-breaking impurity can lead to a richer behaviour of quantum relaxation, including non-monotonic evolution of the entanglement entropy, a modification of the long-time steady state, and persistent subsystem mutual information and entanglement negativity. This behaviour is understood in terms of long-range correlations generated by scattering from the impurity, and facilitated by the ballistic propagation of the conserved charges. 
\end{abstract}
\newpage 

\tableofcontents

\section{Introduction}
A quantum impurity model has two components: the impurity, which consists of a small number of localized degrees of freedom, and the bulk, which contains a large number of mobile constituents coupled to the impurity. Standard constructions of such models lead, at low energies, to a $1+1$-dimensional conformal field theory perturbed by a local operator which is marginal or relevant. This structure of  a CFT with a localized perturbation is seen in many well-known quantum impurity problems, such as the Kondo~\cite{kondo1964resistance} or Kane-Fisher models~\cite{kane1992transport}. It is complex enough to capture many interesting strongly correlated phenomena, yet simple enough to facilitate exact analysis in certain cases, using e.g. boundary CFT methods~\cite{affleck1991kondo,affleck1993exact} or the Bethe ansatz~\cite{andrei1983solution,rylands2016quantum}.  These methods work exceedingly well in equilibrium but lose some of their ubiquity in far-from-equilibrium scenarios, such as quantum quenches. 

Local relaxation of closed many-body quantum systems that have been quenched out of equilibrium has been studied intensely for many years~\cite{Deutsch1991,srednicki1,Rigol:2007juv,polkovnikov,DAlessio:2015qtq,Gogolin_2016,Essler_quench,Calabrese_2016,rylands2020nonequilibrium}. The typical scenario involves translationally invariant systems, or perhaps a translationally invariant time evolution operator and a non-translationally-invariant initial state. However, systems in which the time evolution operator breaks translational invariance, for instance via a quantum impurity, have also been studied and exhibited distinct nonequilibrium behaviours, particularly in their quantum information properties; see for example Refs.~\cite{santos2014local,gruber2020time,fritzsch2022boundary,fritzsch2023boundary,capizzi2023entanglement,capizzi2023domain,rylands2023transport,fraenkel2021entanglement,fraenkel2023extensive,fraenkel2024entanglement,li2025dynamics}. In this paper, we study the local relaxation of a quenched many-body quantum system that contains a spatially fixed quantum impurity which breaks a global symmetry of the bulk.  Our goal is to understand how local relaxation is modified by the presence of the symmetry-breaking impurity outside of the subsystem of interest. 

We take as our setting brickwork quantum circuits and, inspired by the CFT plus perturbation structure, consider a dual-unitary circuit that is perturbed by a non-dual-unitary gate acting on a spatially fixed set of sites. Dual-unitary circuits possess many features that are reminiscent of CFTs, such as strict light-cone propagation, and, for a certain class of initial states, facilitate exact solvability in a non-equilibrium setting, even when their dynamics is non-integrable~\cite{bertini2018exact,gopalakrishnan2019unitary,prosen2021many,claeys2021ergodic,kos2021correlations,kos2023circuits,foligno2023temporal,piroli2020scrambling,bertini2020operator,bertini2020operator2,holdendye2023fundamental,borsi2022construction,bertini2019exact,piroli2020exact,giudice2022temporal,vernier2023integrable,bertini2025exactly}. Previously, in Ref.~\cite{fraenkel2024entanglement}, we studied the impact of an impurity on entanglement entropy dynamics in dual-unitary circuits, finding that it can significantly alter these dynamics. We observed qualitatively different dynamics between circuits with a non-interacting bulk and circuits with a chaotic bulk, imposing no symmetry on the latter. We showed that the dynamics can be exactly captured by two distinct effective models of the bulk: the quasiparticle picture~\cite{calabrese2005evolution,alba2017entanglement} in the non-interacting case and the entanglement membrane picture~\cite{jonay2018coarse,nahum2017quantum,nahum2020entanglement} in the chaotic case.

In this paper, we consider charged dual-unitary circuits that have a global Abelian symmetry, namely particle-number conservation, which is broken by the impurity gate. In this setting, we compute the dynamics of the bipartite entanglement entropy, mutual information, and logarithmic negativity for subsystems that do not contain the impurity. Using a combination of exact methods, rigorous bounds, and the effective quasiparticle and membrane pictures, we derive analytic expressions for our quantities of interest and check them against previously known results. On our way there, we complement the results of Ref.~\cite{fraenkel2024entanglement} by deriving the dynamics of mutual information and entanglement negativity in uncharged dual-unitary circuits with impurities. We furthermore confirm analytically that the quasiparticle-picture dynamics holds not only for a non-interacting bulk, but also for a bulk corresponding to the Floquet XXZ chain, a paradigmatic interacting integrable model.

Similar to Ref.~\cite{fraenkel2024entanglement}, we find that a symmetry-breaking impurity can significantly impact the entanglement dynamics in a system with a charge-conserving bulk. In particular, the impurity can lead to non-monotonic behaviour of the bipartite entanglement entropy and, in turn, a longer relaxation time for the subsystem. In contrast to the uncharged circuit, where such non-monotonicity does not arise within the membrane-picture prediction, here it is predicted by both the quasiparticle and membrane pictures. Moreover, we find that the stationary value of the entanglement for a symmetry-breaking impurity is lower-bounded by that of a symmetric one. These features can be attributed to the creation of charge correlations between different parts of the subsystem, induced by the impurity. 

Our study of the dynamics of the mutual information and logarithmic negativity focuses on the correlation between subsystems on opposite sides of the impurity. For an uncharged circuit, we find that the mutual information displays the typical rise and fall behaviour, with the rate of increase and decrease dependent on the impurity. The quasiparticle and membrane pictures provide qualitatively similar predictions. Surprisingly, however, if the impurity is reflective enough, the membrane picture predicts a higher peak value. At late times, the exact stationary state can be found and shows vanishing mutual information. A similar structure is seen in the negativity dynamics, although we obtain a membrane-picture prediction only at early and late times. When the system has a bulk charge whose conservation is broken by the impurity, the behaviour has some striking differences; most notably, the symmetry-breaking impurity is able to sustain a non-vanishing stationary value of the mutual information and negativity. The impurity can therefore create persistent quantum charge correlations between different parts of the system. 

The rest of the paper is structured as follows: In Section~\ref{sec:setup}, we introduce our model and discuss some of its properties and the quantities we wish to study. In Section~\ref{sec:ent_uncharged}, we review the results from Ref.~\cite{fraenkel2024entanglement} for the entanglement entropy in an uncharged circuit and provide details on the quasiparticle and entanglement membrane pictures. In Section~\ref{sec:Two_Intervals} we determine the mutual information and negativity for an uncharged circuit. We then switch to considering a circuit with a charged bulk, i.e., one possessing a $U(1)$ symmetry, which is broken by the impurity. In Section~\ref{sec:ent_charged}, we obtain exact early-time results for the entanglement entropy, prove a bound on its late-time behaviour, and derive the predictions of the quasiparticle and entanglement membrane pictures. In Section \ref{sec:mut_info_charged} we examine the mutual information and negativity dynamics for the charged-bulk circuit, and then summarize and conclude in Section~\ref{sec:conclusions}. 

\section{Setup and notation}\label{sec:setup}

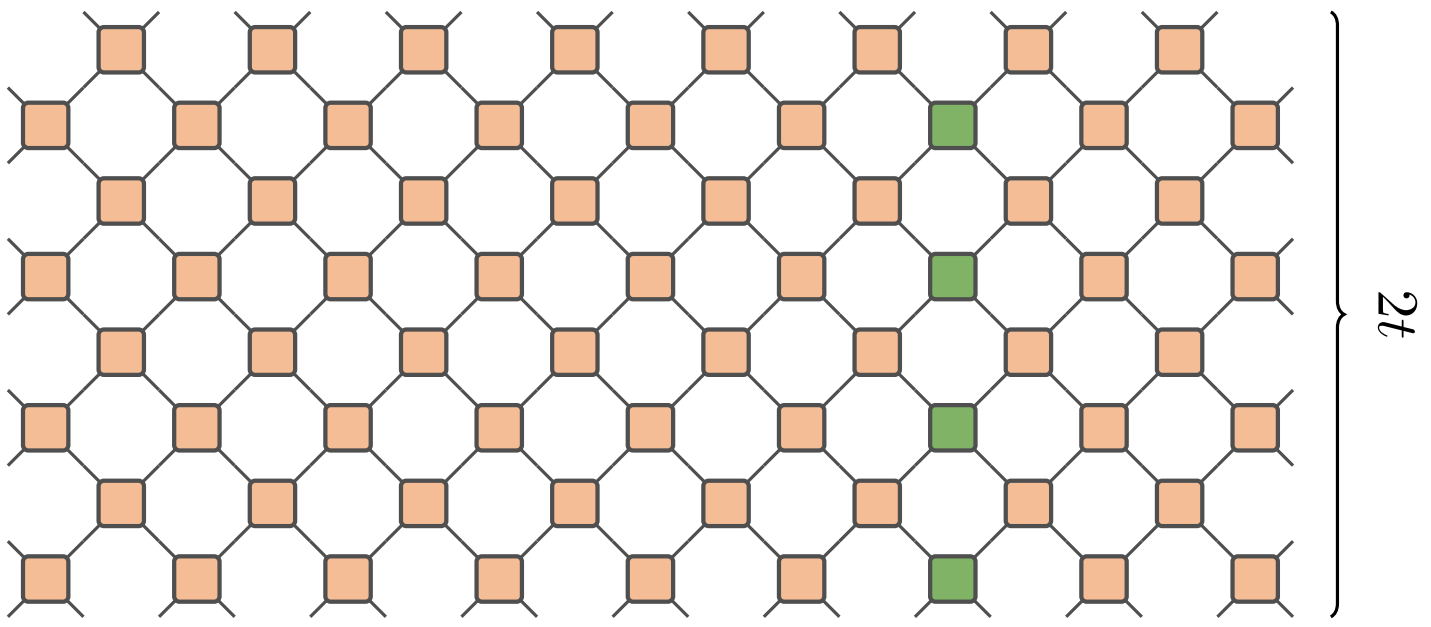
\begin{figure}\centering
\begin{tikzpicture}[baseline={([yshift=0.6ex]current bounding box.center)},scale=0.5]
    \foreach \x in {-5,-3,...,5}{\prop{\x}{-3}{colSt}{}}
     \foreach \x in {7,9}{\prop{\x}{-3}{colSt}{}}
      \foreach \x in {-5,-3,...,5}{\prop{\x}{-1}{colSt}{}}
           \foreach \x in {7,9}{\prop{\x}{-1}{colSt}{}}
        \foreach \x in {-5,-3,...,5}{\prop{\x}{1}{colSt}{}}
             \foreach \x in {7,9}{\prop{\x}{1}{colSt}{}}
          \foreach \x in {-5,-3,...,5}{\prop{\x}{3}{colSt}{}}
               \foreach \x in {7,9}{\prop{\x}{3}{colSt}{}}
    \foreach \x in {-6,-4,...,6}{\prop{\x}{-4}{colSt}{}}
         \foreach \x in {8,10}{\prop{\x}{-4}{colSt}{}}
     \foreach \x in {-6,-4,...,6}{\prop{\x}{-2}{colSt}{}}
          \foreach \x in {8,10}{\prop{\x}{-2}{colSt}{}}
    \foreach \x in {-6,-4,...,6}{\prop{\x}{0}{colSt}{}} 
         \foreach \x in {8,10}{\prop{\x}{0}{colSt}{}}
     \foreach \x in {-6,-4,...,6}{\prop{\x}{2}{colSt}{}}
         \foreach \x in {8,10}{\prop{\x}{2}{colSt}{}}
      \foreach \x in {-4,-2,0,2} {\prop{6}{\x}{colimp}{}}
  \draw[semithick,decorate,decoration={brace}] (11,3.5) -- (11,-4.5) node[midway,xshift=12.5pt,rotate= 270] {$2t$};
  \end{tikzpicture}
  \caption{\label{fig:impurity_circuit}The general circuit construction studied in this paper, shown for evolution time $t=4$. Orange boxes denote local dual-unitary bulk gates, while green boxes denote non-dual-unitary impurity gates.  }
\end{figure}

We consider a one-dimensional chain of $2L$ qudits, each with local Hilbert space dimension $q$, located at coordinates $x=1,\dots, 2L$.   The dynamics of the system are generated by a brickwork-pattern quantum circuit, such that the state of the system after $t$ time steps is given by 
\be\label{eq:time_evo_op}
\ket{\Psi(t)}=\mathbb{U}(t)\ket{\Psi(t-1)},
\ee
and where the time-evolution operator is 
\be 
 \mathbb{U}(t)=
\bigotimes_{x\,{\rm odd}}U_{x,x+1}(x,t)
\bigotimes_{x'\,{\rm even}} U_{x'\!\!,x'+1}(x',t)~.
\ee
Here, $U_{x,x+1}(x,t)$ are unitaries acting non-trivially at sites $x,x+1$ as $U(x,t)$, and as the identity on the rest of the chain. The $q^2\times q^2$ unitary matrices $U(x,t)$ are referred to as local gates. The circuit is built from two types of local gates, namely bulk gates and impurity gates, which differ in particular properties that we detail below. The impurity gates are located only along the line $x=L,~\forall t$, and the bulk gates comprise the rest of the circuit; see Fig.~\ref{fig:impurity_circuit}. We take the system to have periodic boundary conditions, i.e., $x+2L=x$. 

The bulk gates are always chosen to be dual-unitary gates, which means that they are unitary when viewed in both the time and space directions. Specifically, we have that
\be 
U(x,t)U^\dag(x,t)=\tilde{U}(x,t)\tilde{U}^\dag(x,t)=\1,~\forall~t, ~\forall~x\neq L,
\ee
where the space-time swapped gate is defined as
\begin{eqnarray}
~\!\bra{ij}\tilde{U}(x,t)\ket{kl}=~\!\bra{jl}U(x,t)\ket{ik}.
\end{eqnarray}
In addition, we focus on scenarios in which the bulk gates are charged, meaning that they are symmetric with respect to a group $\mathcal{G}$. For symmetric dual-unitary circuits and continuous groups $\mathcal{G}$,  one can show that the symmetry decomposes into independent left- and right-moving components, $\mathcal{G}=\mathcal{G}^{(l)}\otimes \mathcal{G}^{(r)}$,  which need not be the same~\cite{foligno2024nonequilibrium,bertini2020operator2}.  Here we consider the simplest possible case wherein $\mathcal{G}$  is composed of a single left-moving $U(1)$ symmetry and a single right-moving $U(1)$ symmetry, 
\be
\mathcal{G}=U(1)\otimes U(1)~,
\ee
although multiple independent $U(1)$ symmetries can be treated in a similar fashion.
These chiral $U(1)$ symmetries are generated by either a left- or right-moving charge, $Q^{(l)}$ or $Q^{(r)}$, with an integer spectrum, such that for $g_\lambda^{(\sigma)}\in \mathcal{G}^{(\sigma)}$ 
 ($\sigma=l,r$),
\be 
g_\lambda^{(\sigma)}=e^{i\lambda Q^{(\sigma)}},~\lambda\in (-\pi,\pi].
\ee
In turn, the charge can be expressed as a sum of local charge densities, each acting only on a single site. For the right-moving charge, the densities live on the even sites, 
\be 
Q^{(r)}=\sum_{x=1}^{L}Q^{(r)}_{2x}~,
\ee while the left-movers are instead supported on the odd sites,
\be 
Q^{(l)}=\sum_{x=1}^{L}Q^{(l)}_{2x-1}~.
\ee 

We assume for concreteness that the local operator $Q^{(\sigma)}_x$, which acts at site $x$,  has eigenvalues $0,1$ with degeneracy $d$, such that $2d=q$.  This allows us to express the group elements as
\begin{eqnarray}
\label{eq:group_elements}
g_\lambda^{(r)}\!\!&=&\bigotimes_{x=1}^{L}\big(\Pi^{(r)}_{0,2x}+e^{i\lambda}\Pi_{1,2x}^{(r)} \big)~,\nonumber\\
g_\lambda^{(l)}\!\!&=&\bigotimes_{x=1}^{L}\big(\Pi^{(l)}_{0,2x-1}+e^{i\lambda}\Pi_{1,2x-1}^{(l)} \big)~,
\end{eqnarray}
where $\Pi^{(\sigma)}_{0,x},~\Pi^{(\sigma)}_{1,x}$ are orthogonal projectors onto the $d$-dimensional eigenspaces of the local left- or right-moving charge densities at site $x$. They obey the completeness relation
\be \label{eq:completeness}
\1_x=\Pi^{(\sigma)}_{0,x}+\Pi^{(\sigma)}_{1,x},~~\sigma=l,r,
\ee
where $\1_x$ is the $q$-dimensional identity, and so the element $g_\lambda^{(\sigma)}$ at $\lambda=0$ is the identity operator. As the name suggests, these projectors are propagated ballistically to the left or right by the local gates, i.e., 
\begin{eqnarray}\label{eq:ballistic_charges}
U(x,t)\big[\Pi^{(r)}_{j,x}\otimes \1_{x+1}\big]U^\dagger(x,t)&=& \big[\1_{x}\otimes \Pi^{(r)}_{j,x+1}\big],\nonumber\\
     U(x,t)\big[\1_{x}\otimes \Pi_{j,x+1}^{(l)}\big]U^\dagger(x,t)&=&\big[\Pi_{j,x}^{(l)}\otimes \1_{x+1}\big],
\end{eqnarray}
for $j=0,1$. This shifts the right-moving projectors to odd sites and the left-moving ones to even sites. Taking into account the brickwork pattern of the time evolution, however, one sees that in a symmetric dual-unitary circuit the projectors are shifted by two sites in a single time step, $\mathbb{U}(t)Q_x^{(\sigma)}\mathbb{U}^\dagger(t)=Q_{x\pm 2}^{(\sigma)}$ with the plus (minus) sign for right (left) movers. Therefore, left-movers remain on odd sites and right-movers on even ones. As a consequence, 
\be\label{eq:global_charge_symmetry} 
\mathbb{U}(t)Q^{(\sigma)}\mathbb{U}^\dagger(t)=Q^{(\sigma)}~,
\ee 
in agreement with the fact that these charges generate symmetries of the dynamics. We shall refer to the local projectors onto charge sectors as solitons.

Furthermore, when the bulk gates are $\mathcal{G}$-symmetric, they can be decomposed into their action on the eigenspaces of the local charge densities as 
\be\label{eq:local_gate_decomp}
U(x,t)=\bigoplus_{j,k=0}^1U_{j,k}(x,t)~,
\ee 
where $U_{j,k}(x,t)$ is given by
\be 
 U_{j,k}(x,t)=\Pi_{k,x}^{(l)}\Pi_{j,x+1}^{(r)}U(x,t)\Pi_{j,x}^{(r)}\Pi_{k,x+1}^{(l)},
\ee 
and acts only in the $j$ eigenspace on site $x$ and the $k$ eigenspace on site $x+1$. Inside each charge block, the local gate is still dual-unitary. For $q=2$, any homogeneous circuit built from a two-site $U(1)$ symmetric unitary gate is integrable~\cite{znidaric2025intergability}, but for $q>2$ charged dual-unitary gates can exhibit chaotic behaviour as well~\cite{foligno2024nonequilibrium}. 

To complete our setup, we choose the impurity gate $U(L,t)$ to break from the properties attributed to the bulk gates. In particular, while we assume $U(L,t)$ to be unitary, it is chosen to be neither dual-unitary nor symmetric unless stated otherwise. This entails that the overall time-evolution operator $\mathbb{U}(t)$ is not dual-unitary. When the bulk gates are symmetric, while the local conservation laws in Eq.~\eqref{eq:ballistic_charges} still hold for $x\neq L$, the global symmetry in Eq.~\eqref{eq:global_charge_symmetry} is broken due to the inclusion of the impurity gate.

\textbf{Initial states:} The system is initialized in a two-site product state which we write as 
\be \label{eq:initial_state}
\ket{\Psi(0)}=\bigotimes_{x=1}^L\ket{m},\,\, 
\ket{m} \!=\sum_{i,j=1}^{q}[m]_{ij}\ket{ij}_x,
\ee
where $\ket{ij}_x=\ket{i}_{2x-1}\otimes\ket{j}_{2x}$ with $\ket{i}_y,i=1,\dots,q$ spanning the local space of the qudit at site $y$, and $m$ is a $q\times q$ matrix characterizing the initial state. We work with a special set of initial states, called charged solvable~\cite{foligno2024nonequilibrium}, which are compatible with the charged dual-unitary bulk. They are chosen to satisfy the following conditions
\begin{eqnarray}\label{eq:charge_solvable_1}
\tr_{2x-1}[\ketbra{m}{m}]&=&\sum_{i,j=1}^q [m^\dagger m]_{ij}\ket{j}\!\bra{i}_{2x}=\frac{1-\theta}{d}\Pi^{(r)}_{0}+\frac{\theta}{d}\Pi^{(r)}_{1}~,\nonumber\\
\tr_{2x}[\ketbra{m}{m}]&=&\sum_{i,j=1}^q [mm^\dagger]_{ij}\ket{i}\!\bra{j}_{2x-1}=\frac{1-\theta}{d}\Pi^{(l)}_{0}+\frac{\theta}{d}\Pi^{(l)}_{1}~,
\end{eqnarray}
with $\theta\in[0,1]$ being the expectation value of the local left- or right-moving charge density,
\begin{eqnarray} 
\theta&=&\frac{1}{L}\bra{\Psi(0)}{Q^{(r)}}\ket{\Psi(0)}=\tr [m\Pi^{(r)}_{1}m^\dagger]\nonumber\\
&=&\frac{1}{L}\bra{\Psi(0)}{Q^{(l)}}\ket{\Psi(0)}=\tr [m^\dagger\Pi^{(l)}_{1}m].
\end{eqnarray}
These states are the natural generalizations of the solvable states in uncharged dual-unitary circuits~\cite{piroli2020exact}, in that their evolution under \textit{symmetric} dual-unitary circuits admits exact analysis, enabling the analytical computation of information quantities such as entanglement entropies~\cite{foligno2024nonequilibrium,foligno2024entanglement}. They are not eigenstates of the charges $Q^{(\sigma)}$  unless $\theta=0,1$, and so the initial state generically breaks the symmetry.  
When considering non-symmetric dual-unitary dynamics, the initial states above are only solvable for $\theta=1/2$. In that case, $\ket{m}$ is a Bell state.

We note that although here we focus on charged solvable states, one can also consider product states or matrix product states that do not have this property, and still obtain analytic predictions for certain quantities~\cite{foligno2024nonequilibrium}. Generalizations of the charged solvable states in Eq.~\eqref{eq:initial_state} to more general matrix product states also exist and can be studied using the formalism developed here; however, we focus on the simplest case for clarity. 

\textbf{Quantities of interest: }Our aim is to study the dynamics of entanglement and information for the initial states described above, when evolved using our impure dual-unitary circuit. In particular, we define a subsystem $A$, which we take to consist of $2\ell$ contiguous qudits located within a distance $2r$ to the left of the impurity,
\be\label{eq:single_interval_def} 
A=(L-2\ell-2r,L-2r]~.
\ee 
The state within the subsystem is described by the reduced density matrix
\be 
\rho_A(t)=\tr_{\bar{A}}[\ketbra{\Psi(t)}{\Psi(t)}],
\ee
where $\bar{A}$ is the complement of $A$. The bipartite entanglement between $A$ and $\bar{A}$ can be measured through the R\'enyi entanglement entropy, 
\be 
S^{(n)}[\rho_A(t)]=\frac{1}{1-n}\log \tr_A[\rho^n_A(t)]
\ee 
from which one can obtain the von Neumann entanglement entropy by taking the limit $n\to 1$~\cite{horodecki2009entanglement}: 
\be S[\rho_A(t)]=\lim_{n\to 1}S^{(n)}[\rho_A(t)]~.
\ee
We shall be interested in both the R\'enyi entropy and its von Neumann limit.  

We also investigate the dynamics of the bipartite mutual information across the impurity. For this purpose, we take the subsystem to consist of two subregions $A=A_1
\cup A_2$ that lie on opposite sides of the impurity, so that 
\begin{eqnarray}\label{eq:two_intervals_def}
A_1&=&(L-2\ell-2r,L-2r]~,\nonumber\\
A_2&=&(L+2r,L+2r+2\ell]~. 
\end{eqnarray}
The R\'enyi mutual information is defined as 
\begin{eqnarray}
    \mathcal{I}^{(n)}_{A_1:A_2}(t)=S^{(n)}[\rho_{A_1}(t)]+S^{(n)}[\rho_{A_2}(t)]-S^{(n)}[\rho_{A}(t)]
\end{eqnarray}
with the  von Neumann mutual information found upon taking the replica limit, 
\be 
 \mathcal{I}_{A_1:A_2}(t)=\lim_{n\to 1} \mathcal{I}^{(n)}_{A_1:A_2}(t).
\ee 
This quantity captures both classical and quantum correlations between the two parts of subsystem $A$~\cite{groisman2005correlations}. To pinpoint the quantum entanglement between the two subsystems, we can instead look at the logarithmic negativity~\cite{VidalPRA2002,Plenio2005}. Like in the case of the entanglement entropy, the negativity can be computed using a replica trick via the R\'enyi logarithmic negativities, defined as
\begin{eqnarray}
    {\cal E}^{(n)}_{A_1:A_2}(t)=\log {\rm tr}\left[\left(\rho_A^{T_2}\right)^{n}\right],
\end{eqnarray}
where $T_2$ is the partial transpose applied to the region $A_2$. From this, the logarithmic negativity is obtained via the analytic continuation of even-integer R\'enyi negativities, namely
\begin{eqnarray}
    \mathcal{E}_{A_1:A_2}(t)=\lim_{n\to \frac{1}{2}}{\cal E}^{(2n)}_{A_1:A_2}(t).
\end{eqnarray}
   The negativity was predicted to be equivalent to $ \frac{1}{2}\mathcal{I}^{(1/2)}_{A_1:A_2}(t)$~\cite{alba2019quantum} in integrable quenches. This was later proven to hold more generally, at short times, using space-time duality~\cite{bertini2022entanglement}. 

The length scales $\ell$ and $r$, as well as the time $t$ along which we study the dynamics, are all assumed to be much smaller than $L$. This allows us to neglect boundary effects, and renders the overall system effectively infinite (which is also the regime where unperturbed dual-unitary circuits are solvable~\cite{piroli2020exact}). We also assume throughout this paper that $r\le\ell/2$, since this regime gives rise to the richest dynamics of quantum information measures~\cite{fraenkel2024entanglement}; the extension to $r>\ell/2$ is straightforward, and the qualitative behaviour can be easily inferred from the effective quasiparticle and membrane pictures that we describe below.

\textbf{Graphical notation:} We can represent our quantum circuit and the evolved state pictorially using the standard diagrammatic notation of tensor networks, which is also very convenient for performing calculations. To this end, we depict the folded, replicated local gates as
\bea\label{eq:gate_notation}
\left[U(x,t)\otimes U^{\star}(x,t)\right]^{\otimes_{\rm{r}} n}&=&
  \begin{tikzpicture}[baseline={([yshift=-0.6ex]current bounding box.center)},scale=0.65]
   \prop{0}{0}{colSt}{}
 \end{tikzpicture},\nonumber\\
 \left[U(L,t)\otimes U^{\star}(L,t)\right]^{\otimes_{\rm{r}} n}&=&
  \begin{tikzpicture}[baseline={([yshift=-0.6ex]current bounding box.center)},scale=0.65]
   \prop{0}{0}{colimp}{}
 \end{tikzpicture},
\eea
where we have denoted by $\otimes_{\rm{r}}$ the tensor product over different replicas, and where the two different copies inside a single replica, $U(x,t)$ and $U^\star(x,t)$, represent forward and backward time evolution, respectively. Occasionally, we will make the number of replicas explicit by using the notation  $\begin{tikzpicture}[baseline={([yshift=-0.6ex]current bounding box.center)},scale=0.5]
   \prop{0}{0}{colSt}{n}
 \end{tikzpicture},
  \begin{tikzpicture}[baseline={([yshift=-0.6ex]current bounding box.center)},scale=0.5]
   \prop{0}{0}{colimp}{n}
 \end{tikzpicture}$, but for the most part we use~\eqref{eq:gate_notation}. 
The folded, replicated, initial state is represented as
 \begin{align}
\left(\ket{m}\otimes\ket{m^\star}\right)^{\otimes_{\rm r}n}=\begin{tikzpicture}[baseline={([yshift=-0.6ex]current bounding box.center)},scale=0.65]
  \tgridLine{0}{0}{-0.25}{0.25}
  \tgridLine{1}{0}{1.25}{0.25}
  \istate{0}{0}{colSt}~
  \end{tikzpicture}~,~
\left(\bra{m}\otimes\bra{m^\star}\right)^{\otimes_{\rm r}n}=\begin{tikzpicture}[baseline={([yshift=-0.6ex]current bounding box.center)},scale=0.65]
  \tgridLine{0}{0}{-0.25}{-0.25}
  \tgridLine{1}{0}{1.25}{-0.25}
  \istateC{0}{0}{colSt}~
  \end{tikzpicture}~.
 \end{align}   
 
 We also introduce three special states in the replicated Hilbert space and their corresponding graphical notation. The first, which is denoted by an empty circle, couples together the forward and backward time evolution within each replica, 
\be \label{eq:circle}
\left[\sum_{i=1}^q\ket{i}_x\otimes\ket{i}_x\right]^{\otimes_{\rm r} n}\equiv\begin{tikzpicture}
[baseline={([yshift=-0.6ex]current bounding box.center)},scale=0.65]
   \gridLine{0}{0}{0}{0.75}
   \circle{0}{0}
\end{tikzpicture}~,~\left[\sum_{i=1}^q\bra{i}_x\otimes\bra{i}_x\right]^{\otimes_{\rm r} n}\equiv\begin{tikzpicture}
[baseline={([yshift=-0.6ex]current bounding box.center)},scale=0.65]
   \gridLine{0}{0}{0}{-0.75}
   \circle{0}{0}
\end{tikzpicture}.
\ee
This state will be used to implement a partial trace on a single site. The second, denoted with an empty square,  instead couples the forward time evolution on one replica to the backward time evolution of the previous replica,
\begin{eqnarray} \label{eq:square}
\sum_{i_1,\dots,i_n=1}^q\left[\ket{i_1}_x\otimes\ket{i_2}_x\right]\otimes_{\rm r}\left[\ket{i_2}_x\otimes\ket{i_3}_x\right]\otimes_{\rm r}\dots \otimes_{\rm r}\left[\ket{i_n}_x\otimes\ket{i_1}_x\right]&\equiv&\begin{tikzpicture}
[baseline={([yshift=-0.6ex]current bounding box.center)},scale=0.65]
   \gridLine{0}{0}{0}{0.75}
   \square{0}{0}
\end{tikzpicture}~,\\\sum_{i_1,\dots,i_n=1}^q\left[\bra{i_1}_x\otimes\bra{i_2}_x\right]\otimes_{\rm r}\left[\bra{i_2}_x\otimes\bra{i_3}_x\right]\otimes_{\rm r}\dots \otimes_{\rm r}\left[\bra{i_n}_x\otimes\bra{i_1}_x\right]&\equiv&\begin{tikzpicture}
[baseline={([yshift=-0.6ex]current bounding box.center)},scale=0.65]
   \gridLine{0}{0}{0}{-0.75}
   \square{0}{0}
\end{tikzpicture}~.
\end{eqnarray}
This will be used to represent matrix multiplication of density matrices followed by a trace. Lastly, we introduce the triangle state,
\begin{equation}    \sum_{i_1,\dots,i_n=1}^q\left[\bra{i_1}_x\otimes\bra{i_n}_x\right]\otimes_{\rm r}\left[\bra{i_2}_x\otimes\bra{i_1}_x\right]\otimes_{\rm r}\left[\bra{i_3}_x\otimes\bra{i_2}_x\right]\otimes_{\rm r}\dots \otimes_{\rm r}\left[\bra{i_n}_x\otimes\bra{i_{n-1}}_x\right]\equiv\begin{tikzpicture}
[baseline={([yshift=-0.6ex]current bounding box.center)},scale=0.65]
   \gridLine{0}{0}{0}{-0.75}
   \triangleDown{0}{0}
\end{tikzpicture}~
\end{equation}
along with its ket version, $\begin{tikzpicture}
[baseline={([yshift=-0.6ex]current bounding box.center)},scale=0.45]
   \gridLine{0}{0}{0}{-0.75}
   \triangleUp{0}{-0.75}
\end{tikzpicture}$, which is defined similarly. This state, like the square state, couples different replicas, but now couples forward time evolution in one replica to backward time evolution in the next replica.  This last state is used to depict matrix multiplication after a transposition. It will be employed when considering the dynamics of the negativity, while for the entanglement entropy and mutual information, only the first two states are needed. 

Using these three states, we introduce the basic diagrammatic rules that will be used in our calculations, which we derive from the unitarity of the local gate. These are given by
 \begin{eqnarray}
\nonumber
 \begin{tikzpicture}[baseline={([yshift=-0.6ex]current bounding box.center)},scale=0.65]
   \prop{0}{0}{colSt}{}
   \circle{0.5}{0.5}
    \circle{-0.5}{0.5}
 \end{tikzpicture}=
\begin{tikzpicture}[baseline={([yshift=-0.6ex]current bounding box.center)},scale=0.65]
    \gridLine{0}{0}{0}{0.75}
   \circle{0}{0.75}
    \gridLine{.5}{0}{.5}{0.75}
   \circle{.5}{0.75}
 \end{tikzpicture}\quad,\quad
 \begin{tikzpicture}[baseline={([yshift=-0.6ex]current bounding box.center)},scale=0.65]
   \prop{0}{0}{colSt}{}
   \square{0.5}{0.5}
    \square{-0.5}{0.5}
 \end{tikzpicture}=
 \begin{tikzpicture}[baseline={([yshift=-0.6ex]current bounding box.center)},scale=0.65]
    \gridLine{0}{0}{0}{0.75}
   \square{0}{0.75}
    \gridLine{.5}{0}{.5}{0.75}
   \square{.5}{0.75}
 \end{tikzpicture}
 \quad,\quad
  \begin{tikzpicture}[baseline={([yshift=-0.6ex]current bounding box.center)},scale=0.65]
   \prop{0}{0}{colSt}{}
   \triangleDown{0.5}{0.5}
    \triangleDown{-0.5}{0.5}
 \end{tikzpicture}=
\begin{tikzpicture}[baseline={([yshift=-0.6ex]current bounding box.center)},scale=0.65]
    \gridLine{0}{0}{0}{0.75}
   \triangleDown{0}{0.75}
    \gridLine{.5}{0}{.5}{0.75}
   \triangleDown{.5}{0.75}
 \end{tikzpicture}\quad,\\\label{eq:local_gate_rels_1}
 \begin{tikzpicture}[baseline={([yshift=-0.6ex]current bounding box.center)},scale=0.65]
   \prop{0}{0}{colSt}{}
   \circle{0.5}{-0.5}
    \circle{-0.5}{-0.5}
 \end{tikzpicture}=
\begin{tikzpicture}[baseline={([yshift=-0.6ex]current bounding box.center)},scale=0.65]
    \gridLine{0}{0}{0}{-0.75}
   \circle{0}{-0.75}
    \gridLine{.5}{0}{.5}{-0.75}
   \circle{.5}{-0.75}
 \end{tikzpicture}\quad,\quad\begin{tikzpicture}[baseline={([yshift=-0.6ex]current bounding box.center)},scale=0.65]
   \prop{0}{0}{colSt}{}
   \square{0.5}{-0.5}
    \square{-0.5}{-0.5}
 \end{tikzpicture}=
\begin{tikzpicture}[baseline={([yshift=-0.6ex]current bounding box.center)},scale=0.65]
    \gridLine{0}{0}{0}{-0.75}
   \square{0}{-0.75}
    \gridLine{.5}{0}{.5}{-0.75}
   \square{.5}{-0.75}
 \end{tikzpicture}\quad,\quad\begin{tikzpicture}[baseline={([yshift=-0.6ex]current bounding box.center)},scale=0.65]
   \prop{0}{0}{colSt}{}
   \triangleUp{0.5}{-0.5}
    \triangleUp{-0.5}{-0.5}
 \end{tikzpicture}=
\begin{tikzpicture}[baseline={([yshift=-0.6ex]current bounding box.center)},scale=0.65]
    \gridLine{0}{0}{0}{-0.75}
   \triangleUp{0}{-0.75}
    \gridLine{.5}{0}{.5}{-0.75}
   \triangleUp{.5}{-0.75}
 \end{tikzpicture}\quad.
 \end{eqnarray}
Similarly, for a unitary impurity gate,
 \begin{eqnarray}
\nonumber
 \begin{tikzpicture}[baseline={([yshift=-0.6ex]current bounding box.center)},scale=0.65]
   \prop{0}{0}{colimp}{}
   \circle{0.5}{0.5}
    \circle{-0.5}{0.5}
 \end{tikzpicture}=
\begin{tikzpicture}[baseline={([yshift=-0.6ex]current bounding box.center)},scale=0.65]
    \gridLine{0}{0}{0}{0.75}
   \circle{0}{0.75}
    \gridLine{.5}{0}{.5}{0.75}
   \circle{.5}{0.75}
 \end{tikzpicture}\quad,\quad
 \begin{tikzpicture}[baseline={([yshift=-0.6ex]current bounding box.center)},scale=0.65]
   \prop{0}{0}{colimp}{}
   \square{0.5}{0.5}
    \square{-0.5}{0.5}
 \end{tikzpicture}=
 \begin{tikzpicture}[baseline={([yshift=-0.6ex]current bounding box.center)},scale=0.65]
    \gridLine{0}{0}{0}{0.75}
   \square{0}{0.75}
    \gridLine{.5}{0}{.5}{0.75}
   \square{.5}{0.75}
 \end{tikzpicture}
 \quad,\quad
  \begin{tikzpicture}[baseline={([yshift=-0.6ex]current bounding box.center)},scale=0.65]
   \prop{0}{0}{colimp}{}
   \triangleDown{0.5}{0.5}
    \triangleDown{-0.5}{0.5}
 \end{tikzpicture}=
\begin{tikzpicture}[baseline={([yshift=-0.6ex]current bounding box.center)},scale=0.65]
    \gridLine{0}{0}{0}{0.75}
   \triangleDown{0}{0.75}
    \gridLine{.5}{0}{.5}{0.75}
   \triangleDown{.5}{0.75}
 \end{tikzpicture}\quad,\\\label{eq:local_gate_rels_2}
 \begin{tikzpicture}[baseline={([yshift=-0.6ex]current bounding box.center)},scale=0.65]
   \prop{0}{0}{colimp}{}
   \circle{0.5}{-0.5}
    \circle{-0.5}{-0.5}
 \end{tikzpicture}=
\begin{tikzpicture}[baseline={([yshift=-0.6ex]current bounding box.center)},scale=0.65]
    \gridLine{0}{0}{0}{-0.75}
   \circle{0}{-0.75}
    \gridLine{.5}{0}{.5}{-0.75}
   \circle{.5}{-0.75}
 \end{tikzpicture}\quad,\quad\begin{tikzpicture}[baseline={([yshift=-0.6ex]current bounding box.center)},scale=0.65]
   \prop{0}{0}{colimp}{}
   \square{0.5}{-0.5}
    \square{-0.5}{-0.5}
 \end{tikzpicture}=
\begin{tikzpicture}[baseline={([yshift=-0.6ex]current bounding box.center)},scale=0.65]
    \gridLine{0}{0}{0}{-0.75}
   \square{0}{-0.75}
    \gridLine{.5}{0}{.5}{-0.75}
   \square{.5}{-0.75}
 \end{tikzpicture}\quad,\quad\begin{tikzpicture}[baseline={([yshift=-0.6ex]current bounding box.center)},scale=0.65]
   \prop{0}{0}{colimp}{}
   \triangleUp{0.5}{-0.5}
    \triangleUp{-0.5}{-0.5}
 \end{tikzpicture}=
\begin{tikzpicture}[baseline={([yshift=-0.6ex]current bounding box.center)},scale=0.65]
    \gridLine{0}{0}{0}{-0.75}
   \triangleUp{0}{-0.75}
    \gridLine{.5}{0}{.5}{-0.75}
   \triangleUp{.5}{-0.75}
 \end{tikzpicture}\quad.
 \end{eqnarray}
For the bulk gates, we have the additional dual-unitarity conditions
 \begin{eqnarray}\nonumber
 \begin{tikzpicture}[baseline={([yshift=-0.6ex]current bounding box.center)},scale=0.65]
   \prop{0}{0}{colSt}{}
   \circle{-0.5}{-0.5}
    \circle{-0.5}{0.5}
 \end{tikzpicture}=
\begin{tikzpicture}[baseline={([yshift=-0.6ex]current bounding box.center)},scale=0.65]
  \circle{-0.5}{-0.5}
    \circle{-0.5}{0.5}
    \gridLine{-0.38}{-0.5}{0.3}{-0.5}
      \gridLine{-0.38}{0.5}{0.3}{0.5}
 \end{tikzpicture}\quad,\quad
 \begin{tikzpicture}[baseline={([yshift=-0.6ex]current bounding box.center)},scale=0.65]
   \prop{0}{0}{colSt}{}
   \square{-0.5}{-0.5}
    \square{-0.5}{0.5}
 \end{tikzpicture}=
 \begin{tikzpicture}[baseline={([yshift=-0.6ex]current bounding box.center)},scale=0.65]
      \square{-0.5}{-0.5}
    \square{-0.5}{0.5}
        \gridLine{-0.38}{-0.5}{0.3}{-0.5}
      \gridLine{-0.38}{0.5}{0.3}{0.5}
 \end{tikzpicture}
 \quad,\quad
 \begin{tikzpicture}[baseline={([yshift=-0.6ex]current bounding box.center)},scale=0.65]
   \prop{0}{0}{colSt}{}
   \triangleUp{-0.5}{-0.5}
    \triangleDown{-0.5}{0.5}
 \end{tikzpicture}=
 \begin{tikzpicture}[baseline={([yshift=-0.6ex]current bounding box.center)},scale=0.65]
      \triangleUp{-0.5}{-0.5}
    \triangleDown{-0.5}{0.5}
        \gridLine{-0.38}{-0.5}{0.3}{-0.5}
      \gridLine{-0.38}{0.5}{0.3}{0.5}
 \end{tikzpicture}\quad,\\\label{eq:local_gate_rels_3}
\begin{tikzpicture}[baseline={([yshift=-0.6ex]current bounding box.center)},scale=0.65]
   \prop{0}{0}{colSt}{}
   \circle{0.5}{-0.5}
    \circle{0.5}{0.5}
 \end{tikzpicture}=
\begin{tikzpicture}[baseline={([yshift=-0.6ex]current bounding box.center)},scale=0.65]
  \circle{0.5}{-0.5}
    \circle{0.5}{0.5}
    \gridLine{-0.38}{-0.5}{0.3}{-0.5}
      \gridLine{-0.38}{0.5}{0.3}{0.5}
 \end{tikzpicture}\quad,\quad
 \begin{tikzpicture}[baseline={([yshift=-0.6ex]current bounding box.center)},scale=0.65]
   \prop{0}{0}{colSt}{}
   \square{0.5}{-0.5}
    \square{0.5}{0.5}
 \end{tikzpicture}=
 \begin{tikzpicture}[baseline={([yshift=-0.6ex]current bounding box.center)},scale=0.65]
      \square{0.5}{-0.5}
    \square{0.5}{0.5}
        \gridLine{-0.38}{-0.5}{0.3}{-0.5}
      \gridLine{-0.38}{0.5}{0.3}{0.5}
 \end{tikzpicture}\quad,\quad  \begin{tikzpicture}[baseline={([yshift=-0.6ex]current bounding box.center)},scale=0.65]
   \prop{0}{0}{colSt}{}
   \triangleUp{0.5}{-0.5}
    \triangleDown{0.5}{0.5}
 \end{tikzpicture}=
 \begin{tikzpicture}[baseline={([yshift=-0.6ex]current bounding box.center)},scale=0.65]
      \triangleUp{0.5}{-0.5}
    \triangleDown{0.5}{0.5}
        \gridLine{-0.38}{-0.5}{0.3}{-0.5}
      \gridLine{-0.38}{0.5}{0.3}{0.5}
 \end{tikzpicture}\quad.
 \end{eqnarray}
To depict the charge structure of the dynamics, we also introduce a representation for the solitons,
\be \label{eq:soliton_proj}
[\Pi^{(r)}_{j_i}\otimes \1]^{\otimes_{\rm r}n}=\begin{tikzpicture}[baseline={([yshift=-0.6ex]current bounding box.center)},scale=0.65]
  \tgridLine{0.2}{-0.25}{0.2}{0.6}
  \circledg{0.2}{0.15};
  \end{tikzpicture}\quad,\quad [\Pi^{(l)}_{j_i}\otimes \1]^{\otimes_{\rm r}n}=\begin{tikzpicture}[baseline={([yshift=-0.6ex]current bounding box.center)},scale=0.65]
  \tgridLine{0.2}{-0.25}{0.2}{0.6}
 \circlelg{0.2}{0.15};
  \end{tikzpicture}\quad,
\ee 
where $j_i=0,1$ and $i=1,\dots, n$. The solitons can be different in each replica, i.e., $ j_i\neq j_{i'}$, but they only act on the forward copy. We shall also use the same symbols, but with different colours, to denote linear combinations of solitons. We shall explicitly specify the meaning when we do so and reserve certain colours for particular combinations. In terms of these notations, the ballistic propagation of the charges (see Eq.~\eqref{eq:ballistic_charges}) is expressed as
\be 
 \begin{tikzpicture}[baseline={([yshift=-0.6ex]current bounding box.center)},scale=0.65]
   \prop{0}{0}{colSt}{}
   \tgridLine{0.5}{0.5}{0.75}{0.75}
     \tgridLine{-0.5}{-0.5}{-0.75}{-0.75}
     \tgridLine{0.5}{0.5}{0.75}{0.75}
     \tgridLine{-0.5}{0.5}{-0.75}{0.75}
     \tgridLine{0.5}{0.5}{0.75}{0.75}
     \tgridLine{0.5}{-0.5}{0.75}{-0.75}
    \circledg{-.5}{-.5};
 \end{tikzpicture}= \begin{tikzpicture}[baseline={([yshift=-0.6ex]current bounding box.center)},scale=0.65]
   \prop{0}{0}{colSt}{}
     \tgridLine{-0.5}{-0.5}{-0.75}{-0.75}
     \tgridLine{0.5}{0.5}{0.75}{0.75}
     \tgridLine{-0.5}{0.5}{-0.75}{0.75}
     \tgridLine{0.5}{0.5}{0.75}{0.75}
     \tgridLine{0.5}{-0.5}{0.75}{-0.75}
    \circledg{.5}{.5}
    ;
 \end{tikzpicture}\quad,\quad
  \begin{tikzpicture}[baseline={([yshift=-0.6ex]current bounding box.center)},scale=0.65]
   \prop{0}{0}{colSt}{}
   \tgridLine{0.5}{0.5}{0.75}{0.75}
     \tgridLine{-0.5}{-0.5}{-0.75}{-0.75}
     \tgridLine{0.5}{0.5}{0.75}{0.75}
     \tgridLine{-0.5}{0.5}{-0.75}{0.75}
     \tgridLine{0.5}{0.5}{0.75}{0.75}
     \tgridLine{0.5}{-0.5}{0.75}{-0.75}
    \circlelg{.5}{-.5};
 \end{tikzpicture}= \begin{tikzpicture}[baseline={([yshift=-0.6ex]current bounding box.center)},scale=0.65]
   \prop{0}{0}{colSt}{}
     \tgridLine{-0.5}{-0.5}{-0.75}{-0.75}
     \tgridLine{0.5}{0.5}{0.75}{0.75}
     \tgridLine{-0.5}{0.5}{-0.75}{0.75}
     \tgridLine{0.5}{0.5}{0.75}{0.75}
     \tgridLine{0.5}{-0.5}{0.75}{-0.75}
    \circlelg{-.5}{.5}
    ;
 \end{tikzpicture}~,
\ee
which are also valid for any linear combination of the solitons. 

The charge solvability condition of the initial states is depicted as 
\begin{eqnarray}\label{eq:charge_solvable}
 \begin{tikzpicture}[baseline={([yshift=-0.6ex]current bounding box.center)},scale=0.65]
 \tgridLine{1}{0}{1.25}{0.25}
  \istate{0}{0}{colSt}
    \circle{-0.12}{0.12}
\end{tikzpicture}=\begin{tikzpicture}[baseline={([yshift=-0.6ex]current bounding box.center)},scale=0.65]
   \tgridLine{0}{0}{0.65}{0.65}
  \circleb{0.35}{0.35}
  \circle{0}{0};
;\end{tikzpicture}\quad,\quad\begin{tikzpicture}[baseline={([yshift=-0.6ex]current bounding box.center)},scale=0.65]
 \tgridLine{0}{0}{-0.25}{0.25}
  \istate{0}{0}{colSt}
    \circle{1.12}{0.12}
\end{tikzpicture}=\begin{tikzpicture}[baseline={([yshift=-0.6ex]current bounding box.center)},scale=0.65]
    \tgridLine{0}{0}{-0.65}{0.65}
   \circler{-0.35}{0.35}
  \circle{0.0}{0}
;\end{tikzpicture}
~,
 \end{eqnarray}
with similar relations holding when the square or triangle states are used instead of the circle. In~\eqref{eq:charge_solvable} the projectors on the right-hand side are the specific linear combinations that come from the initial state given in~\eqref{eq:charge_solvable_1}.  To be explicit, on the right-hand side of these equations we mean
\begin{eqnarray}\label{eq:charge_structure_intial}
\begin{tikzpicture}[baseline={([yshift=-0.6ex]current bounding box.center)},scale=0.65]
   \tgridLine{0}{0}{0.65}{0.65}
  \circleb{0.35}{0.35}
  \circle{0}{0};
;\end{tikzpicture}&=&\Big[\frac{1-\theta}{d}\Pi^{(r)}_{0}+\frac{\theta}{d}\Pi^{(r)}_{1}\Big]^{\otimes_{\rm r}n}~,\nonumber\\
\begin{tikzpicture}[baseline={([yshift=-0.6ex]current bounding box.center)},scale=0.65]
    \tgridLine{0}{0}{-0.65}{0.65}
   \circler{-0.35}{0.35}
  \circle{0.0}{0}
;\end{tikzpicture}&=&\Big[\frac{1-\theta}{d}\Pi^{(l)}_{0}+\frac{\theta}{d}\Pi^{(l)}_{1}\Big]^{\otimes_{\rm r}n}~.
\end{eqnarray}
We reserve the blue and red colouring for these special combinations of right- (blue) and left- (red) moving solitons. In Secs.~\ref{sec:ent_uncharged} and \ref{sec:Two_Intervals}, where we study the dynamics without a charge-conserving bulk, we restrict the initial state to the value $\theta=1/2$, and the soliton that is produced is merely $\1 /q$. In this case, we omit the symbol for the soliton and have instead the solvability conditions $
 \begin{tikzpicture}[baseline={([yshift=-0.6ex]current bounding box.center)},scale=0.5]
 \tgridLine{1}{0}{1.25}{0.25}
  \istate{0}{0}{colSt}
    \circle{-0.12}{0.12}
\end{tikzpicture}=\frac{1}{q^n}\begin{tikzpicture}[baseline={([yshift=-0.6ex]current bounding box.center)},scale=0.5]
   \tgridLine{0}{0}{0.65}{0.65}
  \circle{0}{0};
;\end{tikzpicture}$, $\begin{tikzpicture}[baseline={([yshift=-0.6ex]current bounding box.center)},scale=0.5]
  \tgridLine{1}{0}{1.25}{0.25}
  \istate{0}{0}{colSt}
    \square{-0.15}{0.15}
\end{tikzpicture}=\frac{1}{q^n}\begin{tikzpicture}[baseline={([yshift=-0.6ex]current bounding box.center)},scale=0.5]
  \tgridLine{0}{0}{0.65}{0.65}
  \square{0}{0};
\end{tikzpicture}$, and $\begin{tikzpicture}[baseline={([yshift=-0.6ex]current bounding box.center)},scale=0.5]
  \tgridLine{1}{0}{1.25}{0.25}
  \istate{0}{0}{colSt}
    \triangleDown{-0.15}{0.15}
\end{tikzpicture}=\frac{1}{q^n}\begin{tikzpicture}[baseline={([yshift=-0.6ex]current bounding box.center)},scale=0.5]
  \tgridLine{0}{0}{0.65}{0.65}
  \triangleDown{0}{0};
\end{tikzpicture}$, as well as the other three diagrams obtained via reflection about the vertical axis. 

The normalization of the initial state amounts to the rule 
\be 
1=\begin{tikzpicture}[baseline={([yshift=-0.6ex]current bounding box.center)},scale=0.65]
   \tgridLine{0}{0}{1}{0}
  \circleb{0.5}{0}
  \circle{0}{0};
  \circle{1}{0};
;\end{tikzpicture}=
\begin{tikzpicture}[baseline={([yshift=-0.6ex]current bounding box.center)},scale=0.65]
   \tgridLine{0}{0}{1}{0}
  \circler{0.5}{0}
  \circle{0}{0};
  \circle{1}{0};
;\end{tikzpicture}
=
\begin{tikzpicture}[baseline={([yshift=-0.6ex]current bounding box.center)},scale=0.65]
   \tgridLine{0}{0}{1}{0}
  \circler{0.5}{0}
  \square{0}{0};
  \square{1}{0};
;\end{tikzpicture}=
\begin{tikzpicture}[baseline={([yshift=-0.6ex]current bounding box.center)},scale=0.65]
   \tgridLine{0}{0}{1}{0}
  \circleb{0.5}{0}
  \square{0}{0};
  \square{1}{0};
;\end{tikzpicture}~,
\ee 
while another useful expression involves the contraction of the initial state with a circle state and a square state,
\begin{eqnarray}\nonumber
\begin{tikzpicture}[baseline={([yshift=-0.6ex]current bounding box.center)},scale=0.65]
  \istate{0}{0}{colSt}
    \circle{-0.12}{0.12}
    \square{1.12}{0.12}
\end{tikzpicture}&=&\begin{tikzpicture}[baseline={([yshift=-0.6ex]current bounding box.center)},scale=0.65]
   \tgridLine{0}{0}{1}{0}
  \circleb{0.5}{0}
  \circle{0}{0};
  \square{1}{0};
;\end{tikzpicture}\\\nonumber
&=&\frac{(1-\theta)^{n}}{d^{n-1}}+\frac{\theta^{n}}{d^{n-1}}\\\label{eq:local_renyi_diagram}
&=&\begin{tikzpicture}[baseline={([yshift=-0.6ex]current bounding box.center)},scale=0.65]
   \tgridLine{0}{0}{1}{0}
  \circler{0.5}{0}
  \square{0}{0};
  \circle{1}{0};
;\end{tikzpicture}=\begin{tikzpicture}[baseline={([yshift=-0.6ex]current bounding box.center)},scale=0.65]
  \istate{0}{0}{colSt}
    \square{-0.12}{0.12}
    \circle{1.12}{0.12}
\end{tikzpicture}~.
\end{eqnarray}
We will see below that this quantity is related to the R\'enyi entropy of the subsystem.  Lastly, we have the contractions involving the triangle state, restricting to even values of $n$. For these, we observe that the contraction of a circle and a triangle through a soliton is given by
\begin{eqnarray}\label{eq:negativity_1}
   \begin{tikzpicture}[baseline={([yshift=-0.6ex]current bounding box.center)},scale=0.65]
   \tgridLine{0}{0}{1}{0}
  \circleb{0.5}{0}
  \circle{0}{0};
  \triangleDown{1}{0};\end{tikzpicture}=\begin{tikzpicture}[baseline={([yshift=-0.6ex]current bounding box.center)},scale=0.65]
   \tgridLine{0}{0}{1}{0}
  \circler{0.5}{0}
  \circle{0}{0};
  \triangleDown{1}{0};
;\end{tikzpicture}=\frac{(1-\theta)^{n}}{d^{n-1}}+\frac{\theta^{n}}{d^{n-1}},
\end{eqnarray}
whereas the contraction between a square and a triangle is given by
\begin{eqnarray}\label{eq:negativity_2}
\begin{tikzpicture}[baseline={([yshift=-0.6ex]current bounding box.center)},scale=0.65]
   \tgridLine{0}{0}{1}{0}
  \circleb{0.5}{0}
  \square{0}{0};
  \triangleDown{1}{0};
;\end{tikzpicture}=
\begin{tikzpicture}[baseline={([yshift=-0.6ex]current bounding box.center)},scale=0.65]
   \tgridLine{0}{0}{1}{0}
  \circler{0.5}{0}
  \square{0}{0};
  \triangleDown{1}{0};
  \end{tikzpicture}  = \left[\frac{\theta^{n/2}+(1-\theta)^{n/2}}{d^{n/2-1}}\right]^2.
\end{eqnarray}
 These properties will be used when determining the logarithmic negativity. 

\textbf{Impurity gates:} While our analytical results will always be formulated for an arbitrary choice of the impurity gate, we will illustrate them throughout the paper using two specific families of impurity gates. The first family is parametrized by $c\in{\mathbb R}$, and defined as
\begin{equation}\label{eq:ISWAP_impurity_def}
    U_{\rm Id}(c)=\frac{1}{\sqrt{1+c^2}}\left(P+ic\1\right),
\end{equation}
where $P$ is the swap gate. This family interpolates between the fully transparent dual-unitary $P$ ($c=0$) and the fully reflective identity gate ($c\to\infty$). This example was already considered in Ref.~\cite{fraenkel2024entanglement} in the context of general dual-unitary circuits without any charge conservation. Note that $U_{\rm Id}(c)$ conserves the global $U(1)$ symmetry, but nevertheless breaks the chiral $U(1)$ symmetries that emerge in the dual-unitary setting.

We will also consider a family of impurity gates that explicitly break the global $U(1)$ symmetry. These gates, which will be parametrized by $\gamma\in\left[0,1\right]$ and denoted by $U_{\rm XY}(\gamma)$, are defined through their action on two-qudit standard basis states. It will be convenient to express this action by writing the single-qudit basis states as $\ket{j,k}\equiv\ket{jd+k}$, where $j\in\left\{0,1\right\}$ (designating the charge sector) and $k\in\left\{1,\ldots,d\right\}$. The gate $U_{\rm XY}(\gamma)$ is then defined as follows:
\begin{equation}\label{eq:XY_impurity_def}
    U_{\rm XY}(\gamma)\ket{j_1,k_1}\ket{j_2,k_2} = \sum_{j_1',j_2'=0}^1\left(e^{i\frac{\pi}{4}\left[(1+\gamma)X^{\otimes 2}+(1-\gamma)Y^{\otimes 2}\right]}\right)_{(j_1,j_2),(j_1',j_2')}\ket{j_1',k_2}\ket{j_2',k_1}.
\end{equation}
That is, it acts as a two-qubit XY gate on the charge labels $j_1$ and $j_2$ and swaps between the internal labels $k_1$ and $k_2$. The gate $U_{\rm XY}(\gamma)$ is $U(1)$-symmetric and dual-unitary for $\gamma=0$, but it violates both properties for any $\gamma\neq 0$.

\begin{figure}\centering
\begin{tikzpicture}[baseline={([yshift=0.6ex]current bounding box.center)},scale=0.5]
    \foreach \x in {-5,-3,...,5}{\prop{\x}{-3}{colSt}{}}
     \foreach \x in {7,9}{\prop{\x}{-3}{colSt}{}}
      \foreach \x in {-5,-3,...,5}{\prop{\x}{-1}{colSt}{}}
           \foreach \x in {7,9}{\prop{\x}{-1}{colSt}{}}
        \foreach \x in {-5,-3,...,5}{\prop{\x}{1}{colSt}{}}
             \foreach \x in {7,9}{\prop{\x}{1}{colSt}{}}
          \foreach \x in {-5,-3,...,5}{\prop{\x}{3}{colSt}{}}
               \foreach \x in {7,9}{\prop{\x}{3}{colSt}{}}
    \foreach \x in {-6,-4,...,6}{\prop{\x}{-4}{colSt}{}}
         \foreach \x in {8,10}{\prop{\x}{-4}{colSt}{}}
     \foreach \x in {-6,-4,...,6}{\prop{\x}{-2}{colSt}{}}
          \foreach \x in {8,10}{\prop{\x}{-2}{colSt}{}}
    \foreach \x in {-6,-4,...,6}{\prop{\x}{0}{colSt}{}} 
         \foreach \x in {8,10}{\prop{\x}{0}{colSt}{}}
     \foreach \x in {-6,-4,...,6}{\prop{\x}{2}{colSt}{}}
         \foreach \x in {8,10}{\prop{\x}{2}{colSt}{}}
      \foreach \x in {-4,-2,0,2} {\prop{6}{\x}{colimp}{}}
    \foreach \x in {-5.5,-3.5,...,5.5}{\istate{\x}{-4.5}{colSt}}
        \foreach \x in {6.5,8.5}{\istate{\x}{-4.5}{colSt}}
    \foreach \x in {-3,-1,...,1}{\tgridLine{\x+0.75}{3.75}{\x+0.5}{3.5}
    \square{\x+0.75}{3.75}
    }
    \foreach \x in {-4,-2,...,1}{\tgridLine{\x+0.25}{3.75}{\x+.5}{3.5}\square{\x+0.25}{3.75}}
     \foreach \x in {3,5,...,9}{\tgridLine{\x+0.75}{3.75}{\x+0.5}{3.5}
     \circle{\x+0.75}{3.75}}
     \foreach \x in {2,4,...,8}{\tgridLine{\x+0.25}{3.75}{\x+.5}{3.5}\circle{\x+0.25}{3.75}}
       \foreach \x in {-5}{\tgridLine{\x+0.75}{3.75}{\x+0.5}{3.5}\circle{\x+0.75}{3.75}}
       \foreach \x in {-6}{\tgridLine{\x+0.25}{3.75}{\x+.5}{3.5}\circle{\x+0.25}{3.75}}
 \draw[semithick,decorate,decoration={brace}] (-3.85,4.25) -- (1.85,4.25) node[midway,yshift=7.5pt] {$2\ell$};
  \draw[semithick,decorate,decoration={brace}] (2.15,4.25) -- (5.85,4.25) node[midway,yshift=7.5pt] {$2r$};
  \draw[semithick,decorate,decoration={brace}] (11,3) -- (11,-4.5) node[midway,xshift=12.5pt,rotate= 270] {$2t$};
  \end{tikzpicture}
  \caption{\label{fig:rho_circuit}The diagrammatic representation of $\tr[\rho^n_A(t)]$ with $\ell=3$, $r=2$, and $t=4$, for a single interval to the left of the impurity.  Orange boxes denote bulk dual-unitary operators while green represents the impurity. Orange circles at the bottom denote the initial state, and empty circles and squares at the top denote the boundary conditions imposed by the different orders of trace.  }
\end{figure}
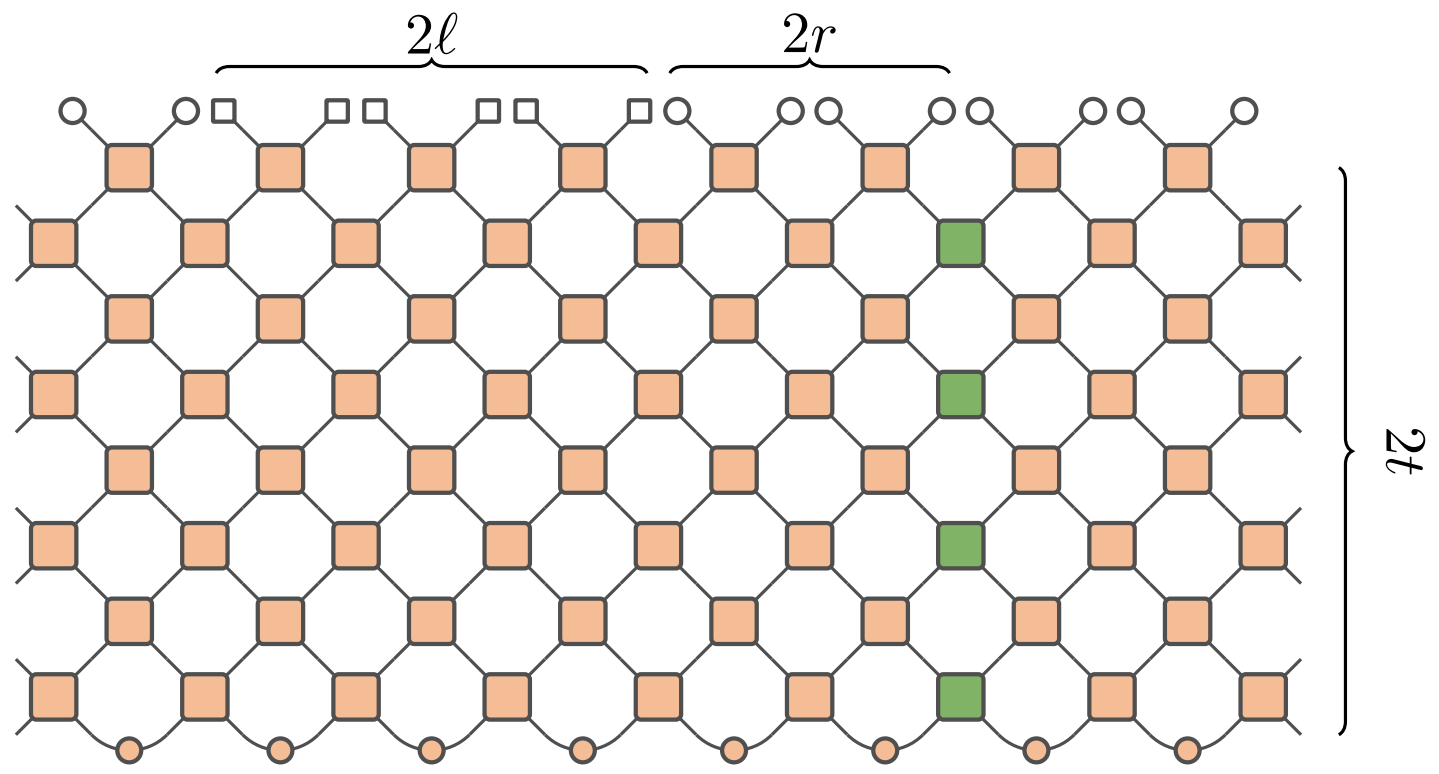

\section{Entanglement dynamics in an uncharged circuit}\label{sec:ent_uncharged}
The entanglement entropy for a single connected subsystem $A=(L-2\ell-2r,L-2r]$ in an uncharged circuit was previously studied in Ref.~\cite{fraenkel2024entanglement}. In this section, we briefly review those results with a view to extending them in the subsequent sections. The starting point of the calculation is $\tr[\rho^n_A(t)]$, which can be depicted graphically using the notation of the previous section as shown in Fig.~\ref{fig:rho_circuit}. We break the calculation up into early, intermediate, and late time regimes. The first  and last of these are universal, meaning that the result is independent of the specifics of the bulk circuit, while the intermediate time regime is not universal. For the intermediate time regime we employ two effective theories, the quasiparticle and entanglement membrane pictures, which will also be used in the charged case. Below we explain the details of these two effective theories.

\subsection{Universal behaviour}
The early time regime is defined to be $t\leq \ell/2$, that is, times for which the two edges of the subsystem are not in causal contact. Utilizing the local gate unitarity relations, Eqs.~\eqref{eq:local_gate_rels_1} and~\eqref{eq:local_gate_rels_2}, we can reduce the circuit shown in Fig.~\ref{fig:rho_circuit}  to a product of two disconnected pieces as follows: 
\begin{align}
\label{eq:D_def}
\tr[\rho^n_A(t)]=\hspace{-0.8cm}\begin{tikzpicture}[baseline={([yshift=1.2ex]current bounding box.center)},scale=0.5]
{\prop{2}{2}{colSt}{}}
  \foreach \x in {1,3}{\prop{\x}{1}{colSt}{}}   
    \foreach \x in {0,2,...,4}{\prop{\x}{0}{colSt}{}}
    \foreach \x in {-1,1,...,5}{\prop{\x}{-1}{colSt}{}}
    \foreach \x in {-2,0,...,6}{\prop{\x}{-2}{colSt}{}}
    \foreach \x in {-3.5,-1.5,...,6.5}{\istate{\x}{-2.5}{colSt}}
    \foreach \x in {-4,...,1}{\tgridLine{\x+0.25}{\x+1.75}{\x+0.5}{\x+1.5}
    \circle{\x+0.25}{\x+1.75}}
    \foreach \x in {-4,...,1}{
    \tgridLine {4-\x-0.25}{\x+1.75} {4-\x-0.5}{\x+1.5}\square
    {4-\x-0.25}{\x+1.75}}
  \end{tikzpicture} \,\begin{tikzpicture}[baseline={([yshift=0.6ex]current bounding box.center)},scale=0.5]
{\prop{2}{2}{colSt}{}}
  \foreach \x in {1,3}{\prop{\x}{1}{colSt}{}}   
    \foreach \x in {0,2,...,4}{\prop{\x}{0}{colSt}{}}
    \foreach \x in {-1,1,...,5}{\prop{\x}{-1}{colSt}{}}
    \foreach \x in {-2,0,...,6}{\prop{\x}{-2}{colSt}{}}
   \foreach \x in {-2,0} {\prop{4}{\x}{colimp}{}}
    \foreach \x in {-3.5,-1.5,...,6.5}{\istate{\x}{-2.5}{colSt}}
    \foreach \x in {-4,...,1}{\tgridLine{\x+0.25}{\x+1.75}{\x+0.5}{\x+1.5}
    \square{\x+0.25}{\x+1.75}}
    \foreach \x in {-4,...,1}{
    \tgridLine {4-\x-0.25}{\x+1.75} {4-\x-0.5}{\x+1.5}\circle
    {4-\x-0.25}{\x+1.75}}
     \draw[semithick,decorate,decoration={brace}] (2.7,3.3) -- (8,-2.05) node[midway,xshift=7.5pt,yshift=7.5pt] {$2t$};
  \end{tikzpicture}\!\!,
\end{align}
where we have depicted explicitly the case of $t=3$ and $r=1$.  Each of these two terms represents the backward light cone emanating from the left or right edges of subsystem $A$: the first corresponds to the left edge and is independent of the impurity, while the second corresponds to the right edge and is impurity-dependent provided $t>r$, i.e., provided the backward light cone intersects the impurity line. 

Both terms can then be further reduced 
using the solvability of the initial state and the dual unitarity of the bulk. Starting at the bottom left-hand corner of the first diagram, we can remove the initial state of a qudit pair using~\eqref{eq:charge_solvable}. Recall that in the non-symmetric case, we consider only $\theta=1/2$, hence no soliton is produced, but we do pick up a factor of $1/q^n$.  After removing the initial state pair, the bottom left corner has a bulk gate contracted with two circles from the left. This can be reduced using the dual unitarity condition~\eqref{eq:local_gate_rels_3}. As a result, one obtains another contraction of two circle states with a bulk gate higher up along the left-hand edge. Continuing in this fashion by repeatedly using~\eqref{eq:local_gate_rels_3}, the left-hand diagonal can be removed entirely, leaving behind a factor of $\begin{tikzpicture}[baseline={([yshift=-0.6ex]current bounding box.center)},scale=0.65]
   \tgridLine{0}{0}{.75}{0}
  \circle{0}{0};
  \square{.75}{0};
;\end{tikzpicture}/q^n=q^{1-n}$, which comes from the final contraction between the circle states and the bulk gate at the apex of the diagram, along with the normalization of the initial state. This procedure of contracting along the diagonal can be repeated until the diagram has been fully contracted.

The same can also be applied to the diagram with the impurity gates, with the final result being
\begin{eqnarray}\label{eq:early_time_uncharged}
    S^{(n)}[\rho_A(t)]=\begin{cases}
    4t \log(q)~~&t<r~,\\
    2(t+r)\log(q)+(t-r)s^{(n)}_{\rm imp}~~&r\le t<\ell/2~.\end{cases}
\end{eqnarray}
For $t<r$ this agrees with the result in a pure dual-unitary circuit~\cite{piroli2020exact}, while for $t>r$ the effect of the impurity is felt through the term 
\be \label{eq:Simp}
s^{(n)}_{\rm imp}=\frac{1}{1-n}\log\left(\, \begin{tikzpicture}[baseline={([yshift=-0.6ex]current bounding box.center)},scale=0.5]
   \prop{0}{0}{colimp}{}
   \tgridLine{0.5}{0.5}{0.75}{0.75}
     \tgridLine{-0.5}{-0.5}{-0.75}{-0.75}
     \tgridLine{0.5}{0.5}{0.75}{0.75}
     \tgridLine{-0.5}{0.5}{-0.75}{0.75}
     \tgridLine{0.5}{0.5}{0.75}{0.75}
     \tgridLine{0.5}{-0.5}{0.75}{-0.75}
     \square{-0.75}{-0.75}
      \circle{0.75}{-0.75}
       \square{-0.75}{0.75}
      \circle{0.75}{0.75};
 \end{tikzpicture}~~\big / q^{2n}\right)~,
\ee 
 which can be computed upon specifying a choice of an impurity gate. The impurity contribution to the von Neumann entanglement entropy can then be obtained upon taking the replica limit, $s_{\rm imp} =\lim_{n \to 1} s^{(n)}_{\rm imp}$. This is in fact the operator entanglement entropy of the impurity gate; its maximal possible value $2\log (q)$ is saturated only by dual-unitary operators~\cite{rather2020creating}, and thus, by construction, the impurity operator entanglement is lower than that of the bulk gates.
 
 An analytical calculation of $s_{\rm imp}$ is generally hard to carry out explicitly, as it requires expressing $s^{(n)}_{\rm imp}$ as an analytic function of $n$, a difficult task for any but the simplest choices for an impurity gate (e.g., an identity gate). Nevertheless, it is generally possible to numerically estimate $s_{\rm imp}$ without extrapolating R\'enyi entropies. This is done by viewing the diagram inside the logarithm in Eq.~\eqref{eq:Simp} as the $n$th moment of a reduced density matrix representing a two-qudit subsystem within a four-qudit system. The four-qudit system is in a pure state, obtained by initializing the system as two Bell pairs, choosing one qudit from each pair, and applying the impurity gate to the two chosen qudits. The reduced density matrix is then obtained by tracing out one of the qudit pairs. By diagonalizing this matrix, which is of dimension $q^2$, the desired von Neumann entropy (as well as any R\'enyi entropy) can be directly computed from its spectrum.
 
 One can similarly obtain the exact result for the entropies at late times, $t\ge\ell+r$. Using the dual unitarity of the bulk and the unitarity of the impurity, we find
\begin{eqnarray}\label{eq:long_time_exact}
    S^{(n)}[\rho_A(t)]=2\ell \log (q)~,\quad t\ge\ell+r~,
\end{eqnarray}
which is independent of the impurity and agrees with the pure circuit result~\cite{piroli2020exact}, i.e., the entanglement is unaffected by the presence of the impurity at long times. Indeed, one can show that in an uncharged circuit, the long-time stationary state of subsystem $A$ is the same regardless of whether an impurity is present, a fact that can be attributed to the unitarity of the bulk and impurity gates. This will not be the case when we consider charged circuits. 

 For intermediate times, $\ell/2\le t<\ell+r$, we can use dual unitarity to reduce Fig.~\ref{fig:rho_circuit} to the following expression:
\begin{align}\label{eq:inter}
\!\!\!\tr[\rho^n_A(t)]= \begin{tikzpicture}[baseline={([yshift=0.6ex]current bounding box.center)},scale=0.45]
{\prop{2}{2}{colSt}{}}
   \foreach \x in {1,3}{\prop{\x}{1}{colSt}{}}   
    \foreach \x in {0,2,...,4}{\prop{\x}{0}{colSt}{}}
    \foreach \x in {-1,1,...,5}{\prop{\x}{-1}{colSt}{}}
    \foreach \x in {-2,0,...,6}{\prop{\x}{-2}{colSt}{}}
    \foreach \x in {-3,-1,...,5}{\prop{\x}{-3}{colSt}{}}
        \foreach \x in {7}{\prop{\x}{-3}{colSt}{}}
    \foreach \x in {-4,-2,...,6}{\prop{\x}{-4}{colSt}{}}
     \foreach \x in {8}{\prop{\x}{-4}{colSt}{}}
   \foreach \x in {-4,-2} {\prop{6}{\x}{colimp}{}}
    \foreach \x in {-5.5,-3.5,...,4.5}{\istate{\x}{-4.5}{colSt}}
        \foreach \x in {6.5,8.5}{\istate{\x}{-4.5}{colSt}}
    \foreach \x in {-4,...,1}{\square{\x+0.5}{\x+1.5}}
     \foreach \x in {-6,-5}{\circle{\x+0.5}{\x+1.5}}
    \foreach \x in {-6,...,1}{\circle
    {4-\x-0.5}{\x+1.5}}
      \draw[semithick,decorate,decoration={brace}] (-4,-2.25) -- (1.25,3) node[midway,xshift=-7.5pt,yshift=7.5pt] {$\ell$};
      \draw[semithick,decorate,decoration={brace}] (-6.2,-4.45) -- (-4.65,-2.9) node[midway,xshift=-11.5pt,yshift=7.5pt] {$2t-\ell$};
 \draw[semithick](9.5,-.25) -- (9.5,-.25) node[xshift=-1pt]{$\times \,q^{(1-n)\ell}$};
\draw[semithick,decorate,decoration={brace}] (2.7,3.3) -- (6,-0.05) node[midway,xshift=7.5pt,yshift=7.5pt] {$2r$};
  \end{tikzpicture}. 
\end{align}
At this point, the dual unitarity of the bulk is not enough to obtain a closed exact result for $t<\ell+r$. In order to go beyond this, we employ the quasiparticle and entanglement membrane pictures.

\subsection{Quasiparticle picture}
The quasiparticle picture is an effective theory of quench dynamics for systems that possess a stable set of quasiparticle excitations. It is valid at the ballistic scale, i.e., taking $t,\ell$ to be large with the ratio $t/\ell$ fixed. It posits that the initial state acts as a source of quasiparticles, which are produced locally in entangled pairs (or multiplets) throughout the system and then propagate in opposite directions~\cite{calabrese2005evolution,alba2017entanglement}. The propagation is assumed to be semiclassical and ballistic, while the entanglement among a given pair is fixed by the initial state and does not change over time.  However, since the quasiparticles propagate, they cause entanglement to be generated between different spatial regions. Within the quasiparticle picture, the entanglement entropy is found by summing up the contributions to the entanglement from all the pairs that are shared between $A$ and $\bar A$. 

  After being originally proposed in the context of conformal field theories~\cite{calabrese2005evolution} and then extended to non-interacting spin chains~\cite{fagotti2008evolution}, it was later shown that interacting integrable models can also be captured by this effective theory~\cite{alba2017entanglement}. The key ingredient in the latter was to understand that the velocity of the quasiparticles can be modified in a state-dependent way, which can be calculated using standard techniques in integrability~\cite{bertini2016transport,castro2016emergent}. In our system,  dual unitarity restricts the quasiparticles to travel along the light cone, which cannot be modified, and thus the major effect of the interactions is absent. Therefore, the general properties of interacting integrable models can be understood through the simplest non-interacting system. For this we can take $U(x,t)=P$ for all $t$ and all $x\neq L$, where $P$ is the swap gate, which exchanges the state of the qudits on which it acts,
\be \label{eq:swap_gate}
P\ket{i}\ket{j}=\ket{j}\ket{i}~.
\ee 
In terms of diagrams, we write $P$ as 
\be\label{eq:swap_gate_diagram}
P= \begin{tikzpicture}[baseline={([yshift=-0.6ex]current bounding box.center)},scale=0.35]
   \tgridLine{-0.5}{-0.5}{0.5}{0.5}
   \tgridLine{0.5}{-0.5}{0.1}{-0.1}
   \tgridLine{-0.5}{0.5}{-0.1}{0.1}
 \end{tikzpicture}~.
\ee
Inserting this into the second term on the right-hand side of~\eqref{eq:D_def}, we have the diagram
\begin{align}
\begin{tikzpicture}[baseline={([yshift=0.6ex]current bounding box.center)},scale=0.5]
{\swapprop{2}{2}{colSt}{}}
  \foreach \x in {1,3}{\swapprop{\x}{1}{colSt}{}}   
    \foreach \x in {0,2,...,4}{\swapprop{\x}{0}{colSt}{}}
    \foreach \x in {-1,1,...,5}{\swapprop{\x}{-1}{colSt}{}}
    \foreach \x in {-2,0,...,6}{\swapprop{\x}{-2}{colSt}{}}
   \foreach \x in {-2,0} {\prop{4}{\x}{colimp}{}}
    \foreach \x in {-3.5,-1.5,...,6.5}{\istate{\x}{-2.5}{colSt}}
    \foreach \x in {-4,...,1}{\tgridLine{\x+0.25}{\x+1.75}{\x+0.5}{\x+1.5}
    \square{\x+0.25}{\x+1.75}}
    \foreach \x in {-4,...,1}{
    \tgridLine {4-\x-0.25}{\x+1.75} {4-\x-0.5}{\x+1.5}\circle
    {4-\x-0.25}{\x+1.75}};
  \end{tikzpicture}\!\!,
\end{align}
which can be easily contracted without needing to use the dual unitarity of the model, reproducing the exact result \eqref{eq:early_time_uncharged}. 

The simplicity of the bulk gates now means that we can go beyond the early-time regime and calculate~\eqref{eq:inter} for all times. The result for $t\ge\ell/2$ is then
\begin{equation}\label{eq:ent_swap_gates}
 S^{(n)}[\rho_A(t)]=\!\!\begin{cases}
2(\ell+r-t)\log(q)+(t-r)s^{(n)}_{\rm imp} & \frac{\ell}{2} \le t<\frac{\ell}{2}+r~,\\
2(t-r)\log(q)+(\ell+r-t)s^{(n)}_{\rm imp}&\frac{\ell}{2}+r\le t< \ell+r~,\\
2\ell \log(q)& \ell+r\le t~.
\end{cases}
\end{equation}
Combining this with the exact early-time behaviour~\eqref{eq:early_time_uncharged} gives the full time dynamics of the entanglement entropy. Note that the entanglement entropy decreases for $\ell/2<t<\ell/2+r$ since $s^{(n)}_{\rm imp}\leq 2\log (q), ~\forall n$, with equality holding only if the impurity is dual-unitary. The entropy then grows again until it reaches the long-time limit for $t\ge\ell+r$, as prescribed by the universal result~\eqref{eq:long_time_exact}. The non-monotonic behaviour can be understood through the quasiparticle picture by tracking the scattering of quasiparticles off the impurity and back into the subsystem. Indeed, Eq.~\eqref{eq:ent_swap_gates} can be exactly recovered from a calculation that is based on the quasiparticle picture~\cite{fraenkel2024entanglement}. 

The fact that Eq.~\eqref{eq:ent_swap_gates} can be directly derived via the quasiparticle picture strongly suggests that the non-monotonic behaviour should also arise when the bulk gates are those of an interacting integrable circuit, given the known quasiparticle structure of such circuits~\cite{vernier2023integrable}. In Appendix~\ref{app:XXZ_single_interval}, we rigorously corroborate this expectation for a dual-unitary XXZ bulk. We do so by showing that, within the time regime $\ell/2\le t<\ell/2+r$, the von Neumann entropy differs from the result of Eq.~\eqref{eq:ent_swap_gates} by an additive correction that is bounded (in modulus) by $\log\left(2t-\ell+1\right)$, thus confirming that $S[\rho_A(t)]$ must decrease during this stage of its evolution. Naturally, the interpretation through the quasiparticle picture depends on the existence of such quasiparticles, and in the next subsection we contrast it with an effective description in which they are absent. 

\subsection{Entanglement membrane picture}\label{sec:membrane_uncharged}
For systems that are not integrable, the quasiparticle picture cannot be applied.  In such cases we can, instead, avail of the entanglement membrane picture, which was first developed in the context of random unitary quantum circuits~\cite{jonay2018coarse,nahum2017quantum,nahum2020entanglement}, and later extended to dual-unitary circuits and their generalizations~\cite{rampp2024entanglement,foligno2024quantum,foligno2024nonequilibrium,foligno2024entanglement}. The main idea behind the membrane picture is that, for chaotic systems, the largest contribution to diagrams like~\eqref{eq:D_def} arises from projecting the internal legs onto circle $\begin{tikzpicture}
[baseline={([yshift=-0.6ex]current bounding box.center)},scale=0.45]
   \gridLine{0}{0}{0}{-0.75}
   \circle{0}{0}
\end{tikzpicture}$ or square states $ \begin{tikzpicture}
[baseline={([yshift=-0.6ex]current bounding box.center)},scale=0.45]
   \gridLine{0}{0}{0}{-0.75}
   \square{0}{0}
\end{tikzpicture}$. These states appear in clusters, with domain walls forming between clusters of different types of states. The domain walls are known as entanglement membranes, and calculating  $\tr [\rho^n_A(t)]$ can be viewed as calculating the partition function of a system containing these membranes. The entanglement entropy is then proportional to the corresponding free energy of this partition function.  The core of this approach therefore lies in finding the minimum-energy path of the membranes that emerge from the edges of the subsystem and either end on the initial state or on each other.

To see how this works in practice, let us consider the first diagram on the right-hand side of~\eqref{eq:D_def}. We pick a downward-directed path for the membrane that emanates from the interface between the circle and square states at the apex of the triangle and ends on the initial state, passing through the local gates. One possible path, represented by the red dashed line, is shown below: 
\be \label{eq:right}
\begin{tikzpicture}[baseline={([yshift=1.2ex]current bounding box.center)},scale=0.5]
{\prop{2}{2}{colSt}{}}
  \foreach \x in {1,3}{\prop{\x}{1}{colSt}{}}   
    \foreach \x in {0,2,...,4}{\prop{\x}{0}{colSt}{}}
    \foreach \x in {-1,1,...,5}{\prop{\x}{-1}{colSt}{}}
    \foreach \x in {-2,0,...,6}{\prop{\x}{-2}{colSt}{}}
    \foreach \x in {-3.5,-1.5,...,6.5}{\istate{\x}{-2.5}{colSt}}
    \foreach \x in {-4,...,1}{\tgridLine{\x+0.25}{\x+1.75}{\x+0.5}{\x+1.5}
    \circle{\x+0.25}{\x+1.75}}
    \foreach \x in {-4,...,1}{
    \tgridLine {4-\x-0.25}{\x+1.75} {4-\x-0.5}{\x+1.5}\square
    {4-\x-0.25}{\x+1.75}}
    \draw[very thick,dashed,red] (2,2.65)--(2,2)--(6,-2)--(6,-3);
  \end{tikzpicture}~.
\ee 
The energy of this membrane is then computed by projecting onto the relevant domain (circle or square) on either side of the membrane path. This is achieved by replacing the gates along this path according to 
\begin{equation}\label{eq:membrane_replacements}
    \begin{tikzpicture}[baseline={([yshift=-.6ex]current bounding box.center)},scale=0.5]
        \prop{0}{0}{colSt}{}
         \draw[very thick,dashed,red] (-0.5,0.5)--(0.5,-0.5);
    \end{tikzpicture}
    \to \frac{1}{q^{2n}}\begin{tikzpicture}[baseline={([yshift=-.6ex]current bounding box.center)},scale=0.5]
        \prop{0}{0}{colSt}{}
          \tgridLine{1.25}{1.25}{0.9}{0.9}
            \tgridLine{-1.25}{-1.25}{-0.9}{-0.9} 
        \square{.85}{.85}
        \circle{-.85}{-.85}
        \square{.5}{.5}
        \circle{-.5}{-.5};
    \end{tikzpicture}~, 
\end{equation}
while at the beginning and end we perform 
\begin{equation}\label{eq:membrane_replacements_top_bottom}
    \begin{tikzpicture}[baseline={([yshift=-.6ex]current bounding box.center)},scale=0.5]
        \prop{0}{0}{colSt}{}
         \draw[very thick,dashed,red] (0,0.5)--(-0.,0.)--(0.5,-0.5);
    \end{tikzpicture}
    \to \frac{1}{q^{3n}} \begin{tikzpicture}[baseline={([yshift=-.6ex]current bounding box.center)},scale=0.5]
        \prop{0}{0}{colSt}{}
          \tgridLine{1.25}{1.25}{0.9}{0.9}
            \tgridLine{-1.25}{-1.25}{-0.9}{-0.9} 
             \tgridLine{-1.25}{1.25}{-0.9}{0.9} 
        \square{.85}{.85}
        \circle{-.85}{-.85}
         \circle{-.85}{.85}
        \square{.5}{.5}
        \circle{-.5}{-.5}
        \circle{-.5}{.5};
    \end{tikzpicture} ~,\quad\quad\quad  \begin{tikzpicture}[baseline={([yshift=-.6ex]current bounding box.center)},scale=0.5]
        \prop{0}{0}{colSt}{}
         \draw[very thick,dashed,red] (0,-0.5)--(0.,0.)--(-0.5,0.5);
    \end{tikzpicture}
    \to \frac{1}{q^{3n}} \begin{tikzpicture}[baseline={([yshift=-.6ex]current bounding box.center)},scale=0.5]
        \prop{0}{0}{colSt}{}
          \tgridLine{1.25}{1.25}{0.9}{0.9}
            \tgridLine{-1.25}{-1.25}{-0.9}{-0.9} 
             \tgridLine{1.25}{-1.25}{0.9}{-0.9} 
        \square{.85}{.85}
        \circle{-.85}{-.85}
         \square{.85}{-.85}
        \square{.5}{.5}
        \circle{-.5}{-.5}
        \square{.5}{-.5};
    \end{tikzpicture}~. 
\end{equation}
 Here, the coefficients are normalization factors for the projectors. These ensure that only the interface between different domains, and not the domains themselves, has an energy cost. We are then left with a diagram of the following form:
\be \label{eq:membrane_left}
\frac{1}{q^{4nt}}\times\hspace{-.74cm}
\begin{tikzpicture}[baseline={([yshift=1.2ex]current bounding box.center)},scale=0.5]

  \foreach \x in {1}{\prop{\x}{1}{colSt}{}} 
    \foreach \x in {0,2}{\prop{\x}{0}{colSt}{}}
    \foreach \x in {-1,1,...,3}{\prop{\x}{-1}{colSt}{}}
    \foreach \x in {-2,0,...,4}{\prop{\x}{-2}{colSt}{}}
    \foreach \x in {-3.5,-1.5,...,4.5,8.25}{\istate{\x}{-2.5}{colSt}}
    \foreach \x in {-4,...,0}{\tgridLine{\x+0.25}{\x+1.75}{\x+0.5}{\x+1.5}
    \circle{\x+0.25}{\x+1.75}}
    \foreach \x in {-3,...,1}{
    \tgridLine {3-\x-0.25}{\x+.75} {3-\x-0.5}{\x+.5}\circle
    {3-\x-0.25}{\x+.75}}
    \foreach \x in{1,...,5} {\prop{2+\x}{4-\x}{colSt}{} 
    \circle{1.5+\x}{3.5-\x} 
    \square{2.5+\x}{4.5-\x}
    \tgridLine{3+\x}{5-\x}{3.5+\x}{5.5-\x}
    \tgridLine{9.5}{-2.25}{9.25}{-2.5}
    \tgridLine{8.25}{-2.5}{8}{-2.25}
    \square{3+\x}{5-\x}
     \square{3.5+\x}{5.5-\x} }
     \square{7.5}{-1.5}
     \circle{2.5}{3.5}
     \tgridLine{2}{4}{1.5}{4.5}
     \circle{2}{4}
     \circle{1.5}{4.5}
     \square{8}{-2.25}
    \square{9.5}{-2.25}
      
  \end{tikzpicture}~.
\ee   We interpret the part of this diagram that contains contractions of square and circle states through the gates as the energy cost of the domain wall. 
 This diagram can be contracted completely to obtain the same result one finds by using only dual unitarity, namely $q^{(1-n)t}$. 
  
  Alternatively, we could have chosen the path to lie along the left diagonal instead, 
  
\be \label{eq:left}
\begin{tikzpicture}[baseline={([yshift=1.2ex]current bounding box.center)},scale=0.5]
{\prop{2}{2}{colSt}{}}
  \foreach \x in {1,3}{\prop{\x}{1}{colSt}{}}   
    \foreach \x in {0,2,...,4}{\prop{\x}{0}{colSt}{}}
    \foreach \x in {-1,1,...,5}{\prop{\x}{-1}{colSt}{}}
    \foreach \x in {-2,0,...,6}{\prop{\x}{-2}{colSt}{}}
    \foreach \x in {-3.5,-1.5,...,6.5}{\istate{\x}{-2.5}{colSt}}
    \foreach \x in {-4,...,1}{\tgridLine{\x+0.25}{\x+1.75}{\x+0.5}{\x+1.5}
    \circle{\x+0.25}{\x+1.75}}
    \foreach \x in {-4,...,1}{
    \tgridLine {4-\x-0.25}{\x+1.75} {4-\x-0.5}{\x+1.5}\square
    {4-\x-0.25}{\x+1.75}
    {4-\x-.55}{\x+1.45}}
    \draw[very thick,dashed,red] (2,2.65)--(2,2)--(-2,-2)--(-2,-3);
  \end{tikzpicture} .~
\ee 
In this case we proceed by performing the following projections:
\begin{equation}\label{eq:membrane_replacements_2}
    \begin{tikzpicture}[baseline={([yshift=-.6ex]current bounding box.center)},scale=0.5]
        \prop{0}{0}{colSt}{}
         \draw[very thick,dashed,red] (-0.5,-0.5)--(0.5,0.5);
    \end{tikzpicture}
    \to \frac{1}{q^{2n}}\begin{tikzpicture}[baseline={([yshift=-.6ex]current bounding box.center)},scale=0.5]
        \prop{0}{0}{colSt}{}
          \tgridLine{1.25}{-1.25}{0.9}{-0.9}
            \tgridLine{-1.25}{1.25}{-0.9}{0.9} 
        \square{.85}{-.85}
        \circle{-.85}{.85}
        \square{.5}{-.5}
        \circle{-.5}{.5};
    \end{tikzpicture}~, \quad\quad
    \begin{tikzpicture}[baseline={([yshift=-.6ex]current bounding box.center)},scale=0.5]
        \prop{0}{0}{colSt}{}
         \draw[very thick,dashed,red] (0,0.5)--(-0.,0.)--(-0.5,-0.5);
    \end{tikzpicture}
    \to \frac{1}{q^{3n}} \begin{tikzpicture}[baseline={([yshift=-.6ex]current bounding box.center)},scale=0.5]
        \prop{0}{0}{colSt}{}
          \tgridLine{1.25}{1.25}{0.9}{0.9}
            \tgridLine{1.25}{-1.25}{0.9}{-0.9} 
             \tgridLine{-1.25}{1.25}{-0.9}{0.9} 
        \square{.85}{.85}
        \square{.85}{-.85}
         \circle{-.85}{.85}
        \square{.5}{.5}
        \circle{.5}{-.5}
        \circle{-.5}{.5};
    \end{tikzpicture}~, \quad\quad \begin{tikzpicture}[baseline={([yshift=-.6ex]current bounding box.center)},scale=0.5]
        \prop{0}{0}{colSt}{}
         \draw[very thick,dashed,red] (0,-0.5)--(0.,0.)--(0.5,0.5);
    \end{tikzpicture}
    \to \frac{1}{q^{3n}} \begin{tikzpicture}[baseline={([yshift=-.6ex]current bounding box.center)},scale=0.5]
        \prop{0}{0}{colSt}{}
          \tgridLine{-1.25}{1.25}{-0.9}{0.9}
            \tgridLine{-1.25}{-1.25}{-0.9}{-0.9} 
             \tgridLine{1.25}{-1.25}{0.9}{-0.9} 
        \circle{-.85}{.85}
        \circle{-.85}{-.85}
         \square{.85}{-.85}
        \circle{-.5}{.5}
        \circle{-.5}{-.5}
        \square{.5}{-.5};
    \end{tikzpicture} ~.
\end{equation}
After making these replacements, \eqref{eq:left} can  be completely contracted, returning the same result as in~\eqref{eq:right}. In fact, the dual unitarity of the bulk ensures that all such paths will give the same result~\cite{nahum2020entanglement}. Thus, one needs to consider only a single path of this kind, and evaluate the membrane-picture prediction along it. 

When the impurity gate is present, however, this path-independence ceases to hold, and paths that travel along the impurity are favoured. This is because the impurity will generate less entanglement than the bulk gates, and hence a domain wall travelling along such a configuration will have lower free energy~\cite{fraenkel2024entanglement,nahum2018dynamics}. For the second diagram in~\eqref{eq:D_def}, we should therefore consider the following path:
\be 
\begin{tikzpicture}[baseline={([yshift=1.2ex]current bounding box.center)},scale=0.5]
{\prop{2}{2}{colSt}{}}
  \foreach \x in {1,3}{\prop{\x}{1}{colSt}{}}   
    \foreach \x in {0,2,...,4}{\prop{\x}{0}{colSt}{}}
    \foreach \x in {-1,1,...,5}{\prop{\x}{-1}{colSt}{}}
    \foreach \x in {-2,0,...,6}{\prop{\x}{-2}{colSt}{}} \foreach \x in {-2,0} {\prop{4}{\x}{colimp}{}}
    \foreach \x in {-3.5,-1.5,...,6.5}{\istate{\x}{-2.5}{colSt}}
    \foreach \x in {-4,...,1}{\tgridLine{\x+0.25}{\x+1.75}{\x+0.5}{\x+1.5}
    \square{\x+0.25}{\x+1.75}
    }
    \foreach \x in {-4,...,1}{
    \tgridLine {4-\x-0.25}{\x+1.75} {4-\x-0.5}{\x+1.5}\circle
    {4-\x-0.25}{\x+1.75}
    }
    \draw[very thick,dashed,red] (2,2.65)--(2,2)--(4,0)--(4,-3);
  \end{tikzpicture} .
\ee 
After inserting the relevant projectors on either side of the membrane path, we have

\be \label{eq:membrane_right}
\frac{1}{q^{4nt}}\times\hspace{-.74cm}\begin{tikzpicture}[baseline={([yshift=1.2ex]current bounding box.center)},scale=0.5]
  \foreach \x in {1}{\prop{\x}{1}{colSt}{}}   
    \foreach \x in {0,2}{\prop{\x}{0}{colSt}{}}
    \foreach \x in {-1,1,3,6.5}{\prop{\x}{-1}{colSt}{}}
    \foreach \x in {-2,0,2,7.5}{\prop{\x}{-2}{colSt}{}} 
    \foreach \x in {-2,0} {\prop{4.75}{\x+1}{colimp}{}}
    \foreach \x in {1,2}{\prop{4.75-\x}{1+\x}{colSt}{}} 
    \foreach \x in {-3.5,-1.5,...,2.5,6,8}{\istate{\x}{-2.5}{colSt}}
    \foreach \x in {-4,...,0,1.875}{\tgridLine{\x+0.25}{\x+1.75}{\x+0.5}{\x+1.5}
    \square{\x+0.25}{\x+1.75}
    }
    \foreach \x in {0,1,2}{\square{2.125+\x}{2.375-\x}\square{1.625+\x}{1.625-\x}\circle{3.375+\x}{3.625-\x} \tgridLine{4.375+\x}{4.625-\x}{3.875+\x}{4.125-\x}\circle{4.375+\x}{4.625-\x}\circle{3.875+\x}{4.125-\x}
   \circle{4.375+\x}{4.625-\x}
   \circle{7.125+\x}{-.375-\x}
    }
    \circle{3.375+2}{3.625-3.25}
     \circle{3.375+2}{3.625-4}
      \square{3.375+.75}{3.625-4}
     \circle{3.375+2}{3.625-5.25}
     \circle{3.375+2.5}{3.625-4}
       \circle{3.375+2.5}{3.625-5.25}
       \circle{3.375+2.5}{3.625-6.}
     \square{3.375+.75}{3.625-5.25}
      \square{3.375+.25}{3.625-5.25}
      \square{3.375+.25}{3.625-6}
     \tgridLine{1.625}{4.125}{1.125}{4.625}
     \square{1.625}{4.125}
      \square{1.125}{4.625}
  \end{tikzpicture}~, 
\ee 
which can be contracted completely. Combining this with the expression obtained from the previous diagram, we find agreement with~\eqref{eq:early_time_uncharged}. Thus, we see that the membrane picture, like the quasiparticle picture, is capable of reproducing the exact entanglement entropy in the early-time regime.  Moreover, as we now recap, it also allows us to go beyond this early-time regime.

The starting point for obtaining results for $t\geq \ell/2$ is the diagram depicted in~\eqref{eq:inter}.  In this instance, there are two interfaces between circle and square states, and therefore two membranes will be present, and they can now either end on the initial state or combine to form a single membrane. This entails that there are two paths that should be considered. The first (Path I) is simply the continuation of the early-time behaviour, and it is of the form
\begin{align}\label{eq:path_I}
   \begin{tikzpicture}[baseline={([yshift=0.6ex]current bounding box.center)},scale=0.45]
{\prop{2}{2}{colSt}{}}
   \foreach \x in {1,3}{\prop{\x}{1}{colSt}{}}   
    \foreach \x in {0,2,...,4}{\prop{\x}{0}{colSt}{}}
    \foreach \x in {-1,1,...,5}{\prop{\x}{-1}{colSt}{}}
    \foreach \x in {-2,0,...,6}{\prop{\x}{-2}{colSt}{}}
    \foreach \x in {-3,-1,...,5}{\prop{\x}{-3}{colSt}{}}
        \foreach \x in {7}{\prop{\x}{-3}{colSt}{}}
    \foreach \x in {-4,-2,...,6}{\prop{\x}{-4}{colSt}{}}
     \foreach \x in {8}{\prop{\x}{-4}{colSt}{}}
   \foreach \x in {-4,-2} {\prop{6}{\x}{colimp}{}}
    \foreach \x in {-5.5,-3.5,...,4.5}{\istate{\x}{-4.5}{colSt}}
        \foreach \x in {6.5,8.5}{\istate{\x}{-4.5}{colSt}}
    \foreach \x in {-4,...,1}{\square{\x+0.5}{\x+1.5}}
     \foreach \x in {-6,-5}{\circle{\x+0.5}{\x+1.5}}
    \foreach \x in {-6,...,-3}{\circle
    {4-\x-0.5}{\x+1.5}}
    \foreach \x in {-2,...,1}{
    \tgridLine{4-\x-0.25}{\x+1.75}{4-\x-0.5}{\x+1.5}\circle
    {4-\x-0.25}{\x+1.75}}
\draw[very thick,dashed,red] (2,3) -- (2,2)-- (6,-2) --(6,-5) ;
      \draw[very thick,dashed,red] (-3.75,-3.) -- (-3,-3.)--(-2,-4)--(-2,-5);
  \end{tikzpicture}
  \hspace{-.5cm}. 
\end{align} 
This can then be evaluated as before, using \eqref{eq:membrane_replacements}, \eqref{eq:membrane_replacements_top_bottom} and \eqref{eq:membrane_replacements_2}. This configuration will not be favoured at all times, however, and eventually the following path (Path II), where the membranes have combined,  will dominate:
\begin{align}\label{eq:path_II}
    \begin{tikzpicture}[baseline={([yshift=0.6ex]current bounding box.center)},scale=0.45]
{\prop{2}{2}{colSt}{}}
   \foreach \x in {1,3}{\prop{\x}{1}{colSt}{}}   
    \foreach \x in {0,2,...,4}{\prop{\x}{0}{colSt}{}}
    \foreach \x in {-1,1,...,5}{\prop{\x}{-1}{colSt}{}}
    \foreach \x in {-2,0,...,6}{\prop{\x}{-2}{colSt}{}}
    \foreach \x in {-3,-1,...,5}{\prop{\x}{-3}{colSt}{}}
        \foreach \x in {7}{\prop{\x}{-3}{colSt}{}}
    \foreach \x in {-4,-2,...,6}{\prop{\x}{-4}{colSt}{}}
     \foreach \x in {8}{\prop{\x}{-4}{colSt}{}}
   \foreach \x in {-4,-2} {\prop{6}{\x}{colimp}{}}
    \foreach \x in {-5.5,-3.5,...,4.5}{\istate{\x}{-4.5}{colSt}}
        \foreach \x in {6.5,8.5}{\istate{\x}{-4.5}{colSt}}
    \foreach \x in {-4,...,1}{\tgridLine{\x+.25}{\x+1.75}{\x+.5}{\x+1.5}
    \square{\x+.25}{\x+1.75}}
     \foreach \x in {-6,-5}{\circle{\x+0.5}{\x+1.5}}
    \foreach \x in {-6,...,1}{\circle
    {4-\x-0.5}{\x+1.5}}
\draw[very thick,dashed,red] (2,3)--(2,2)--(-3,-3)--(-4,-3);
  \end{tikzpicture}
  \hspace{-.5cm}. 
\end{align}
 The entanglement entropy is given by the minimum between the expressions obtained from Paths I and II. The resulting prediction of the entanglement membrane picture is 
\begin{eqnarray}\label{eq:ent_mem_uncharged}
S^{(n)}[\rho_A(t)]=\begin{cases} 2\left(t+r\right)\log(q) +  (t-r)s^{(n)}_{\text{imp}}&\ell/2\le t< t_*,\\
2 \ell\log(q)&t\ge t_*,
\end{cases}
\end{eqnarray}
where $t_*$ is the point in time at which Path II becomes favoured over Path I. This is also the saturation time wherein the entanglement entropy becomes time-independent; it is given by
\begin{eqnarray}\label{eq:sat_time}
    t_*=\frac{2(\ell-r) \log(q)+r s^{(n)}_{\rm imp}}{2\log(q)+s^{(n)}_{\rm imp}}~,
\end{eqnarray}
and satisfies $\ell/2\leq t_*\leq \ell-r$. Note that the long-time value of the entropy coincides with the exact result~\eqref{eq:long_time_exact}. Importantly, the entanglement entropy in the membrane picture monotonically increases up to the saturation time. The non-monotonic behaviour of the quasiparticle picture is absent.

Here we have employed the membrane picture as an effective coarse-grained description of entanglement dynamics in the system. As such, it is expected to represent the qualitative behaviour for a chaotic system containing an impurity that has low operator entanglement relative to the bulk. Furthermore, the bulk dual unitarity allows us to find certain limits where the membrane-picture prediction of Eq.~\eqref{eq:ent_mem_uncharged} can be derived exactly from microscopic principles. Indeed, this was shown to be the case for the second R\'enyi entropy in a circuit with a large qudit dimension and bulk gates with maximal entangling power~\cite{fraenkel2024entanglement}. The entangling power $p\!\left(U\right)$~\cite{zanardi2000entangling,rather2020creating,jonnadula2020entangling,foligno2023growth} of a two-qudit gate $U$ is defined based on the state ensemble $\{U\ket{\phi_1}\ket{\phi_2}\}$ that is generated by applying $U$ to Haar-random two-qudit product states, and it measures the average two-qudit entanglement in this ensemble. Letting $\rho_1=\tr_2\!\left[U\ket{\phi_1}\ket{\phi_2}\bra{\phi_1}\bra{\phi_2}U^{\dagger}\right]$ denote the reduced state of the first qudit within such a random state, the entangling power is given by
\begin{equation}\label{eq:entangling_power_def}
    p\!\left(U\right) = \frac{q^2}{q^2-1}\left\langle 1-\tr\!\left[{\rho_1}^2\right]\right\rangle_{\ket{\phi_1},\ket{\phi_2}}~.
\end{equation}
For $q>2$ the entangling power ranges between 0 and 1, and it governs the mixing rate (decay rate of temporal correlations) in random dual-unitary circuits~\cite{arvinda2021bernoulli}. In this sense, Ref.~\cite{fraenkel2024entanglement} showed that the membrane picture holds exactly for the second R\'enyi entropy in an impurity circuit with a maximally mixing ($p=1$) bulk.

The underlying reason that maximal entangling power enables an exact microscopic calculation is that the tensor representing a dual-unitary gate with $p=1$ is a perfect tensor~\cite{Helwig2012AME,Pastawski2015holographic}. That is, it corresponds to the tensor of a pure four-qudit state for which any bipartition is maximally entangled. In particular, for any number of replicas $n$, this implies that
\begin{equation}\label{eq:maximal_p_remaining_contractions}
    \begin{tikzpicture}[baseline={([yshift=-0.6ex]current bounding box.center)},scale=0.65]
   \prop{0}{0}{colSt}{}
   \circle{0.5}{0.5}
    \square{-0.5}{0.5}
    \circle{-0.5}{-0.5}
    \square{0.5}{-0.5}
 \end{tikzpicture}
 =
 \begin{tikzpicture}[baseline={([yshift=-0.6ex]current bounding box.center)},scale=0.65]
   \prop{0}{0}{colSt}{}
   \circle{-0.5}{0.5}
    \square{0.5}{0.5}
    \square{-0.5}{-0.5}
    \circle{0.5}{-0.5}
 \end{tikzpicture}=q^2.
\end{equation}
The value of these contracted diagrams cannot be determined by relying solely on the dual unitarity of the gate, but it becomes fixed when the assumption $p=1$ is made. This result may be contrasted with that of the swap gate, which represents the other extreme with $p=0$, and for which the same diagrams yield $q^{2n}$.

To understand why maximal entangling power is associated with a membrane picture behaviour, consider now a diagram representing the partition function $\tr [\rho^n_A(t)]$, where the bulk satisfies $p=1$. Consider then a component of the form $\begin{tikzpicture}[baseline={([yshift=-0.6ex]current bounding box.center)},scale=0.65]
   \prop{0}{0}{colSt}{}
    \circle{-0.5}{0.5}
    \square{0.5}{0.5}
 \end{tikzpicture}$, such that the open legs are attached to other gates in the diagram. We may observe that 
 \begin{eqnarray}\label{eq:maximal_p_contraction_comparison}
     \begin{tikzpicture}[baseline={([yshift=-0.6ex]current bounding box.center)},scale=0.65]
   \prop{0}{0}{colSt}{}
    \circle{-0.5}{0.5}
    \square{0.5}{0.5}
    \square{-0.5}{-0.5}
    \square{0.5}{-0.5}
 \end{tikzpicture}=
 \begin{tikzpicture}[baseline={([yshift=-0.6ex]current bounding box.center)},scale=0.65]
   \prop{0}{0}{colSt}{}
    \circle{-0.5}{0.5}
    \square{0.5}{0.5}
    \circle{-0.5}{-0.5}
    \circle{0.5}{-0.5}
 \end{tikzpicture}=q^{n+1}>
 \begin{tikzpicture}[baseline={([yshift=-0.6ex]current bounding box.center)},scale=0.65]
   \prop{0}{0}{colSt}{}
   \circle{-0.5}{0.5}
    \square{0.5}{0.5}
    \square{-0.5}{-0.5}
    \circle{0.5}{-0.5}
 \end{tikzpicture}\,.
 \end{eqnarray}
 Thus, if we resolve the open legs of $\begin{tikzpicture}[baseline={([yshift=-0.6ex]current bounding box.center)},scale=0.65]
   \prop{0}{0}{colSt}{}
    \circle{-0.5}{0.5}
    \square{0.5}{0.5}
 \end{tikzpicture}$ in a (multi-replica) basis that contains the circle and square states, the contribution of the terms \begin{tikzpicture}[baseline={([yshift=-0.6ex]current bounding box.center)},scale=0.65]
   \prop{0}{0}{colSt}{}
    \circle{-0.5}{0.5}
    \square{0.5}{0.5}
    \square{-0.5}{-0.5}
    \square{0.5}{-0.5}
 \end{tikzpicture} and \begin{tikzpicture}[baseline={([yshift=-0.6ex]current bounding box.center)},scale=0.65]
   \prop{0}{0}{colSt}{}
    \circle{-0.5}{0.5}
    \square{0.5}{0.5}
    \circle{-0.5}{-0.5}
    \circle{0.5}{-0.5}
 \end{tikzpicture} to the partition function will be larger than that of \begin{tikzpicture}[baseline={([yshift=-0.6ex]current bounding box.center)},scale=0.65]
   \prop{0}{0}{colSt}{}
   \circle{-0.5}{0.5}
    \square{0.5}{0.5}
    \square{-0.5}{-0.5}
    \circle{0.5}{-0.5}
 \end{tikzpicture}, where the latter is negligible in the large $q$ limit. The first two terms are in fact domain-wall configurations, rendering the formation of domains a leading-order effect.

 For $n=2$ and random bulk gates (with $p=1$ fixed), this argument becomes exact, since the averaging over the randomness projects the legs onto a subspace spanned by the circle and square permutation states. For $n>2$, the projected subspace is spanned by a larger set of independent permutation states; it is, however, reasonable to expect that the contribution of the additional permutation states is suppressed, given that the boundary condition of the diagram involves only circle and square states. Therefore, throughout the paper, we support the predictions of the membrane picture via exact calculations for $n=2$ replicas of circuits featuring a random dual-unitary bulk with large $q$ and $p=1$. The extension to $n>2$ in non-universal time regimes is not made rigorously, yet it relies on the simple logic stated above that stems from Eq.~\eqref{eq:maximal_p_contraction_comparison}, which is indeed true for all $n$. For all $n$, the same membrane picture calculation also recovers the exact results obtained for the universal time regimes, which serves as another strong check for the validity of this calculation.


\section{Dynamics of disjoint intervals in an uncharged circuit}\label{sec:Two_Intervals}
After reviewing the results from Ref.~\cite{fraenkel2024entanglement}, we now extend them by examining the dynamics of disjoint subsystems in an uncharged circuit. The subsystem of interest is taken to be $A=A_1\cup A_2$, where each of the intervals $A_1$ and $A_2$ extends across $2\ell$ sites, and is separated from the impurity by $2r$ sites (see Eq.~\eqref{eq:two_intervals_def}).  

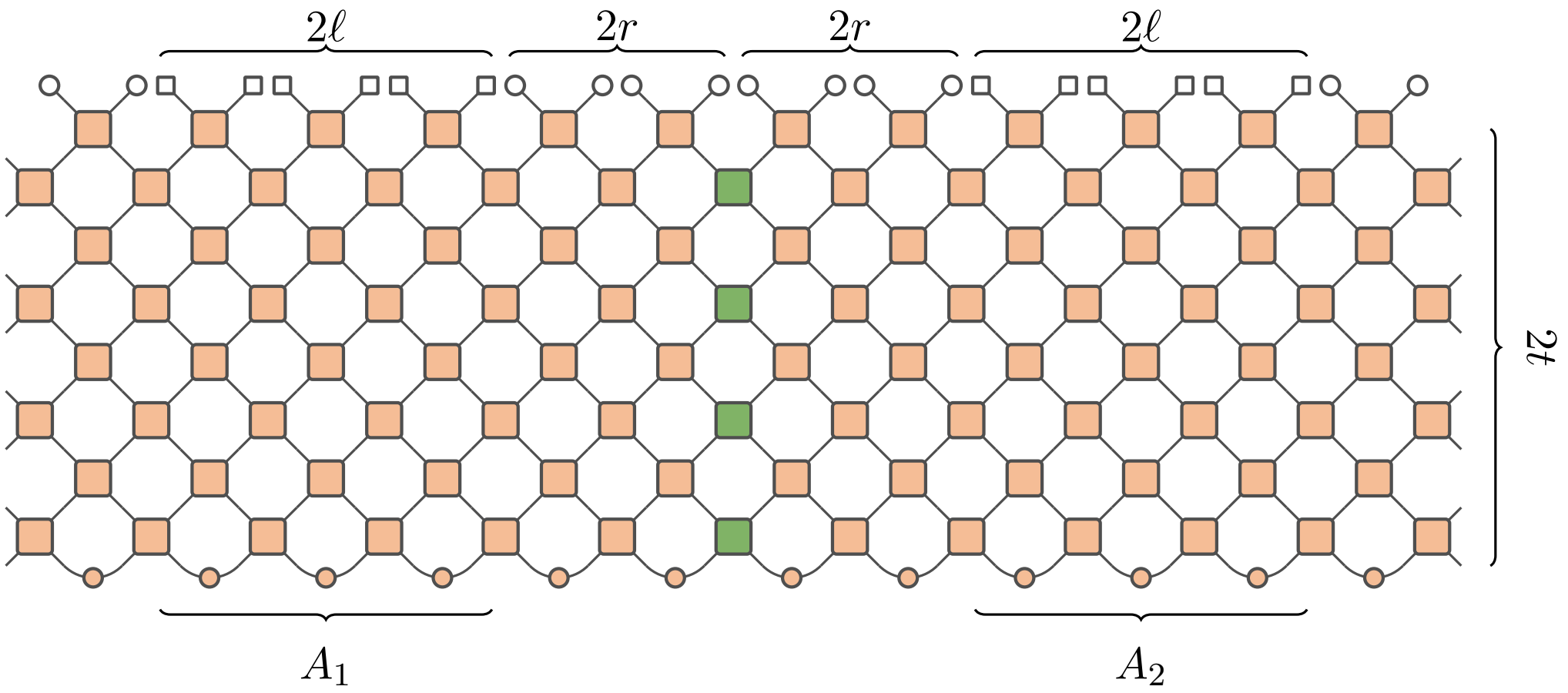
\begin{figure}\centering
    \begin{tikzpicture}[baseline={([yshift=0.6ex]current bounding box.center)},scale=0.5]
    \foreach \x in {-5,-3,...,5}{\prop{\x}{-3}{colSt}{}}
     \foreach \x in {7,9,...,17}{\prop{\x}{-3}{colSt}{}}
      \foreach \x in {-5,-3,...,5}{\prop{\x}{-1}{colSt}{}}
           \foreach \x in {7,9,...,17}{\prop{\x}{-1}{colSt}{}}
        \foreach \x in {-5,-3,...,5}{\prop{\x}{1}{colSt}{}}
             \foreach \x in {7,9,...,17}{\prop{\x}{1}{colSt}{}}
          \foreach \x in {-5,-3,...,5}{\prop{\x}{3}{colSt}{}}
               \foreach \x in {7,9,...,17}{\prop{\x}{3}{colSt}{}}
    \foreach \x in {-6,-4,...,6}{\prop{\x}{-4}{colSt}{}}
         \foreach \x in {8,10,...,18}{\prop{\x}{-4}{colSt}{}}
     \foreach \x in {-6,-4,...,6}{\prop{\x}{-2}{colSt}{}}
          \foreach \x in {8,10,...,18}{\prop{\x}{-2}{colSt}{}}
    \foreach \x in {-6,-4,...,6}{\prop{\x}{0}{colSt}{}} 
         \foreach \x in {8,10,...,18}{\prop{\x}{0}{colSt}{}}
     \foreach \x in {-6,-4,...,6}{\prop{\x}{2}{colSt}{}}
         \foreach \x in {8,10,...,18}{\prop{\x}{2}{colSt}{}}
      \foreach \x in {-4,-2,0,2} {\prop{6}{\x}{colimp}{}}
    \foreach \x in {-5.5,-3.5,...,5.5}{\istate{\x}{-4.5}{colSt}}
        \foreach \x in {6.5,8.5,...,16.5}{\istate{\x}{-4.5}{colSt}}
    \foreach \x in {-3,-1,...,1}{\tgridLine{\x+0.75}{3.75}{\x+0.5}{3.5}
    \square{\x+0.75}{3.75}
    }
    \foreach \x in {-4,-2,...,1}{\tgridLine{\x+0.25}{3.75}{\x+.5}{3.5}\square{\x+0.25}{3.75}}
     \foreach \x in {3,5,...,9}{\tgridLine{\x+0.75}{3.75}{\x+0.5}{3.5}
     \circle{\x+0.75}{3.75}}
     \foreach \x in {2,4,...,8}{\tgridLine{\x+0.25}{3.75}{\x+.5}{3.5}\circle{\x+0.25}{3.75}}
     \foreach \x in {11,13,...,15}{\tgridLine{\x+0.75}{3.75}{\x+0.5}{3.5}
    \square{\x+0.75}{3.75}
    }
    \foreach \x in {10,12,...,15}{\tgridLine{\x+0.25}{3.75}{\x+.5}{3.5}\square{\x+0.25}{3.75}}
       \foreach \x in {-5}{\tgridLine{\x+0.75}{3.75}{\x+0.5}{3.5}\circle{\x+0.75}{3.75}}
       \foreach \x in {-6}{\tgridLine{\x+0.25}{3.75}{\x+.5}{3.5}\circle{\x+0.25}{3.75}}
       \foreach \x in {17}{\tgridLine{\x+0.75}{3.75}{\x+0.5}{3.5}\circle{\x+0.75}{3.75}}
       \foreach \x in {16}{\tgridLine{\x+0.25}{3.75}{\x+.5}{3.5}\circle{\x+0.25}{3.75}}
 \draw[semithick,decorate,decoration={brace}] (10.15,4.25) -- (15.85,4.25) node[midway,yshift=7.5pt] {$2\ell$};
 \draw[semithick,decorate,decoration={brace}] (-3.85,4.25) -- (1.85,4.25) node[midway,yshift=7.5pt] {$2\ell$};
  \draw[semithick,decorate,decoration={brace}]  (15.85,-5.25) --  (10.15,-5.25)
  node[below,midway,yshift=-5.5pt] {$A_2$};
 \draw[semithick,decorate,decoration={brace}] (1.85,-5.25) --  (-3.85,-5.25)node[below,midway,yshift=-5.5pt] {$A_1$};
   \draw[semithick,decorate,decoration={brace}] (2.15,4.25) -- (5.85,4.25) node[midway,yshift=7.5pt] {$2r$};
  \draw[semithick,decorate,decoration={brace}] (6.15,4.25) -- (9.85,4.25) node[midway,yshift=7.5pt] {$2r$};
  \draw[semithick,decorate,decoration={brace}] (19,3) -- (19,-4.5) node[midway,xshift=12.5pt,rotate= 270] {$2t$};
  \end{tikzpicture}
  \caption{\label{fig:TwoIntervalPartitionFunc}The diagrammatic representation of $\tr[\rho^n_A(t)]$ with $\ell=3$, $r=2$, and $t=4$, for a disjoint subsystem $A=A_1\cup A_2$ that straddles the impurity.}
\end{figure}

  In order to calculate the R\'enyi mutual information $\mathcal{I}^{(n)}_{A_1:A_2}$ between the two parts of subsystem $A$, we need $S^{(n)}[\rho_{A_{1,2}}(t)]$, for which the result is given in Sec.~\ref{sec:ent_uncharged}, as well as $S^{(n)}[\rho_{A}(t)]$, which requires a new calculation. The starting point for the latter quantity, $\tr [\rho^n_A(t)]$, is depicted in Fig.~\ref{fig:TwoIntervalPartitionFunc}. Like the previous diagrams we discussed, this diagram also exhibits a universal time regime---namely, $t\le\ell/2$ as well as $t\ge\ell+r$---where it can be contracted fully using the dual unitarity of the bulk. However, as we saw before, the effective descriptions of the quasiparticle and membrane pictures can capture this regime exactly. We therefore perform the calculation of $S^{(n)}[\rho_{A}(t)]$ by directly using these descriptions, and their agreement with the independent calculation within the universal time regime can be straightforwardly checked. We begin by computing $S^{(n)}[\rho_{A}(t)]$ within the quasiparticle picture in Sec.~\ref{sec:MI_uncharged_QP}, and then in Sec.~\ref{sec:MI_uncharged_mem} present the solution within the membrane picture. Finally, in Sec.~\ref{sec:neg_uncharged} we address the dynamics of entanglement negativity between the intervals.

\subsection{Quasiparticle picture}\label{sec:MI_uncharged_QP} 
Replacing the bulk gates in Fig.~\ref{fig:TwoIntervalPartitionFunc} with the swap gate $P$ leads to a completely reducible diagram at all times.  The result for $S^{(n)}[\rho_{A}(t)]$ separates into numerous different time regimes, corresponding to the intersections between the backward light cones emanating from the edges of $A_1,A_2$ and from the impurity. We find that
\begin{equation}\label{eq:S_disjoint_qpp}
    S^{(n)}[\rho_A(t)]=\begin{cases}
        8t\log (q) & t<r,\\
        4(t+r)\log (q) & r\le t <\ell/2,\\
        4(\ell+r-t)\log (q) & \ell/2\le t <\ell/2+r,\\
        4(t-r)\log (q) & \ell/2+r\le t <\ell+r,\\
        4\ell\log (q) & \ell+r\le t.
    \end{cases}
\end{equation}
Note that this does not depend on the impurity gate, and therefore also holds when it is replaced with a dual-unitary gate. We also point out that the entropy decreases in the time window $\ell/2\leq t\leq \ell/2+r$, which is a distinctive feature of 
 the quasiparticle picture; it occurs due to quasiparticle pairs that are initially generated inside the region separating the two intervals, and then become fully contained in $A$ during that time window.

Combining the above expression with the results for the single-interval entropy in \eqref{eq:early_time_uncharged} and \eqref{eq:ent_swap_gates}, we arrive at the R\'enyi mutual information between the two intervals, 
\begin{equation}\label{eq:mut_info_qpp_uncharged}
    \mathcal{I}_{A_1:A_2}^{(n)}(t)=\begin{cases}
        0 & t<r,\\
        2(t-r)s_{\rm imp}^{(n)} & r\le t <\ell/2+r,\\
        2(\ell+r-t)s_{\rm imp}^{(n)} & \ell/2+r\le t <\ell+r,\\
        0 & \ell+r\le t
    \end{cases}~.
\end{equation}
We observe that, as long as the impurity is not fully reflective (i.e., when $s_{\rm imp}^{(n)}>0$), the mutual information between the intervals will exhibit a rise-and-fall stage. 
A non-dual-unitary impurity gate necessarily diminishes the height of this peak compared to a homogeneous swap circuit, where $s_{\rm imp}^{(n)}$ would have been replaced with $2\log (q)$. In fact, we see that \textit{any} dual-unitary gate put in the location of the impurity will lead to the maximal mutual information peak, even if this gate breaks homogeneity and integrability. This is in line with the notion that dual-unitary gates permit perfect transmission of information.

\subsection{Entanglement membrane picture}\label{sec:MI_uncharged_mem} 
We now evaluate the entanglement entropy of subsystem $A$ using the membrane picture, starting again from Fig.~\ref{fig:TwoIntervalPartitionFunc}. For $t<r$ the diagram can be reduced, by unitarity, to four disconnected pieces which do not depend on the impurity, and each resembles \eqref{eq:right}. These can be evaluated exactly using the membrane picture following the method of Sec.~\ref{sec:ent_uncharged}. Beyond this time, the right edge of $A_1$ and the left edge of $A_2$ come into causal contact and so are no longer disconnected, while at later times, the outer edges of $A_{1}$ and $A_2$ come into play. To obtain the prediction of the membrane picture in this regime, we first reduce the diagram in Fig.~\ref{fig:TwoIntervalPartitionFunc}  using unitarity and dual unitarity. After doing so, we find 
\begin{equation}
   \label{eq:inter_two_inter}
\!\!\!\tr[\rho^n_A(t)]=\hspace{-20pt}  \begin{tikzpicture}[baseline={([yshift=0.6ex]current bounding box.center)},scale=0.5]
    \foreach \x in {-3,-1,...,5}{\prop{\x}{-3}{colSt}{}}
     \foreach \x in {5,7,...,15}{\prop{\x}{-3}{colSt}{}}
      \foreach \x in {-1,1,...,5}{\prop{\x}{-1}{colSt}{}}
           \foreach \x in {7,9,...,13}{\prop{\x}{-1}{colSt}{}}
        \foreach \x in {1,3}{\prop{\x}{1}{colSt}{}}
             \foreach \x in {9,11}{\prop{\x}{1}{colSt}{}}

    \foreach \x in {-4,-2,...,6}{\prop{\x}{-4}{colSt}{}}
         \foreach \x in {8,10,...,16}{\prop{\x}{-4}{colSt}{}}
     \foreach \x in {-2,0,...,6}{\prop{\x}{-2}{colSt}{}}
          \foreach \x in {6,8,...,14}{\prop{\x}{-2}{colSt}{}}
    \foreach \x in {0,2,4}{\prop{\x}{0}{colSt}{}} 
         \foreach \x in {8,10,12}{\prop{\x}{0}{colSt}{}}
     \foreach \x in {2,10}{\prop{\x}{2}{colSt}{}}
      \foreach \x in {-4,-2} {\prop{6}{\x}{colimp}{}}
    \foreach \x in {-5.5,-3.5,...,5.5}{\istate{\x}{-4.5}{colSt}}
        \foreach \x in {6.5,8.5,...,16.5}{\istate{\x}{-4.5}{colSt}}
    
    \foreach \x in {-4,...,1}{\tgridLine{\x+0.25}{\x+1.75}{\x+.5}{\x+1.5}\square{\x+0.25}{\x+1.75}}
    \foreach \x in {-6,...,-5}{\tgridLine{\x+0.25}{\x+1.75}{\x+.5}{\x+1.5}\circle{\x+0.25}{\x+1.75}}
    \foreach \x in {6,...,9}{\tgridLine{\x+0.25}{\x+1.75-8}{\x+.5}{\x+1.5-8}\circle{\x+0.25}{\x+1.75-8}}

    \foreach \x in {10,...,15}{\tgridLine{\x+0.75}{-\x+3.75+9}{\x+0.5}{-\x+3.5+9}
    \square{\x+0.75}{-\x+3.75+9}}
    \foreach \x in {16,...,17}{\tgridLine{\x+0.75}{-\x+3.75+9}{\x+0.5}{-\x+3.5+9}
    \circle{\x+0.75}{-\x+3.75+9}}
    \foreach \x in {2,...,5}{\tgridLine{\x+0.75}{-\x+3.75+1}{\x+0.5}{-\x+3.5+1}
    \circle{\x+0.75}{-\x+3.75+1}};
    \draw[semithick,decorate,decoration={brace}] (-4.2,-2.25) -- (1.05,3) node[midway,xshift=-7.5pt,yshift=7.5pt] {$\ell$};
      \draw[semithick,decorate,decoration={brace}] (-6.4,-4.45) -- (-4.85,-2.9) node[midway,xshift=-11.5pt,yshift=7.5pt] {$2t-\ell$};

\draw[semithick,decorate,decoration={brace}] (2.7,3.3) -- (6,-0.05) node[midway,xshift=7.5pt,yshift=7.5pt] {$2r$};

\draw[semithick](16.5,-.25) -- (16.5,-.25) node[xshift=-1pt]{$\times \,q^{2(1-n)\ell}$};
  \end{tikzpicture}.
\end{equation}
We evaluate this diagram by considering two different membrane configurations. First, during the stage $r\le t<\ell-r$, it is energetically favourable for the membranes to arrange themselves in the following configuration:
\begin{equation}\label{eq:early_time_two_interval_uncharged}
    \begin{tikzpicture}[baseline={([yshift=0.6ex]current bounding box.center)},scale=0.5]
    \foreach \x in {-3,-1,...,5}{\prop{\x}{-3}{colSt}{}}
     \foreach \x in {5,7,...,15}{\prop{\x}{-3}{colSt}{}}
      \foreach \x in {-1,1,...,5}{\prop{\x}{-1}{colSt}{}}
           \foreach \x in {7,9,...,13}{\prop{\x}{-1}{colSt}{}}
        \foreach \x in {1,3}{\prop{\x}{1}{colSt}{}}
             \foreach \x in {9,11}{\prop{\x}{1}{colSt}{}}

    \foreach \x in {-4,-2,...,6}{\prop{\x}{-4}{colSt}{}}
         \foreach \x in {8,10,...,16}{\prop{\x}{-4}{colSt}{}}
     \foreach \x in {-2,0,...,6}{\prop{\x}{-2}{colSt}{}}
          \foreach \x in {6,8,...,14}{\prop{\x}{-2}{colSt}{}}
    \foreach \x in {0,2,4}{\prop{\x}{0}{colSt}{}} 
         \foreach \x in {8,10,12}{\prop{\x}{0}{colSt}{}}
     \foreach \x in {2,10}{\prop{\x}{2}{colSt}{}}
      \foreach \x in {-4,-2} {\prop{6}{\x}{colimp}{}}
    \foreach \x in {-5.5,-3.5,...,5.5}{\istate{\x}{-4.5}{colSt}}
        \foreach \x in {6.5,8.5,...,16.5}{\istate{\x}{-4.5}{colSt}}
    
    \foreach \x in {-4,...,1}{\tgridLine{\x+0.25}{\x+1.75}{\x+.5}{\x+1.5}\square{\x+0.25}{\x+1.75}}
    \foreach \x in {-6,...,-5}{\tgridLine{\x+0.25}{\x+1.75}{\x+.5}{\x+1.5}\circle{\x+0.25}{\x+1.75}}
    \foreach \x in {6,...,9}{\tgridLine{\x+0.25}{\x+1.75-8}{\x+.5}{\x+1.5-8}\circle{\x+0.25}{\x+1.75-8}}

    \foreach \x in {10,...,15}{\tgridLine{\x+0.75}{-\x+3.75+9}{\x+0.5}{-\x+3.5+9}
    \square{\x+0.75}{-\x+3.75+9}}
    \foreach \x in {16,...,17}{\tgridLine{\x+0.75}{-\x+3.75+9}{\x+0.5}{-\x+3.5+9}
    \circle{\x+0.75}{-\x+3.75+9}}
    \foreach \x in {2,...,5}{\tgridLine{\x+0.75}{-\x+3.75+1}{\x+0.5}{-\x+3.5+1}
    \circle{\x+0.75}{-\x+3.75+1}}
     \draw[very thick,dashed,red] (2,3)--(2,2) -- (5,-1)--(7,-1)--(10, 2)--(10,3);
      \draw[very thick,dashed,red] (-4,-3) -- (-3,-3)--(-2,-4)--(-2,-5) ;
       \draw[very thick,dashed,red] (16,-3) -- (15,-3)--(14,-4)--(14,-5) ;
    
  \end{tikzpicture}.
\end{equation}  
 This diagram is dominated by the large domain of square states and is evaluated using the rules detailed in Sec.~\ref{sec:membrane_uncharged}. In this configuration, the middle membrane connecting the left edge of $A_2$ to the right edge of $A_1$ remains fixed at length $4r$, while the two outer membranes, each of length $2t$, continue to grow. This continues until the time $t=\ell-r$ when the favourable configuration becomes the one where there are only two membranes, one connecting the left and right edges of $A_1$ and the other doing the same for $A_2$:
\begin{equation}\label{eq:early_time_two_interval_uncharged_II}
    \begin{tikzpicture}[baseline={([yshift=0.6ex]current bounding box.center)},scale=0.5]
    \foreach \x in {-3,-1,...,5}{\prop{\x}{-3}{colSt}{}}
     \foreach \x in {5,7,...,15}{\prop{\x}{-3}{colSt}{}}
      \foreach \x in {-1,1,...,5}{\prop{\x}{-1}{colSt}{}}
           \foreach \x in {7,9,...,13}{\prop{\x}{-1}{colSt}{}}
        \foreach \x in {1,3}{\prop{\x}{1}{colSt}{}}
             \foreach \x in {9,11}{\prop{\x}{1}{colSt}{}}

    \foreach \x in {-4,-2,...,6}{\prop{\x}{-4}{colSt}{}}
         \foreach \x in {8,10,...,16}{\prop{\x}{-4}{colSt}{}}
     \foreach \x in {-2,0,...,6}{\prop{\x}{-2}{colSt}{}}
          \foreach \x in {6,8,...,14}{\prop{\x}{-2}{colSt}{}}
    \foreach \x in {0,2,4}{\prop{\x}{0}{colSt}{}} 
         \foreach \x in {8,10,12}{\prop{\x}{0}{colSt}{}}
     \foreach \x in {2,10}{\prop{\x}{2}{colSt}{}}
      \foreach \x in {-4,-2} {\prop{6}{\x}{colimp}{}}
    \foreach \x in {-5.5,-3.5,...,5.5}{\istate{\x}{-4.5}{colSt}}
        \foreach \x in {6.5,8.5,...,16.5}{\istate{\x}{-4.5}{colSt}}
    
    \foreach \x in {-4,...,1}{\tgridLine{\x+0.25}{\x+1.75}{\x+.5}{\x+1.5}\square{\x+0.25}{\x+1.75}}
    \foreach \x in {-6,...,-5}{\tgridLine{\x+0.25}{\x+1.75}{\x+.5}{\x+1.5}\circle{\x+0.25}{\x+1.75}}
    \foreach \x in {6,...,9}{\tgridLine{\x+0.25}{\x+1.75-8}{\x+.5}{\x+1.5-8}\circle{\x+0.25}{\x+1.75-8}}

    \foreach \x in {10,...,15}{\tgridLine{\x+0.75}{-\x+3.75+9}{\x+0.5}{-\x+3.5+9}
    \square{\x+0.75}{-\x+3.75+9}}
    \foreach \x in {16,...,17}{\tgridLine{\x+0.75}{-\x+3.75+9}{\x+0.5}{-\x+3.5+9}
    \circle{\x+0.75}{-\x+3.75+9}}
    \foreach \x in {2,...,5}{\tgridLine{\x+0.75}{-\x+3.75+1}{\x+0.5}{-\x+3.5+1}
    \circle{\x+0.75}{-\x+3.75+1}}
\draw[very thick,dashed,red] (2,3)--(2,2)--(-3,-3)--(-4,-3) ;
\draw[very thick,dashed,red] (10,3)--(10,2)--(15,-3)--(16,-3) ;
  \end{tikzpicture}.
\end{equation} 
Again, the diagram can be contracted using the projection rules from Sec.~\ref{sec:membrane_uncharged}. The final result has a particularly simple form, 
\begin{equation}\label{eq:2_Renyi_ent_chaotic}
    S^{(n)}[\rho_A(t)]=\begin{cases}
        8t\log (q) & t<r,\\
        4(t+r)\log (q) & r\le t <\ell-r,\\
        4\ell\log (q) & \ell-r\le t.
    \end{cases}
\end{equation}
We notice again that the evolution is independent of the impurity, and would also hold in a fully dual-unitary circuit.  We observe that the entropy evolution of the disjoint subsystem is now monotonic, in stark contrast to what we obtained using the quasiparticle picture in Eq.~\eqref{eq:S_disjoint_qpp}. 

\begin{figure}
    \centering
    \includegraphics[width=\linewidth]{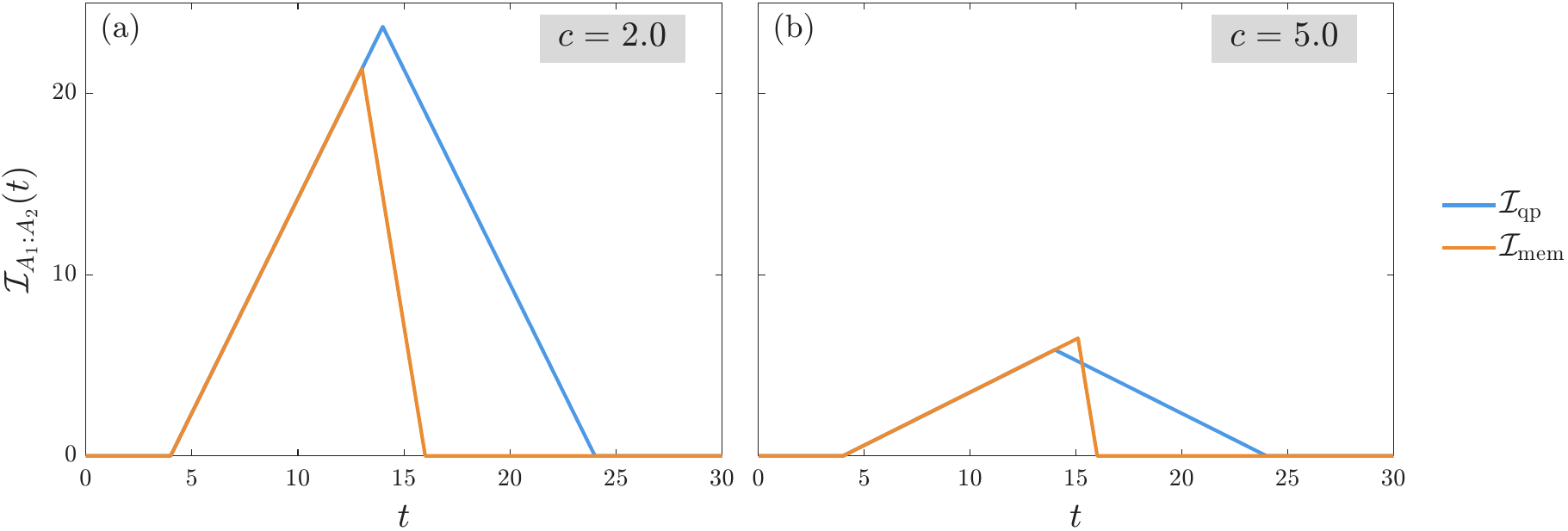}
    \caption{The evolution of the mutual information between two intervals in an uncharged dual-unitary circuit with an impurity $U_{\rm Id}(c)$, as defined in Eq.~\eqref{eq:ISWAP_impurity_def}. The local dimension is $q=6$, and the parameters of the intervals $A_1$ and $A_2$ are fixed at $\ell=20$ and $r=4$. The plot compares the predictions of the quasiparticle (${\mathcal I}_{\rm qp}$) and membrane (${\mathcal I}_{\rm mem}$) pictures for an impurity parameter of (a) $c=2$ and (b) $c=5$; in the former case the mutual information peak is higher for the quasiparticle picture than the membrane picture, while the roles are reversed in the latter case. The quantities ${\mathcal I}_{\rm qp}$ and ${\mathcal I}_{\rm mem}$ were computed by taking the $n\to1$ limit in Eqs.~\eqref{eq:mut_info_qpp_uncharged} and \eqref{eq:renyi_mutual}, respectively.}
    \label{fig:MuInf_Uncharged_ISWAP}
\end{figure}

Combining the above result with \eqref{eq:early_time_uncharged} and \eqref{eq:ent_mem_uncharged}, we find that the prediction of the membrane picture for the R\'enyi mutual information is 
\begin{equation}\label{eq:renyi_mutual}
    \mathcal{I}_{A_1:A_2}^{(n)}(t)=\begin{cases}
        0 & t<r,\\
        2(t-r)s_{\rm imp}^{(n)} & r\le t < t_*,\\
        4(\ell-r-t)\log (q) & t_*\le t <\ell-r,\\
        0 & \ell-r\le t,
    \end{cases}
\end{equation}
where $t_*$ is the saturation time of the single-interval entropy, given by~\eqref{eq:sat_time}. 
Compared to a purely dual-unitary circuit, the impurity not only slows down the growth of mutual information between the two intervals, but, through $t_*$, also affects the time at which it peaks. In contrast, Eq.~\eqref{eq:mut_info_qpp_uncharged} shows that within the quasiparticle picture, the peak time is always fixed at $t=\ell/2+r$.

For a given impurity, the quasiparticle and membrane pictures predict identical dynamics until $t=\min\left\{\ell/2+r,t_*\right\}$. Rather surprisingly, however, if $t_*>\ell/2+r$, which translates to an impurity obeying 
\begin{equation}
    0<s^{(n)}_{\rm imp}<2\Big(1-\frac{4r}{\ell}\Big)\log (q)~,
\end{equation}
then the mutual information peaks at a higher value within the membrane picture than the quasiparticle picture. Since chaotic dynamics is strongly associated with information scrambling, one could have naively anticipated that the mutual information predicted by the membrane picture would always be upper-bounded by the quasiparticle picture prediction~\cite{asplund2015entanglement,alba2019scrambling}, which indeed holds in the absence of an impurity (as can be readily checked by substituting $s^{(n)}_{\rm imp}=2\log (q)$). Evidently, scrambling can also result in enhanced information transfer across a barrier that is sufficiently reflective. This phenomenon is illustrated in Fig.~\ref{fig:MuInf_Uncharged_ISWAP}. Note that, even though the membrane picture allows for a higher peak, the mutual information can diminish much faster than in the quasiparticle picture, since its slope for $t_*\le t<\ell-r$ is determined by the bulk rather than by the impurity. 

As explained in Sec.~\ref{sec:membrane_uncharged}, the membrane picture is generally a coarse-grained approximation that is expected to hold at hydrodynamic scales, but in some cases it can be proven to be exact. This is the case for the early-time and late-time regimes in the model we study. When the bulk gates are random and maximally mixing, it is also true for the average second R\'enyi entropy at all times in the large $q$ limit, as we show in  Appendix \ref{app:chaotic_2_interval_entropy}. There, the exact analytical calculation relies on techniques that were introduced in Refs.~\cite{fraenkel2024entanglement,foligno2023growth}.

\subsection{Negativity dynamics}\label{sec:neg_uncharged}
  
\begin{figure}
    \centering
    \begin{tikzpicture}[baseline={([yshift=0.6ex]current bounding box.center)},scale=0.5]
    \foreach \x in {-5,-3,...,5}{\prop{\x}{-3}{colSt}{}}
     \foreach \x in {7,9,...,17}{\prop{\x}{-3}{colSt}{}}
      \foreach \x in {-5,-3,...,5}{\prop{\x}{-1}{colSt}{}}
           \foreach \x in {7,9,...,17}{\prop{\x}{-1}{colSt}{}}
        \foreach \x in {-5,-3,...,5}{\prop{\x}{1}{colSt}{}}
             \foreach \x in {7,9,...,17}{\prop{\x}{1}{colSt}{}}
          \foreach \x in {-5,-3,...,5}{\prop{\x}{3}{colSt}{}}
               \foreach \x in {7,9,...,17}{\prop{\x}{3}{colSt}{}}
    \foreach \x in {-6,-4,...,6}{\prop{\x}{-4}{colSt}{}}
         \foreach \x in {8,10,...,18}{\prop{\x}{-4}{colSt}{}}
     \foreach \x in {-6,-4,...,6}{\prop{\x}{-2}{colSt}{}}
          \foreach \x in {8,10,...,18}{\prop{\x}{-2}{colSt}{}}
    \foreach \x in {-6,-4,...,6}{\prop{\x}{0}{colSt}{}} 
         \foreach \x in {8,10,...,18}{\prop{\x}{0}{colSt}{}}
     \foreach \x in {-6,-4,...,6}{\prop{\x}{2}{colSt}{}}
         \foreach \x in {8,10,...,18}{\prop{\x}{2}{colSt}{}}
      \foreach \x in {-4,-2,0,2} {\prop{6}{\x}{colimp}{}}
    \foreach \x in {-5.5,-3.5,...,5.5}{\istate{\x}{-4.5}{colSt}}
        \foreach \x in {6.5,8.5,...,16.5}{\istate{\x}{-4.5}{colSt}}
    \foreach \x in {-3,-1,...,1}{\tgridLine{\x+0.75}{3.75}{\x+0.5}{3.5}
    \square{\x+0.75}{3.75}
    }
    \foreach \x in {-4,-2,...,1}{\tgridLine{\x+0.25}{3.75}{\x+.5}{3.5}\square{\x+0.25}{3.75}}
     \foreach \x in {3,5,...,9}{\tgridLine{\x+0.75}{3.75}{\x+0.5}{3.5}
     \circle{\x+0.75}{3.75}}
     \foreach \x in {2,4,...,8}{\tgridLine{\x+0.25}{3.75}{\x+.5}{3.5}\circle{\x+0.25}{3.75}}
     \foreach \x in {11,13,...,15}{\tgridLine{\x+0.75}{3.75}{\x+0.5}{3.5}
    \triangleDown{\x+0.75}{3.75}
    }
    \foreach \x in {10,12,...,15}{\tgridLine{\x+0.25}{3.75}{\x+.5}{3.5}\triangleDown{\x+0.25}{3.75}}
       \foreach \x in {-5}{\tgridLine{\x+0.75}{3.75}{\x+0.5}{3.5}\circle{\x+0.75}{3.75}}
       \foreach \x in {-6}{\tgridLine{\x+0.25}{3.75}{\x+.5}{3.5}\circle{\x+0.25}{3.75}}
       \foreach \x in {17}{\tgridLine{\x+0.75}{3.75}{\x+0.5}{3.5}\circle{\x+0.75}{3.75}}
       \foreach \x in {16}{\tgridLine{\x+0.25}{3.75}{\x+.5}{3.5}\circle{\x+0.25}{3.75}}
 \draw[semithick,decorate,decoration={brace}] (10.15,4.25) -- (15.85,4.25) node[midway,yshift=7.5pt] {$2\ell$};
 \draw[semithick,decorate,decoration={brace}] (-3.85,4.25) -- (1.85,4.25) node[midway,yshift=7.5pt] {$2\ell$};
  \draw[semithick,decorate,decoration={brace}] (2.15,4.25) -- (5.85,4.25) node[midway,yshift=7.5pt] {$2r$};
  \draw[semithick,decorate,decoration={brace}] (6.15,4.25) -- (9.85,4.25) node[midway,yshift=7.5pt] {$2r$};
  \draw[semithick,decorate,decoration={brace}] (19,3) -- (19,-4.5) node[midway,xshift=12.5pt,rotate= 270] {$2t$};
  \draw[semithick,decorate,decoration={brace}]  (15.85,-5.25) --  (10.15,-5.25)
  node[below,midway,yshift=-5.5pt] {$A_2$};
 \draw[semithick,decorate,decoration={brace}] (1.85,-5.25) --  (-3.85,-5.25)node[below,midway,yshift=-5.5pt] {$A_1$};
  \end{tikzpicture}
  \caption{\label{fig:TwoIntervalPartitionFuncNeg} The diagrammatic representation of $\tr\!\left[{\left({\rho_A(t)}^{T_2}\right)}^n\right]$ with $\ell=3$, $r=2$, and $t=4$, for a disjoint subsystem $A=A_1\cup A_2$ that straddles the impurity. The triangle states implement the partial transpose acting on subsystem $A_2$. }
\end{figure}
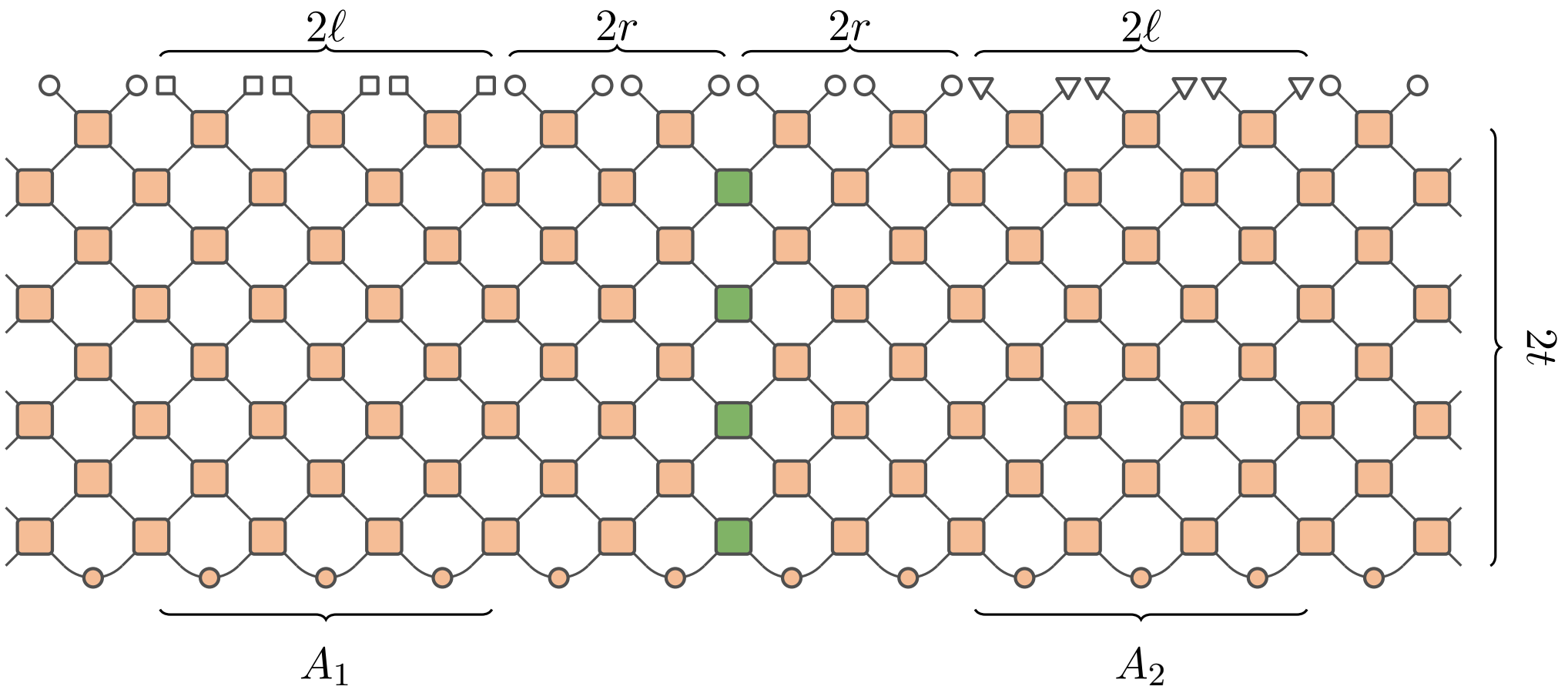

We now examine the entanglement negativity between $A_1$ and $A_2$, to capture the purely quantum correlation that develops between them. For this purpose, we must consider the moments of the partially transposed reduced density matrix, which yield the R\'enyi negativities ${\cal E}^{(n)}_{A_1:A_2}=\log\tr[\left({\rho_A}^{T_2}\right)^{n}]$ for even $n$. The partition function relevant for this case is depicted in Fig.~\ref{fig:TwoIntervalPartitionFuncNeg}. From this starting point, it is possible to contract the diagram for any choice of dual-unitary bulk as long as $t\le\ell/2$. Indeed, using the diagrammatic rules in Eqs.~\eqref{eq:local_gate_rels_1}--\eqref{eq:local_gate_rels_3}, one finds that
\begin{equation}
    {\cal E}^{(n)}_{A_1:A_2}(t)=\begin{cases}
        8t(1-n)\log(q) & t<r,\\
        4(t+r)(1-n)\log(q) + (t-r)(2-n)\,s_{\rm imp}^{(n/2)} & r\le t \le\ell/2,
    \end{cases}
\end{equation}
where we used the fact that
\begin{equation}\label{eq:imp_neg}
    s_{\rm imp}^{(n/2)}=\frac{1}{2-n}\log\left(\, \begin{tikzpicture}[baseline={([yshift=-0.6ex]current bounding box.center)},scale=0.5]
   \prop{0}{0}{colimp}{}
   \tgridLine{0.5}{0.5}{0.75}{0.75}
     \tgridLine{-0.5}{-0.5}{-0.75}{-0.75}
     \tgridLine{0.5}{0.5}{0.75}{0.75}
     \tgridLine{-0.5}{0.5}{-0.75}{0.75}
     \tgridLine{0.5}{0.5}{0.75}{0.75}
     \tgridLine{0.5}{-0.5}{0.75}{-0.75}
     \square{-0.75}{-0.75}
      \triangleUp{0.75}{-0.75}
       \square{-0.75}{0.75}
      \triangleDown{0.75}{0.75};
 \end{tikzpicture}\,/q^{2n}\right).
\end{equation}
From this expression, we see that only the impurity contribution can survive in the replica limit $n\to 1$. Thus, depending on the choice of impurity, non-trivial quantum correlations can be generated between $A_1$ and $A_2$. The impurity contribution $s^{(n/2)}_{\rm imp}$ can be computed for any $n$, including  $n=1$, using the same method as was done for $\eqref{eq:Simp}$, by considering the negativity of a four-qudit pure state generated by acting with the impurity gate on two Bell pairs. We note that since this four-qudit state is pure, the $n$th R\'enyi negativity can be obtained from the  R\'enyi entanglement entropy with $n\to n/2$. This motivates the choice of notation in ~\eqref{eq:imp_neg}.

\begin{figure}
    \centering
    \includegraphics[width=\linewidth]{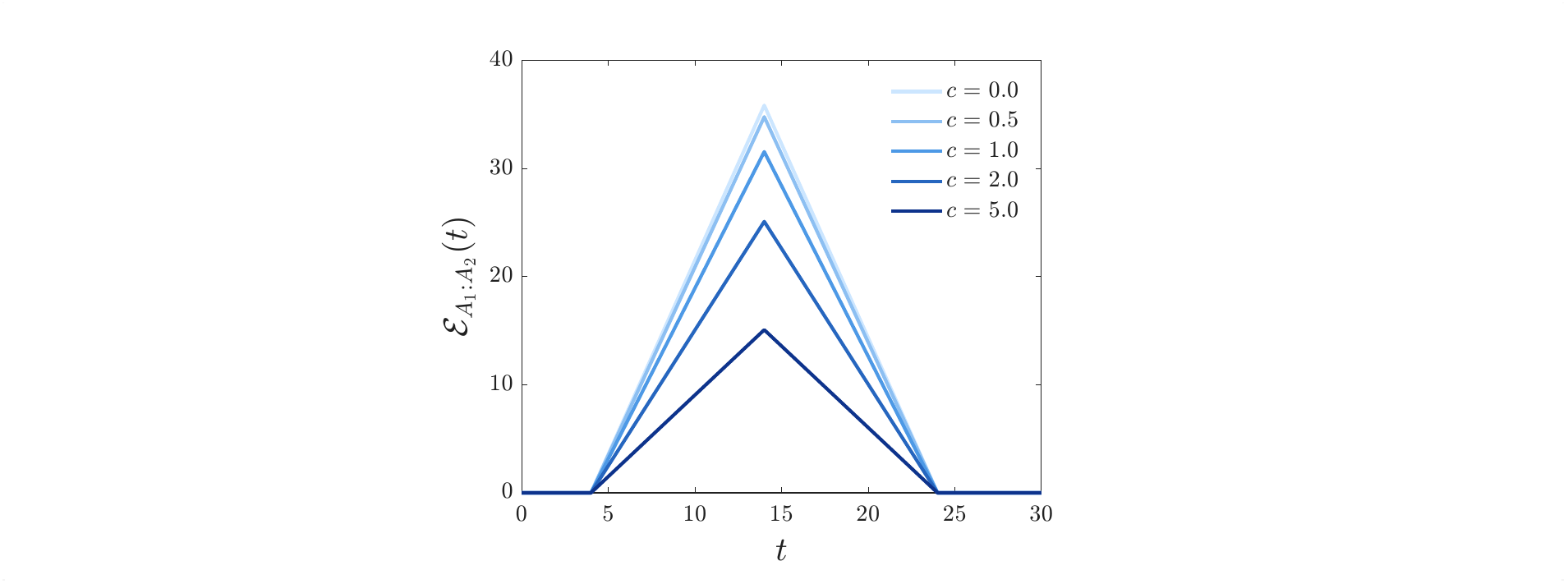}
    \caption{The evolution of the entanglement negativity, ${\cal E}_{A_1:A_2}(t)$, between two intervals in an uncharged dual-unitary circuit with an impurity $U_{\rm Id}(c)$, as defined in Eq.~\eqref{eq:ISWAP_impurity_def}. The local dimension is $q=6$, and the parameters of the intervals $A_1$ and $A_2$ are fixed at $\ell=20$ and $r=4$. The plot shows the prediction of the quasiparticle picture (see Eq.~\eqref{eq:swap_negativity}) for various values of the impurity parameter $c$.}
    \label{fig:Neg_Uncharged_ISWAP}
\end{figure}

As was the case previously, the  calculation of the full dynamics cannot be completed using only dual unitarity. We can easily obtain the prediction of the quasiparticle picture, however, by replacing all bulk gates in Fig.~\ref{fig:TwoIntervalPartitionFuncNeg} with swap gates $P$. A straightforward contraction then yields
\begin{equation}
    {\cal E}^{(n)}_{A_1:A_2}(t)=\begin{cases}
        4(\ell+r-t)(1-n)\log(q) + (t-r)(2-n)\,s_{\rm imp}^{(n/2)} & \ell/2\le t <\ell/2+r,\\
        4(t-r)(1-n)\log(q) + (\ell+r-t)(2-n)\,s_{\rm imp}^{(n/2)} & \ell/2+r\le t <\ell+r,\\
        4\ell(1-n)\log(q) & t\ge \ell+r.
        
    \end{cases}
\end{equation}
Taking $n\to 1$, we obtain in total
\begin{equation}\label{eq:swap_negativity}
    {\cal E}_{A_1:A_2}(t) = \begin{cases}
        0 & t<r,\\
        (t-r)\,s_{\rm imp}^{(1/2)} & r\le t <\ell/2+r,\\
        (\ell+r-t)\,s_{\rm imp}^{(1/2)} & \ell/2+r\le t <\ell+r,\\
        0 & t \ge \ell+r.
    \end{cases}
\end{equation}
Since $s_{\rm imp}^{(1/2)}=2\log(q)$ if the impurity is allowed to be dual-unitary, we observe that the impurity can only diminish the quantum correlations that are temporarily created between the two intervals. These correlations necessarily vanish at long times, since the reduced state of $A$ becomes the maximally mixed state. We furthermore note that the negativity obeys ${\cal E}_{A_1:A_2}(t)=\frac{1}{2}{\cal I}^{(1/2)}_{A_1:A_2}(t)$ for all times (cf.~Eq.~\eqref{eq:mut_info_qpp_uncharged}), a relation that has been well-established for the information dynamics of closed free systems~\cite{alba2019quantum} and at short times more generally~\cite{bertini2022entanglement}. The quasiparticle-picture negativity dynamics are illustrated in Fig.~\ref{fig:Neg_Uncharged_ISWAP} for the impurity defined in Eq.~\eqref{eq:ISWAP_impurity_def}.

While the simplicity of the swap gate allows us to give a quasiparticle prediction for the negativity dynamics at all times, we were not able to do the same for chaotic models. The membrane picture cannot be applied as straightforwardly as before due to the triangle-state boundary conditions in Fig.~\ref{fig:TwoIntervalPartitionFuncNeg}, and so we do not provide a prediction at intermediate times.  Nevertheless, for $r\leq t<\ell/2$ the negativity will increase linearly in time,  in agreement with~\eqref{eq:swap_negativity}, and it must also vanish at late times, $t\geq \ell+r$. In the future, it would be interesting to explore whether the membrane picture can be further developed to enable the full negativity calculation, and to check whether it might give rise to a stage of higher negativity compared to the quasiparticle picture, as was seen for the mutual information in Eq.~\eqref{eq:renyi_mutual}.

 \section{Entanglement entropy with a charged bulk}\label{sec:ent_charged}

We now turn to the calculation of the entanglement entropy for a single contiguous subsystem (defined in Eq.~\eqref{eq:single_interval_def}) in a charged dual-unitary bulk. We again split the calculation into different time regimes. In the case of a charged bulk, only the early-time behaviour ($t\le\ell/2$) is universal; however, the presence of the charge structure allows us to obtain a non-trivial bound on the entanglement at intermediate and late times. As in the previous sections, we can obtain specific predictions in these non-universal regimes by employing the quasiparticle and membrane pictures. 

\subsection{Universal early time regime}
The early time regime, $t\leq \ell/2$, can be treated exactly using the dual unitarity of the bulk gates and the charge solvability of the initial state. The starting point is, once again, the diagram depicted in Fig.~\ref{fig:rho_circuit}, which can be reduced using unitarity to the diagram given in~\eqref{eq:D_def}. This can be completely contracted following the same procedure as in the non-symmetric case, the only difference being that we employ the relations given in~\eqref{eq:charge_solvable}. Thus, removing an initial state produces a soliton with the charge structure of the initial state, and upon completely contracting a diagonal, one obtains a factor of $\begin{tikzpicture}[baseline={([yshift=-0.6ex]current bounding box.center)},scale=0.65]
   \tgridLine{0}{0}{1}{0}
  \circleb{0.5}{0}
  \circle{0}{0};
  \square{1}{0};
;\end{tikzpicture}$ instead of $q^{1-n}$. Reducing the diagram as much as possible, we arrive at 
\be 
\tr[\rho_A^n(t)]=\begin{cases}\left(\begin{tikzpicture}[baseline={([yshift=-0.6ex]current bounding box.center)},scale=0.65]
   \tgridLine{0}{0}{1}{0}
  \circleb{0.5}{0}
  \circle{0}{0};
  \square{1}{0};
;\end{tikzpicture}\right)^{2t}\left(\begin{tikzpicture}[baseline={([yshift=-0.6ex]current bounding box.center)},scale=0.65]
   \tgridLine{0}{0}{1}{0}
  \circler{0.5}{0}
  \square{0}{0};
  \circle{1}{0};
;\end{tikzpicture}\right)^{2t}~~&t<r~,\\
\left(\begin{tikzpicture}[baseline={([yshift=-0.6ex]current bounding box.center)},scale=0.65]
   \tgridLine{0}{0}{1}{0}
  \circleb{0.5}{0}
  \circle{0}{0};
  \square{1}{0};
;\end{tikzpicture}\right)^{2t}\left(\begin{tikzpicture}[baseline={([yshift=-0.6ex]current bounding box.center)},scale=0.65]
   \tgridLine{0}{0}{1}{0}
  \circler{0.5}{0}
  \square{0}{0};
  \circle{1}{0};
;\end{tikzpicture}\right)^{2r}
\left( \begin{tikzpicture}[baseline={([yshift=-0.6ex]current bounding box.center)},scale=0.5]
   \prop{0}{0}{colimp}{}
   \tgridLine{0.5}{0.5}{0.75}{0.75}
     \tgridLine{-0.5}{-0.5}{-0.75}{-0.75}
     \tgridLine{0.5}{0.5}{0.75}{0.75}
     \tgridLine{-0.5}{0.5}{-0.75}{0.75}
     \tgridLine{0.5}{0.5}{0.75}{0.75}
     \tgridLine{0.5}{-0.5}{0.75}{-0.75}
    \circleb{-.45}{-.45}
     \circler{.45}{-.45}
     \square{-0.75}{-0.75}
      \circle{0.75}{-0.75}
       \square{-0.75}{0.75}
      \circle{0.75}{0.75};
 \end{tikzpicture}\right)^{t-r}~~&r\le t<\ell/2~,\end{cases}
\ee
where the charge structure should be read as that coming from the initial state as in~\eqref{eq:charge_structure_intial}.  Using the result of~\eqref{eq:local_renyi_diagram} we then find, for our particular choice~\eqref{eq:charge_structure_intial},  that
\begin{eqnarray}\label{eq:early_time}
    S^{(n)}[\rho_A(t)]=\begin{cases}
    4t s^{(n)}(\theta)~~&t<r~,\\
    2(t+r)s^{(n)}(\theta)+(t-r)s^{(n)}_{\rm imp}(\theta)~~&r\le t<\ell/2~.\end{cases}
\end{eqnarray}
This contains two types of entropy densities: $s^{(n)}(\theta)$ which is the contribution of the bulk gates, and  $s^{(n)}_{\rm imp}(\theta)$ which is the impurity contribution. The bulk contribution can be split into two distinct parts, 
\begin{eqnarray} \label{eq:Sbulk}
s^{(n)}(\theta)=s^{(n)}_{\rm num}(\theta)+s_{\rm conf},
\end{eqnarray} 
which are given explicitly by
\begin{eqnarray} \label{eq:Sbulk2}
s^{(n)}_{\rm num}(\theta)=\frac{1}{1-n}\log[(1-\theta)^n+\theta^n],\quad\quad s_{\rm conf}=\log (d).
\end{eqnarray} 
Here $s^{(n)}_{\rm num}(\theta)$ is the number entropy, which represents the contribution coming from the charge fluctuations between different charge sectors, while $s_{\rm conf}$ is the configuration entropy, which arises from the fluctuations within the charge sectors~\cite{goldstein2018symmetry,xavier2018equipartition}. Typically, it is only possible to decompose the entanglement into number and configuration parts if the state is symmetric and, moreover, do so only for the von Neumann entropy. However, charged dual-unitary circuits are a special case where this is possible even when the symmetry is broken and for general R\'enyi entropies. 
One can easily take the replica limit to find, for the number entropy, that
\be 
s_{\rm num}(\theta)=-\theta\log \theta-(1-\theta)\log(1-\theta)~.
\ee 
The configuration entropy, on the other hand, is the same for all R\'enyi indices in the cases that we consider. The  impurity contribution to the entanglement in Eq.~\eqref{eq:early_time} is defined as\be \label{eq:Simp_charged}
s^{(n)}_{\rm imp}(\theta)=\frac{1}{1-n}\log\left(\, \begin{tikzpicture}[baseline={([yshift=-0.6ex]current bounding box.center)},scale=0.5]
   \prop{0}{0}{colimp}{}
   \tgridLine{0.5}{0.5}{0.75}{0.75}
     \tgridLine{-0.5}{-0.5}{-0.75}{-0.75}
     \tgridLine{0.5}{0.5}{0.75}{0.75}
     \tgridLine{-0.5}{0.5}{-0.75}{0.75}
     \tgridLine{0.5}{0.5}{0.75}{0.75}
     \tgridLine{0.5}{-0.5}{0.75}{-0.75}
    \circleb{-.45}{-.45}
     \circler{.45}{-.45}
     \square{-0.75}{-0.75}
      \circle{0.75}{-0.75}
       \square{-0.75}{0.75}
      \circle{0.75}{0.75};
 \end{tikzpicture}\,\right)~,
\ee 
and for the replica limit we define $ 
s_{\rm imp}(\theta)=\lim_{n\to 1}s^{(n)}_{
\rm imp
}(\theta)
$. 
These quantities will naturally depend on the initial state parameter $\theta$,  because of the presence of the solitons on the right-hand side of~\eqref{eq:Simp_charged}.  If we set $\theta=1/2$, the result in~\eqref{eq:early_time} reduces to that of the uncharged circuit~\eqref{eq:early_time_uncharged}. In that case, the solitons produced by the initial state are proportional to the identity, and the symmetry is not reflected in the entanglement entropy at early times. We can evaluate~\eqref{eq:Simp_charged} in the same way as described below~\eqref{eq:Simp} but now considering the entanglement of a four-qudit state constructed by acting with the impurity gate on $\ket{m}\ket{m}$ instead of two Bell pairs.

\subsection{Bound on the entanglement entropy}

Before determining the predictions for the entanglement entropy using the two effective theories, we prove a general upper bound on the entropy. The bound is valid at all times, but proves particularly useful in providing intuition into the dynamics at intermediate and late times. 

Our starting point for this analysis is the replicated density matrix for $t>\ell/2$. This matrix can be expressed diagrammatically as 
 \be \label{eq:long_time_rho}
[\rho_A(t)]^{\otimes_{\rm r} n}=\,\,\begin{tikzpicture}[baseline={([yshift=-1.25ex]current bounding box.center)},scale=0.45]  
    \foreach \x in {-1,1,3}{\prop{\x}{1}{colSt}{}}
    \foreach \x in {0,2,4}{\prop{\x}{0}{colSt}{}}
    \foreach \x in {1,3,5}{\prop{\x}{-1}{colSt}{}}
    \foreach \x in {2,4,6}{\prop{\x}{-2}{colSt}{}}
    \foreach \x in {3,5}{\prop{\x}{-3}{colSt}{}}
    \foreach \x in {4,6}{\prop{\x}{-4}{colSt}{}}
   \foreach \x in {-4,-2} {\prop{6}{\x}{colimp}{}}
      \circle{6.5}{-3.5}
    \foreach \x in {-1,...,4}  {\tgridLine{-.75+\x}{-.75-\x}{-.5+\x}{-.5-\x}
    \circleb{-.45+\x}{-.45-\x}
    \circle{-.75+\x}{-.75-\x}}
    \foreach \x in {-3,-2,-1}{\circle
    {4-\x-0.5}{\x+1.5}}
     \foreach \x in {-6,-4}{
     \tgridLine{6.75}{\x+1.25}{6.5}{\x+1.5}\circle
    {6.75}{\x+1.25}
    \circler
    {6.45}{\x+1.5}}
  \istate{4.5}{-4.5}{colSt}
       \draw[semithick,decorate,decoration={brace}] (-1.9,2) -- (3.5,2) node[midway,xshift=0pt,yshift=7.5pt] {$\ell$};
           \draw[semithick,decorate,decoration={brace}] (4.4,1.1) -- (6.1,-0.6) node[midway,xshift=7.5pt,yshift=7.5pt] {$2r$}   ;
            \draw[semithick,decorate,decoration={brace}] (5.75,-5.5) -- (4.25,-5.5) node[midway,xshift=-5pt,yshift=-7.5pt] {$\ell+r-t$}   ;
  \end{tikzpicture}~~~,
 \ee
where the diagram has been reduced using dual unitarity and charged solvability. We now follow~\cite{foligno2024entanglement} and apply a particular quantum channel $\mathcal{K}[\cdot]$ to this state; this channel was shown to be unital, meaning that under its action the entropy cannot decrease~\cite{foligno2024entanglement}. The entropy of the new state $\mathcal{K}[\rho_A(t)]$ then provides an upper bound on the entropy of our actual state $\rho_A(t)$,
\be \label{eq:general_bound}
S[\rho_A(t)]\leq S\big[\mathcal{K}[\rho_A(t)]\big].
\ee
The quantum channel in question is defined to be factorized with respect to the different sites, i.e.,
\begin{eqnarray}\label{eq:symm_channel}
    \mathcal{K}[\cdot]=\bigotimes_{x=1}^{2\ell}\mathcal{K}_x[\cdot ],
\end{eqnarray}
with its action on each site being equivalent to the symmetrization of the local reduced state. Explicitly, 
\begin{eqnarray}
    \mathcal{K}_x[\cdot]=\sum_{j=0,1}\Pi^{(\sigma)}_{j,x}  \frac{{\rm tr}[\,\cdot \,\Pi^{(\sigma)}_{j,x}]}{{\rm tr}[ \Pi^{(\sigma)}_{j,x}]}~,
\end{eqnarray}
where $\sigma=r$ for even $x$ and $\sigma=l$ for odd $x$.  The resulting state is represented by the following diagram: 
\begin{eqnarray}
\mathcal{K}\big [\rho_A(t)\big ]=\frac{1}{d^{2\ell}}\,\sum_{j^{(l)}_{1}\!\!,\dots ,j^{(l)}_{\ell}}\sum_{j^{(r)}_{1}\!\!,\dots ,j^{(r)}_{\ell}}
\,\,\begin{tikzpicture}[baseline={([yshift=-1.25ex]current bounding box.center)},scale=0.45]  
    \foreach \x in {-1,1,3}{\prop{\x}{1}{colSt}{1}}
    \foreach \x in {0,2,4}{\prop{\x}{0}{colSt}{1}}
    \foreach \x in {1,3,5}{\prop{\x}{-1}{colSt}{1}}
    \foreach \x in {2,4,6}{\prop{\x}{-2}{colSt}{1}}
    \foreach \x in {3,5}{\prop{\x}{-3}{colSt}{1}}
    \foreach \x in {4,6}{\prop{\x}{-4}{colSt}{1}}
   \foreach \x in {-4,-2} {\prop{6}{\x}{colimp}{1}}
    \foreach \x in {-1,...,4}  {\tgridLine{-.75+\x}{-.75-\x}{-.5+\x}{-.5-\x}
    \circleb{-.45+\x}{-.45-\x}
    \circle{-.75+\x}{-.75-\x}}
    \foreach \x in {-3,-2,-1}{\circle
    {4-\x-0.5}{\x+1.5}}
     \foreach \x in {-6,-4}{
     \tgridLine{6.75}{\x+1.25}{6.5}{\x+1.5}\circle
    {6.75}{\x+1.25}
    \circler
    {6.45}{\x+1.5}}
  \istate{4.5}{-4.5}{colSt}
   \foreach \x in {-1,1,3}{
   \tgridLine{\x-0.75}{1.75}{\x-.25}{1.25}
   \circlelg{\x-0.45}{1.45} \circle{\x-0.75}{1.75}\tgridLine{\x-0.75}{2.25}{\x-0.75}{3.}\circle{\x-0.75}{2.25}
   \circlelg{\x-0.75}{2.65}}
    \foreach \x in {0,2,4}{\tgridLine{\x-0.75}{1.25}{\x-0.25}{1.75}
    \circledg{\x-0.55}{1.45}
    \circle{\x-0.25}{1.75}
     \tgridLine{\x-0.25}{2.25}{\x-0.25}{3.}
     \circle{\x-0.25}{2.25}
      \circledg{\x-0.25}{2.65}} \circle{6.5}{-3.5}
  \end{tikzpicture}~,
\end{eqnarray}
where the sum is taken over all combinations of $j_{x}^{(\sigma)}=0,1$, which indicate the possible insertions of the projectors $\Pi^{(\sigma)}_{j,x}$ on each site. Recall that the $1$ drawn inside the gates signifies that we consider only a single replica. 
The entanglement entropy of $\mathcal{K}[\rho_A(t)]$ can now be explicitly computed by a straightforward contraction of the diagram. The result is found to be 
\begin{align}\label{eq:channel_ent}
    &S^{(n)}\big[\mathcal{K}[\rho_A(t)]\big]\nonumber\\
    &\quad=\begin{cases}(2\ell+r-t)s^{(n)}(\theta)+(t-r)s^{(n)}(\theta_{\rm imp})-(t-r)I^{(n)}(\vartheta)& \ell/2\leq t< \ell/2+r\, ,\\
(2\ell+r-t)s^{(n)}(\theta)+(t-r)s^{(n)}(\theta_{\rm imp})-(\ell+r-t)I^{(n)}(\vartheta)& \ell/2+r\leq t < \ell+r\, ,\\\ell s^{(n)}(\theta)+\ell s^{(n)}(\theta_{\rm imp})& \ell+r\leq t \, ,
\end{cases}
\end{align}
where $s^{(n)}(\theta)$ was defined in~\eqref{eq:Sbulk}, while $s^{(n)}(\theta_{\rm imp})$ and $I^{(n)}(\vartheta)$ are new. The first of these quantities is simply the bulk contribution to the entropy evaluated with an impurity-dependent value, $\theta_{\rm imp}$, which is defined through
\begin{eqnarray}\label{eq:theta_imp}
    \theta_{\rm imp}=\begin{tikzpicture}[baseline={([yshift=-0.6ex]current bounding box.center)},scale=0.5]
   \prop{0}{0}{colimp}{1}
   \tgridLine{0.5}{0.5}{0.75}{0.75}
     \tgridLine{-0.5}{-0.5}{-0.75}{-0.75}
     \tgridLine{0.5}{0.5}{0.75}{0.75}
     \tgridLine{-0.5}{0.5}{-0.75}{0.75}
     \tgridLine{0.5}{0.5}{0.75}{0.75}
     \tgridLine{0.5}{-0.5}{0.75}{-0.75}
    \circleb{-.45}{-.45}
     \circlelg{-.45}{.45}
     \circler{.45}{-.45}
     \circle{-0.75}{-0.75}
      \circle{0.75}{-0.75}
       \circle{-0.75}{0.75}
      \circle{0.75}{0.75};
 \end{tikzpicture}~.
\end{eqnarray}
Here the soliton in the upper left corner is a projector onto the  $j=1$ charge sector, $\Pi^{(l)}_1$. This quantity is the probability that left- and right-moving solitons produced by the initial state and scattered through the impurity produce a $j=1$ left-moving soliton. The remaining term in Eq.~\eqref{eq:channel_ent} is a classical R\'enyi mutual information, 
\begin{eqnarray}\label{eq:class_mut_info}
    I^{(n)}(\vartheta)=\frac{1}{n-1}\left[\log\!\sum_{j,k=0,1}\!\vartheta_{j,k}^n-\log\sum_{j=0,1}\!\Big(\!\sum_{k=0,1}\vartheta_{j,k}\Big)^n-\log\sum_{k=0,1}\!\Big(\!\sum_{j=0,1}\vartheta_{j,k}\Big)^n\,\right],
\end{eqnarray}
where $\vartheta_{j,k}$ is the joint probability distribution defined by the diagram
\begin{eqnarray}\label{eq:vartheta_imp}
    \vartheta_{j,k}=\begin{tikzpicture}[baseline={([yshift=-0.6ex]current bounding box.center)},scale=0.5]
   \prop{0}{0}{colimp}{1}
   \tgridLine{0.5}{0.5}{0.75}{0.75}
     \tgridLine{-0.5}{-0.5}{-1}{-1}
     \tgridLine{0.5}{0.5}{0.75}{0.75}
     \tgridLine{-0.5}{0.5}{-0.75}{0.75}
     \tgridLine{0.5}{0.5}{0.75}{0.75}
     \tgridLine{0.5}{-0.5}{0.75}{-0.75}
    \circledg{-.45}{-.45}
     \circlelg{-.45}{.45}
     \circler{.45}{-.45}
     \circleb{-0.75}{-0.75}
     \circle{-1.1}{-1.1}
      \circle{0.75}{-0.75}
       \circle{-0.75}{0.75}
      \circle{0.75}{0.75};
 \end{tikzpicture}~.
\end{eqnarray}
In this case, the soliton on the top left is a projector onto the $j$ charge sector, $\Pi^{(l)}_j$, while the soliton on the bottom left is a projector onto the $k$ charge sector, $\Pi^{(r)}_k$. From the definition, we have that marginals of this distribution satisfy $\sum_{k=0,1}\vartheta_{1,k}=\theta_{\rm imp}$ and $\sum_{k=0,1}\vartheta_{0,k}=1-\theta_{\rm imp}$, while $\sum_{j=0,1}\vartheta_{j,1}=\theta$ and $\sum_{j=0,1}\vartheta_{j,0}=1-\theta$. The distribution describes the probability of producing a charge $j$ by scattering a left-moving soliton emerging from the initial state with a right-moving soliton of charge $k$ that was produced by the initial state. We emphasize that if the impurity is symmetric, then the distribution $\vartheta_{j,k}$ factorizes and the mutual information vanishes. 

We take the replica limit in the above expressions and obtain a bound on the entanglement entropy of the state:
\begin{equation}\label{eq:bound}
    S\big[\rho_A(t)\big]\leq\begin{cases}(2\ell+r-t)s(\theta)+(t-r)s(\theta_{\rm imp})-(t-r)I(\vartheta)& \ell/2\leq t < \ell/2+r\,,\\
(2\ell+r-t)s(\theta)+(t-r)s(\theta_{\rm imp})-(\ell+r-t)I(\vartheta)& \ell/2+r\leq t < \ell+r\,,\\\ell s(\theta)+\ell s(\theta_{\rm imp})& \ell+r\leq t\,,
\end{cases}
\end{equation}
with $I(\vartheta)=\lim_{n\to 1}I^{(n)}(\vartheta)$. This expression has several important features that are worth discussing. We see that prior to $t=\ell+r$, the impurity gives rise to two different terms in the bound. The first of these is $s(\theta_{\rm imp})$, which we can interpret as a bound on the entanglement that can be generated by solitons that scatter through the impurity. The second term $I(\vartheta)$, being a mutual information measure, quantifies the correlations that are created within the subsystem by solitons scattering back into it due to the impurity. Since $I(\vartheta)\ge 0$, this contribution can only reduce the upper bound on the entanglement. 

This reduction arises from the fact that the impurity generates correlations, and therefore entanglement, between different parts of $A$; by the monogamy of entanglement, this reduces the extent to which $A$ can be entangled with $\bar A$. 
This non-positive contribution eventually disappears at $t=\ell+r$, as the correlations created between the different parts of $A$ cannot be sustained indefinitely, given that the solitons scattered back into $A$ must necessarily exit it. As mentioned above, this contribution to the bound is contingent on the impurity breaking charge symmetry; if the impurity is symmetric, then $I(\vartheta)=0$ and also $\theta_{\rm imp}=\theta$, so that the bound simplifies to 
\begin{eqnarray}
    S[\rho_A(t)]\leq 2\ell s(\theta)~,
\end{eqnarray}
for all times. 

The bound~\eqref{eq:bound} shows that some richer structure of entanglement evolution may be found in the intermediate-time regime, where the finite size of the subsystem can be felt, provided the impurity is symmetry-breaking. Indeed, the non-monotonic contribution to the bound suggests that the entropy itself might evolve non-monotonically. This behaviour was already seen within the quasiparticle picture in the uncharged case~\cite{fraenkel2024entanglement}. Evidently, in the presence of a bulk symmetry it can occur more generally, and does not require the quasiparticle picture. This mirrors the non-monotonic entropy bound observed in Ref.~\cite{foligno2024entanglement} for a disjoint subsystem of two well-separated intervals in a charged dual-unitary circuit (without an impurity). The bound was shown there to be exactly saturated for generic symmetric gates with no integrable structure.

\subsection{Quasiparticle picture}
Here we determine the prediction of the quasiparticle picture beyond the universal early-time regime. The procedure is the same as in the uncharged case, simply requiring the replacement of the bulk gates by the swap gate $P$. The diagram in Fig.~\ref{fig:rho_circuit} can then be completely reduced at all times. For $t\leq \ell/2$, the exact result is given in~\eqref{eq:early_time}, whereas at later times we find 
\begin{equation}\label{eq:inter_time}
    S^{(n)}[\rho_A(t)]=\begin{cases}
    (2\ell+2r-2t) s^{(n)}(\theta)+(t-r)s^{(n)}_{\rm imp}(\theta)~~&\ell/2\le t<\ell/2+r,\\
    \ell s^{(n)}(\theta)+(\ell+r-t)s^{(n)}_{\rm imp}(\theta)+(2t-\ell-2r)\bar{s}^{(n)}_{\rm imp}(\theta)~~&\ell/2+r\le t<\ell+r,\\
    \ell s^{(n)}(\theta)+\ell \bar{s}^{(n)}_{\rm imp}(\theta)&\ell+r\le t,\end{cases}
\end{equation}
where we have introduced
\be \label{eq:S_imp_inf_swap}
\bar{s}^{(n)}_{\rm imp}(\theta)=\frac{1}{1-n}\log  \left( \begin{tikzpicture}[baseline={([yshift=-0.6ex]current bounding box.center)},scale=0.5]
   \prop{0}{0}{colimp}{}
   \tgridLine{0.5}{0.5}{0.75}{0.75}
     \tgridLine{-0.5}{-0.5}{-0.75}{-0.75}
     \tgridLine{0.5}{0.5}{0.75}{0.75}
     \tgridLine{-0.5}{0.5}{-0.75}{0.75}
     \tgridLine{0.5}{0.5}{0.75}{0.75}
     \tgridLine{0.5}{-0.5}{0.75}{-0.75}
    \circleb{-.45}{-.45}
     \circler{.45}{-.45}
     \circle{-0.75}{-0.75}
      \circle{0.75}{-0.75}
       \square{-0.75}{0.75}
      \circle{0.75}{0.75};
 \end{tikzpicture}\right)~,
\ee 
which is the late-time impurity contribution to the stationary entanglement entropy. The expression~\eqref{eq:inter_time} has several features in common with the uncharged case (see Eq.~\eqref{eq:ent_swap_gates}). It can exhibit non-monotonic behaviour, and the time scales for the changes of slope, as well as for saturation, are the same as in the uncharged case. A new feature introduced by the presence of the charge is that the long-time entropy is affected by the impurity through the term $\bar{s}^{(n)}_{\rm imp}(\theta)$. In the replica limit, we define $\bar s_{\rm imp}(\theta)=\lim_{n\to 1} \bar s^{(n)}_{\rm imp}(\theta)$  and note that it obeys
\begin{equation}\label{eq:imp_bound}
    s(\theta)\leq\bar s_{\rm imp}(\theta)\leq s(\theta_{\rm imp})~.
\end{equation}
 The upper bound can be shown by comparing the prediction of the quasiparticle picture~\eqref{eq:inter_time} for $t\ge\ell+r$ to the universal upper bound~\eqref{eq:bound}, while the lower bound comes from considering the mutual information in the presence of a charged bulk. The latter will be discussed in Sec.~\ref{sec:mut_info_charged}.  
Thus, a symmetry-breaking impurity will generally result in a larger stationary entanglement entropy. A simple example of this is given by the case where the initial state has no charge fluctuations, $\theta\in\{0,1\}$. The impurity will then increase the entanglement by introducing charge fluctuations, thereby increasing the number entropy. This feature will also be present if the symmetry-breaking impurity is dual-unitary. 

\subsection{Entanglement membrane picture in a charged circuit}\label{sec:membrane_charged} 

We now turn to the prediction of the entanglement membrane picture, which, due to the presence of the charge, requires a modification of the approach described in Sec.~\ref{sec:membrane_uncharged}. The effective membrane theory is implemented on a specific diagram by first decomposing it into a sum over different charge sectors, applying the membrane picture to each sector separately, and then summing up the contributions from the different sectors~\cite{foligno2024nonequilibrium}. To achieve this, a resolution of the identity~\eqref{eq:completeness} is inserted on each leg of the diagram. For the first term in~\eqref{eq:D_def}, this looks like
\begin{align}
\label{eq:membrane_charge_decomposition}
\sum_{\textbf{j}^{(l)}_{1},\dots ,\,\textbf{j}^{(l)}_{2t}}\sum_{\textbf{j}^{(r)}_{1}\!,\dots, \,\textbf{j}^{(r)}_{2t}}\hspace{-0.5cm}
\begin{tikzpicture}[baseline={([yshift=1.2ex]current bounding box.center)},scale=0.5]
{\prop{2}{2}{colSt}{}}
  \foreach \x in {1,3}{\prop{\x}{1}{colSt}{}}   
    \foreach \x in {0,2,...,4}{\prop{\x}{0}{colSt}{}}
    \foreach \x in {-1,1,...,5}{\prop{\x}{-1}{colSt}{}}
    \foreach \x in {-2,0,...,6}{\prop{\x}{-2}{colSt}{}}
    \foreach \x in {-3.5,-1.5,...,6.5}{\istate{\x}{-2.5}{colSt}}
    \foreach \x in {-4,...,1}{\tgridLine{\x+0.25}{\x+1.75}{\x+0.5}{\x+1.5}
    \circle{\x+0.25}{\x+1.75}
    \circlelg{\x+0.55}{\x+1.45}}
    \foreach \x in {-4,...,1}{
    \tgridLine {4-\x-0.25}{\x+1.75} {4-\x-0.5}{\x+1.5}\square
    {4-\x-0.25}{\x+1.75}
    \circledg
    {4-\x-.55}{\x+1.45}}
  \end{tikzpicture} 
\end{align}
where we sum over $\textbf{j}^{(\sigma)}_{x}=(j^{(\sigma)}_{x,1},j^{(\sigma)}_{x,2},\dots,j^{(\sigma)}_{x,n})$ which indicates all possible insertions of projectors onto charge sectors on each site and on each replica (cf.~\eqref{eq:soliton_proj}). Explicitly,  $j^{(\sigma)}_{x,m}=0,1$ labels the projector inserted into the forward copy of the $m$th replica at site $x$. Each diagram in this sum can be considered to be coming from a circuit with no conserved quantities. Hence, we can apply the membrane picture to each diagram separately. 

For this, we proceed as before and pick a downward-directed path emanating from the interface between the circle and square states. For example,
\be \label{eq:membrane_charged}
\begin{tikzpicture}[baseline={([yshift=1.2ex]current bounding box.center)},scale=0.5]
{\prop{2}{2}{colSt}{}}
  \foreach \x in {1,3}{\prop{\x}{1}{colSt}{}}   
    \foreach \x in {0,2,...,4}{\prop{\x}{0}{colSt}{}}
    \foreach \x in {-1,1,...,5}{\prop{\x}{-1}{colSt}{}}
    \foreach \x in {-2,0,...,6}{\prop{\x}{-2}{colSt}{}}
    \foreach \x in {-3.5,-1.5,...,6.5}{\istate{\x}{-2.5}{colSt}}
    \foreach \x in {-4,...,1}{\tgridLine{\x+0.25}{\x+1.75}{\x+0.5}{\x+1.5}
    \circle{\x+0.25}{\x+1.75}
    \circlelg{\x+0.55}{\x+1.45}}
    \foreach \x in {-4,...,1}{
    \tgridLine {4-\x-0.25}{\x+1.75} {4-\x-0.5}{\x+1.5}\square
    {4-\x-0.25}{\x+1.75}
    \circledg
    {4-\x-.55}{\x+1.45}}
    \draw[very thick,dashed,red] (2,2.65)--(2,2)--(6,-2)--(6,-3) ;
  \end{tikzpicture}~.
\ee 
We then project either side of the membrane onto their relevant domains, taking into account the presence of the projector onto a particular charge sector. This is achieved by the following replacement:
\begin{eqnarray}\label{eq:proj_cricle_charge}
    \begin{tikzpicture}[baseline={([yshift=-.6ex]current bounding box.center)},scale=0.5]
        \prop{0}{0}{colSt}{}
         \draw[very thick,dashed,red] (-0.5,0.5)--(0.5,-0.5);
          \tgridLine{0.5}{0.5}{.75}{.75}
            \circledg{.5}{.5};
    \end{tikzpicture}
    \to \frac{1}{d^{2n}}\begin{tikzpicture}[baseline={([yshift=-.6ex]current bounding box.center)},scale=0.5]
        \prop{0}{0}{colSt}{}
         \tgridLine{1.25}{1.25}{1.75}{1.75} 
            \tgridLine{-1.25}{-1.25}{-1.75}{-1.75} 
            \tgridLine{-0.5}{-0.5}{-0.75}{-0.75}
              \tgridLine{0.5}{0.5}{0.75}{0.75}
        \square{.85}{.85}
        \circle{-.85}{-.85}
        \circle{-1.25}{-1.25}
        \square{.85}{.85}
        \square{1.25}{1.25}
        \circledg{-.5}{-.5}
         \circledg{.5}{.5};
    \end{tikzpicture}\, , 
\end{eqnarray}
and similarly for the beginning and end of the membrane, see \eqref{eq:membrane_replacements_top_bottom}.  Doing this for the full diagram, we obtain that \eqref{eq:membrane_charge_decomposition} is equivalent to 
\begin{align}
\frac{1}{d^{2nt}}\sum_{\textbf{j}^{(l)}_{1},\dots ,\,\textbf{j}^{(l)}_{2t}}\sum_{\textbf{j}^{(r)}_{1},\dots ,\,\textbf{j}^{(r)}_{2t}}
\hspace{-.4cm}\begin{tikzpicture}[baseline={([yshift=1.2ex]current bounding box.center)},scale=0.5]
{\foreach \x in {0,...,4}{\prop{3+\x}{3-\x}{colSt}{}}}
  \foreach \x in {1}{\prop{\x}{1}{colSt}{}}   
    \foreach \x in {0,2}{\prop{\x}{0}{colSt}{}}
    \foreach \x in {-1,1,...,3}{\prop{\x}{-1}{colSt}{}}
    \foreach \x in {-2,0,...,4}{\prop{\x}{-2}{colSt}{}}
    \foreach \x in {-3.5,-1.5,...,4.5,8.5}{\istate{\x}{-2.5}{colSt}}
    \tgridLine{8.25}{-2.25}{8.5}{-2.5}
    \square{8.15}{-2.15}
    \tgridLine{9.5}{-2.5}{9.75}{-2.25}
    \circledg{9.5}{-2.5}
    \circlelg{8.5}{-2.5}
     \square{9.85}{-2.15}
     \square{7.65}{-1.65}
    \foreach \x in {-4,...,0,2}{\tgridLine{\x+0.25}{\x+1.75}{\x+0.5}{\x+1.5}
    \circle{\x+0.25}{\x+1.75}
    \circlelg{\x+0.55}{\x+1.45}}
    \tgridLine{1.25}{4.75}{1.85}{4.15} \circle{1.25}{4.75}
        \circle{1.85}{4.15}
    \circlelg{1.55}{4.45}
    \foreach \x in {-3,...,1}{
    \tgridLine {4-\x-0.25}{\x+1.75} {4-\x-0.5}{\x+1.5}
    \tgridLine {3-\x-0.25}{\x+.75} {3-\x-0.5}{\x+.5}
     \tgridLine {3.5-\x-0.25}{\x+1.25} {4-\x-0.5}{\x+1.5}
    \circle
    {3.45-\x-0.25}{\x+1.2}
    \circle
    {3-\x-0.25}{\x+0.75}
    \circledg
    {4.1-\x-.55}{\x+1.55}
     \circledg
    {3-\x-.55}{\x+.45}
    \tgridLine {5-\x-0.25}{\x+2.75} {5-\x-0.5}{\x+2.5}\square
    {5-\x-0.25}{\x+2.75}
    \tgridLine{5.5-\x-0.25}{\x+3.25} {6.25-\x-0.25}{\x+4.}\square
    {5.5-\x-0.25}{\x+3.25}
     \square
    {6.25-\x-0.25}{\x+4.}
     \circledg
    {5.875-\x-0.25}{\x+3.625}
    \circledg
    {5-\x-.55}{\x+2.45}
    } ;
  \end{tikzpicture},
\end{align}
  where the coefficient $1/d^{2nt}$ comes from the normalization of the inserted projectors. Note that, within a given replica, the solitons are constrained to be identical along the diagonal on which they propagate, but each diagonal and each replica are summed independently. 
Contracting these diagrams, we arrive at,
\be 
\left(\sum^1_{\,\,\textbf{j}^{(l)},\textbf{j}^{(r)}=0}\!\!\begin{tikzpicture}[baseline={([yshift=-0.6ex]current bounding box.center)},scale=0.65]
  \istate{0}{0}{colSt}
  \circlelg{0.14}{-0.1}
    \circle{-0.12}{0.12}
    \circledg{0.85}{-0.1}
    \circle{1.12}{0.12}
\end{tikzpicture}\,\frac{\begin{tikzpicture}[baseline={([yshift=-0.6ex]current bounding box.center)},scale=0.65]
   \tgridLine{0}{0}{1}{0}
  \circledg{0.5}{0}
  \circle{0}{0};
  \square{1}{0};
;\end{tikzpicture}}{\begin{tikzpicture}[baseline={([yshift=-0.6ex]current bounding box.center)},scale=0.65]
   \tgridLine{0}{0}{1}{0}
  \circledg{0.5}{0}
  \circle{0}{0};
  \circle{1}{0};
;\end{tikzpicture}}\,\right)^{2t}=\left(\begin{tikzpicture}[baseline={([yshift=-0.6ex]current bounding box.center)},scale=0.65]
   \tgridLine{0}{0}{1}{0}
  \circleb{0.5}{0}
  \circle{0}{0};
  \square{1}{0};
;\end{tikzpicture}\right)^{2t}\, ,
\ee
where on the right-hand side the charge structure is that of the initial state. To obtain this equality, one simply sums over the left-moving solitons that are a resolution of the identity, and then uses the charge solvability of the initial state.  Depending on the choice of membrane path, we may be required to project instead via
\begin{equation}
    \begin{tikzpicture}[baseline={([yshift=-.6ex]current bounding box.center)},scale=0.5]
        \prop{0}{0}{colSt}{}
         \draw[very thick,dashed,red] (-0.5,-0.5)--(0.5,0.5);
          \tgridLine{-0.5}{0.5}{-.75}{.75}
            \circlelg{-.5}{.5};
    \end{tikzpicture}
    \to \frac{1}{d^{2n}}\begin{tikzpicture}[baseline={([yshift=-.6ex]current bounding box.center)},scale=0.5]
        \prop{0}{0}{colSt}{}
         \tgridLine{1.25}{-1.25}{1.75}{-1.75} 
            \tgridLine{-1.25}{1.25}{-1.75}{1.75} 
            \tgridLine{-0.5}{0.5}{-0.75}{0.75}
              \tgridLine{0.5}{-0.5}{0.75}{-0.75}
        \square{.85}{-.85}
        \circle{-.85}{.85}
        \circle{-1.25}{1.25}
        \square{.85}{-.85}
        \square{1.25}{-1.25}
        \circlelg{-.5}{.5}
         \circlelg{.5}{-.5};
    \end{tikzpicture} ~.
\end{equation}
However, for the diagram in~\eqref{eq:membrane_charged}, any downward path will give the same result. Using this approach for both diagrams appearing in~\eqref{eq:D_def}, we find agreement with the exact result for $t\le\ell/2$ that is given in~\eqref{eq:early_time}. Therefore, as was the case for the uncharged system, the membrane picture correctly reproduces the exact early-time result (and does so on any microscopic scale, not only after coarse-graining).

We can now go beyond this early-time regime by applying the same charged membrane picture to the diagram appearing in~\eqref{eq:inter}. As in Sec.~\ref{sec:membrane_uncharged}, there are two configurations that need to be considered. The membranes in these configurations follow the same paths as those in the uncharged case, i.e., Path I depicted in~\eqref{eq:path_I}  and Path II depicted in \eqref{eq:path_II}. The final prediction for the entanglement will then be given by the minimum between these two configurations,
\begin{equation}\label{eq:mem_minimum}
    S^{(n)}[\rho_A(t)]={\rm min}\{S^{(n)}_{\rm I}(t), S^{(n)}_{\rm II}(t)\},
\end{equation}
where $S^{(n)}_{\rm I}(t)$ and $S^{(n)}_{\rm II}(t)$ are the contributions from Paths I and II, respectively. For Path I, we find 
\begin{align}
S^{(n)}_{\rm I}(t)=\begin{cases}
    (\ell+2r)s^{(n)}(\theta)+(2t-\ell)s_{\rm conf}+(t-r)s^{(n)}_{\rm imp}(\theta) & \frac{\ell}{2}\le t<\frac{\ell}{2}+r,\\
    (\ell+2r) s^{(n)}(\theta)+2r s_{\rm conf}+(\ell+r-t)s^{(n)}_{\rm imp}(\theta)&\\
    \,\,\,\,\,+(2t-\ell-2r)\tilde{s}^{(n)}_{\rm imp}(\theta)&\frac{\ell}{2}+r\le t,
    \end{cases}
\end{align}
where we defined
\begin{align}\label{eq:S_imp_inf_chaotic}
    \tilde{s}^{(n)}_{\rm imp}(\theta)=\frac{1}{1-n}\log\Bigg( \sum_{j=0,1}\frac{\begin{tikzpicture}[baseline={([yshift=-0.6ex]current bounding box.center)},scale=0.5]
   \tgridLine{0}{0}{1}{0}
  \circle{-0.1}{0};
   \circleb{0.7}{0};
   \circledg{0.3}{0};
  \circle{1.1}{0};
\end{tikzpicture}\,\begin{tikzpicture}[baseline={([yshift=-0.6ex]current bounding box.center)},scale=0.5]
   \tgridLine{0}{0}{1}{0}
  \circle{0}{0};
  \circledg{0.5}{0};
  \square{1}{0};
\end{tikzpicture}}{\begin{tikzpicture}[baseline={([yshift=-0.6ex]current bounding box.center)},scale=0.5]
   \tgridLine{0}{0}{1}{0}
  \circle{0}{0};
  \circledg{0.5}{0};
  \circle{1}{0};
\end{tikzpicture}\,\begin{tikzpicture}[baseline={([yshift=-0.6ex]current bounding box.center)},scale=0.5]
   \tgridLine{0}{0}{1}{0}
  \square{0}{0};
  \circledg{0.5}{0};
  \square{1}{0};
\end{tikzpicture}}\,\begin{tikzpicture}[baseline={([yshift=-0.6ex]current bounding box.center)},scale=0.5]
   \prop{0}{0}{colimp}{}
   \tgridLine{0.5}{0.5}{0.75}{0.75}
     \tgridLine{-0.5}{-0.5}{-0.75}{-0.75}
     \tgridLine{0.5}{0.5}{0.75}{0.75}
     \tgridLine{-0.5}{0.5}{-0.75}{0.75}
     \tgridLine{0.5}{0.5}{0.75}{0.75}
     \tgridLine{0.5}{-0.5}{0.75}{-0.75}
    \circledg{-.45}{-.45}
     \circler{.45}{-.45}
     \square{-0.75}{-0.75}
      \circle{0.75}{-0.75}
       \square{-0.75}{0.75}
      \circle{0.75}{0.75};
 \end{tikzpicture}\Bigg)~.
\end{align}
The contribution from Path II can be determined by noting that the effect of decomposing the diagram into charge sectors and performing the projections along the membrane path is equivalent to passing the state through the channel $\mathcal{K}$ given in \eqref{eq:symm_channel} and then calculating the entropy. Accordingly, we have 
\begin{eqnarray}
    S_{\rm II}^{(n)}(t)= S^{(n)}\big[\mathcal{K}[\rho_A(t)]\big]~,
\end{eqnarray}
where $ S^{(n)}\big[\mathcal{K}[\rho_A(t)]\big]$ is given by~\eqref{eq:channel_ent}. 

Let us now comment on these results. Path I can be viewed as a continuation of the paths taken by the membranes at early times.  It is natural, therefore, that Path I continues to minimize the entropy beyond $t=\ell/2$, at least initially. Interestingly, the functional form of $S^{(n)}_{\rm I}(t)$ changes at $t=\ell/2+r$, which can be contrasted with its persistent functional form in the uncharged case (cf.~\eqref{eq:ent_mem_uncharged}). Here, the two-stage behaviour arises from the fact that the ballistic propagation of the solitons can establish long-range correlations between the two different membranes, forcing the membranes to have the same charge fluctuations. As a result, in the regime $t<\ell/2+r$ the bulk contribution is reduced from $(2t +2r)s^{(n)}(\theta)$, which one might expect in the absence of symmetry, to $ (\ell+2r)s^{(n)}(\theta)+(2t-\ell)s_{\rm conf}$. 

Since $S^{(n)}_{\rm I}(t)$ increases with time, the dominant path must eventually switch to Path II. Indeed, since $S^{(n)}\big[\mathcal{K}[\rho_A(t)]\big]$ upper-bounds the entanglement, we are assured that Path II is preferred at long times, and thus the membrane picture predicts a saturation of the bound. The time at which Path II takes over depends both on the impurity gate and on $\theta$; it may occur relatively early, as the examples we present in Sec.~\ref{sec:compare_qp_mem_ent} show. Consequently, in the presence of a charged bulk, the membrane picture allows for a non-monotonic evolution of the entanglement entropy. This is in contrast to the case without a charge-conserving bulk, and can be attributed to the buildup of long-range charge correlations within the subsystem.

While we applied the membrane picture within each charge sector simply by following the same membrane paths as in the uncharged circuit, there is an important difference between the cases that needs to be addressed. Namely, while the bulk gates remain dual-unitary when projected onto any charge sector by construction, the impurity gate does not necessarily remain unitary under such projection. In other words, we have implicitly relied on the assumption that Path I will not change in an uncharged circuit if the impurity gate is non-unitary. We can corroborate this assumption by turning again to the exactly solvable scenario of $n=2$ replicas, maximally mixing bulk gates (in each charge sector), and large $d$. In Appendix~\ref{app:Non_unitary_imp} we show that, even with a non-unitary impurity gate, the same membrane picture gives the leading-order result for the subsystem purity (which can no longer be directly associated with entanglement for a non-unitary evolution, though the summation over charge sectors eventually fixes this). The calculation is performed by modifying the techniques of Ref.~\cite{fraenkel2024entanglement}, which are recalled in Appendix~\ref{app:chaotic_2_interval_entropy}.

\subsection{Comparing the quasiparticle and membrane pictures}\label{sec:compare_qp_mem_ent}
\begin{figure}
    \centering
    \includegraphics[width=\linewidth]{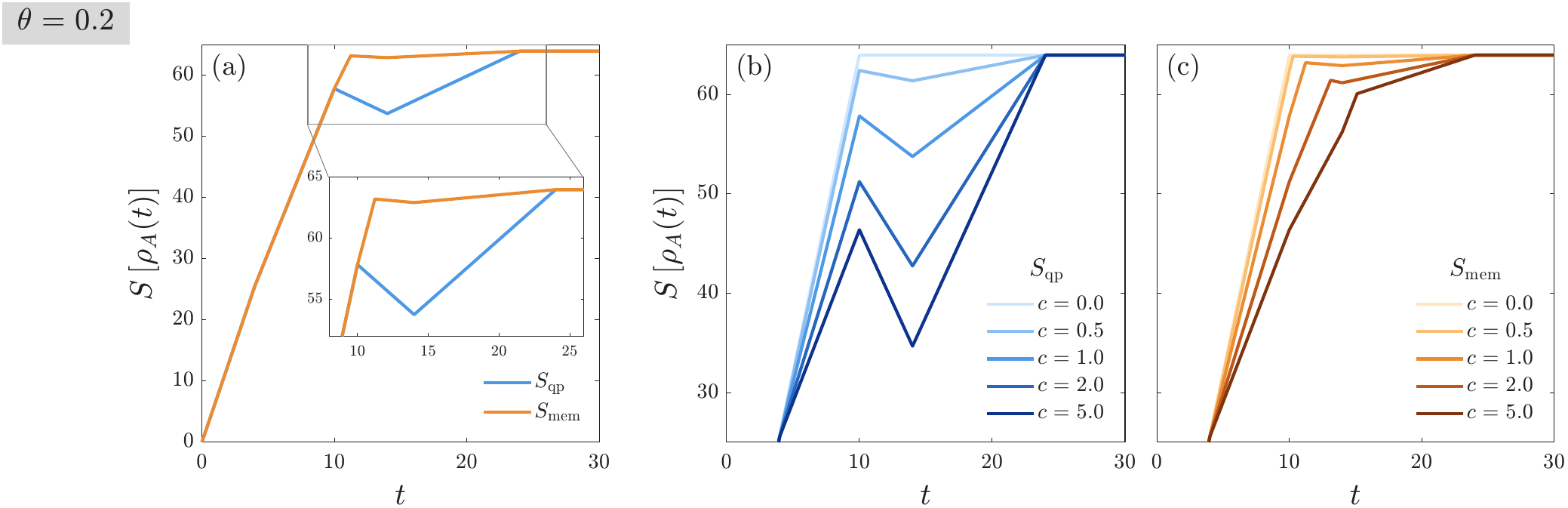}
    \includegraphics[width=\linewidth]{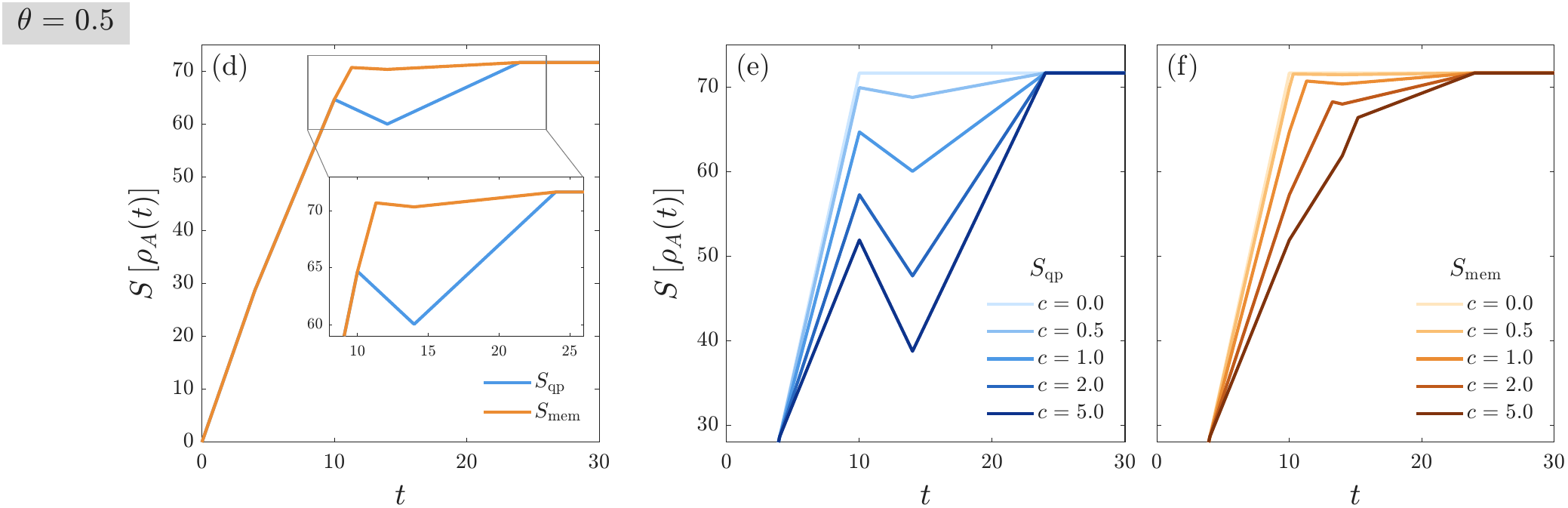}
    \caption{The single-interval entanglement entropy evolution within a charged dual-unitary circuit with an impurity $U_{\rm Id}(c)$, as defined in Eq.~\eqref{eq:ISWAP_impurity_def}. The local charge-sector dimension is $d=3$, and the parameters of the interval $A$ are fixed at $\ell=20$ and $r=4$. Top panels: Results for a symmetry-broken initial state with $\theta=0.2$; Panel (a) compares the predictions of the quasiparticle and membrane pictures for an impurity parameter $c=1$, while Panels (b) and (c) show the quasiparticle picture and membrane picture results (respectively) for various values of $c$. Bottom panels: Similar to top panels, but with $\theta=0.5$.}
    \label{fig:Single_interval_entropy_ISWAP}
\end{figure}

\begin{figure}
    \centering
    \includegraphics[width=\linewidth]{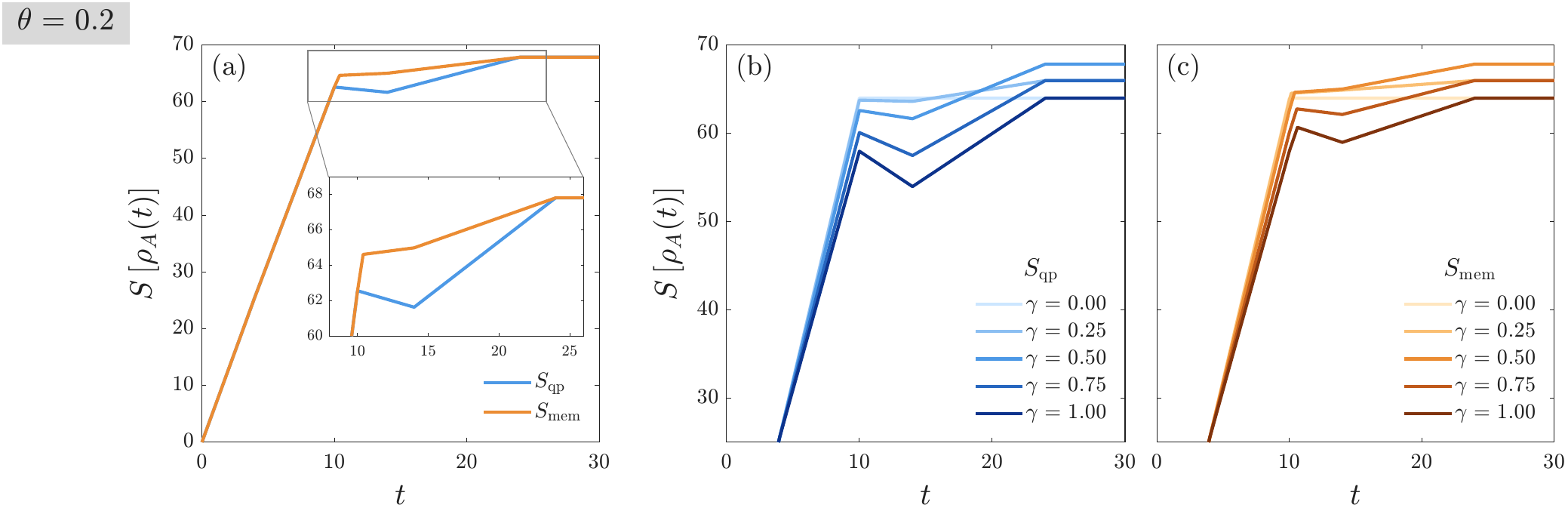}
    \includegraphics[width=\linewidth]{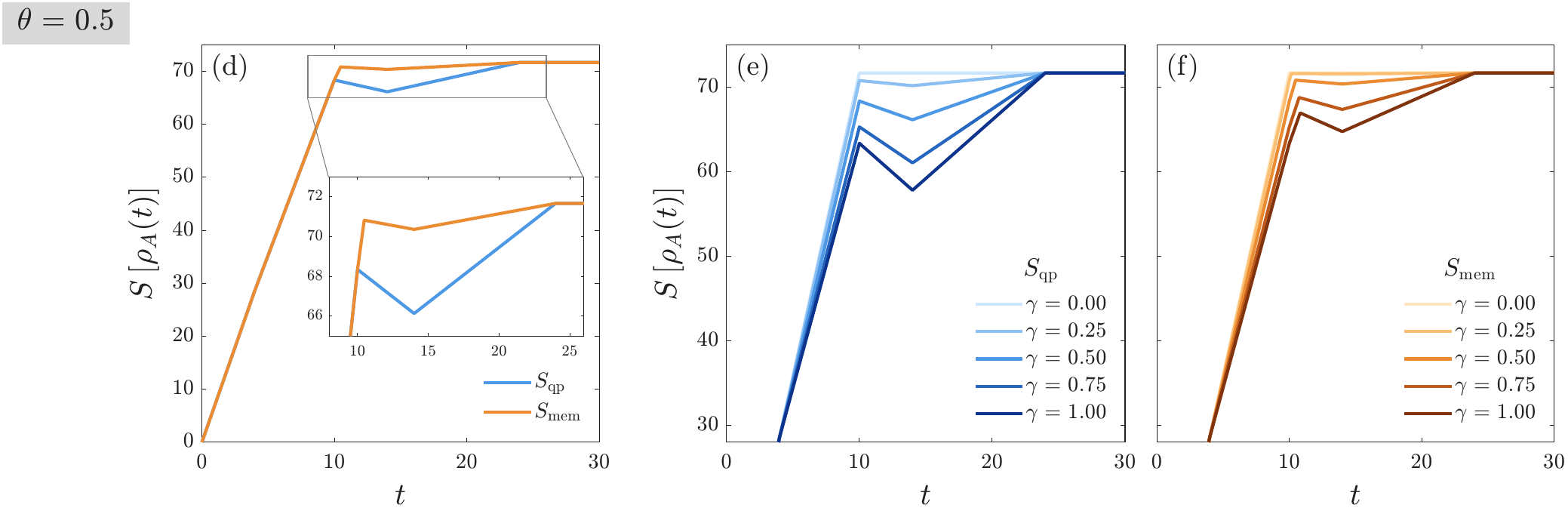}
    \caption{The single-interval entanglement entropy evolution within a charged dual-unitary circuit with an impurity $U_{\rm XY}(\gamma)$, as defined in Eq.~\eqref{eq:XY_impurity_def}. The local charge-sector dimension is $d=3$, and the parameters of the interval $A$ are fixed at $\ell=20$ and $r=4$. Top panels: Results for an initial state with $\theta=0.2$; Panel (a) compares the predictions of the quasiparticle and membrane pictures for an impurity parameter $\gamma=0.5$, while Panels (b) and (c) show the quasiparticle picture and membrane picture results (respectively) for various values of $\gamma$. Bottom panels: Similar to top panels, but with a maximally symmetry-breaking initial state (i.e., $\theta=0.5$).}
    \label{fig:Single_interval_entropy_XY}
\end{figure}

The predictions of the quasiparticle and membrane pictures are presented in Figs.~\ref{fig:Single_interval_entropy_ISWAP} and \ref{fig:Single_interval_entropy_XY} for the two types of impurities defined in Eqs.~\eqref{eq:ISWAP_impurity_def} and \eqref{eq:XY_impurity_def}, respectively. Recall that in an uncharged dual-unitary circuit with an impurity, the quasiparticle picture predicts a non-monotonic evolution of the entanglement entropy, while within the membrane picture the entropy must grow monotonically. This qualitative distinction breaks down in a charged circuit, as can be seen for both impurity types. In both cases, there are parameter values for which the membrane picture yields an evolution stage of decreasing entropy. Within the membrane picture, it can be checked that the non-monotonic behaviour occurs as a result of the contribution from Path II, \eqref{eq:path_II}, whose value coincides with the bound derived earlier, \eqref{eq:bound}. Thus, we attribute this behaviour to the buildup of charge correlations between different parts of the subsystem due to solitons scattering off the impurity. 

Our results further indicate that, when comparing the two effective pictures for the same choice of impurity, the entanglement entropy yielded by the membrane picture is larger than or equal to the prediction of the quasiparticle picture. This behaviour is similar to our findings in the uncharged circuits. The predictions of the two pictures also seem to converge to similar saturation values. While in uncharged circuits the exact saturation value was indeed independent of the bulk, in charged circuits the quasiparticle and membrane pictures formally predict different saturation values (see Eqs.~\eqref{eq:channel_ent} and~\eqref{eq:inter_time}). Nevertheless, the similarity between these two saturation values is remarkably consistent among the cases we checked; even when we examined a Haar-random unitary impurity gate, the membrane-picture saturation entropy was only negligibly larger than the saturation entropy computed within the quasiparticle picture. 

A clear difference between the two types of impurities studied in Figs.~\ref{fig:Single_interval_entropy_ISWAP} and \ref{fig:Single_interval_entropy_XY} is apparent when examining the entropy saturation values for different values of the impurity parameter. For the impurity gate $U_{\rm Id}(c)$, Fig.~\ref{fig:Single_interval_entropy_ISWAP} shows that this saturation value is independent of the parameter $c$, which encodes the interpolation between the swap gate and the identity gate. In contrast, for the impurity gate $U_{\rm XY}(\gamma)$, the saturation value does in fact depend on $\gamma$, unless the initial state maximally breaks the symmetry (i.e., unless $\theta=1/2$); this is exemplified in the top panels of Fig.~\ref{fig:Single_interval_entropy_XY}, and was verified for additional values of $\theta\neq1/2$ (for $\theta=1/2$ the saturation value is $\gamma$-independent, as seen in the bottom panels of Fig.~\ref{fig:Single_interval_entropy_XY}). This qualitative difference between the two families of impurity gates presumably arises from the fact that, while $U_{\rm XY}(\gamma)$ generally breaks the global $U(1)$ symmetry altogether, the impurity gate $U_{\rm Id}(c)$ preserves this symmetry and only breaks the emergent local chiral conservation laws.

 \section{Dynamics of disjoint intervals in a charged-bulk circuit}\label{sec:mut_info_charged}
We now re-examine the behaviour of the mutual information and negativity for a charged-bulk circuit.  Like in the case of the uncharged circuit, these quantities have a universal time regime that can be accessed by relying on the dual unitarity of the bulk, along with the charge solvability of the initial state.  This regime extends up to the time when the finite size of the subsystems $A_1$ and $A_2$ can be felt (i.e., when the two edges of each interval become causally connected). Proceeding as in the uncharged case but taking into account the relations~\eqref{eq:charge_solvable_1}, we find that the mutual information in the early-time regime is 
\begin{equation}\label{eq:mut_info_early_time}
    \mathcal{I}_{A_1:A_2}^{(n)}(t)=\begin{cases}
        0 & t<r,\\
        2(t-r)s_{\rm imp}^{(n)}(\theta) & r\le t <\ell/2.
    \end{cases}
\end{equation}
The result is similar to~\eqref{eq:mut_info_qpp_uncharged} and~\eqref{eq:renyi_mutual} at early times, but with an appropriate replacement of the impurity term to account for the charge structure.  Correlations between the two subsystems can be created at a rate that depends on the impurity. If the impurity is purely reflective, i.e., $s^{(n)}_{\rm imp}(\theta)=0$, these correlations are obstructed, and the mutual information vanishes.  On the other hand, if $s_{\rm imp}^{(n)}(\theta)>2s^{(n)}(\theta)$, the impurity enhances the mutual information between the subsystems relative to a circuit without an impurity, a scenario which is not possible without the bulk charge. This can be achieved by choosing a symmetric initial state, such that the impurity then creates correlated charge fluctuations between $A_1$ and $A_2$.

\subsection{Quasiparticle picture}
The prediction of the quasiparticle picture is easily obtained by replacing any bulk gates with $P$ and then contracting the associated diagram fully.  This reproduces the exact early-time behaviour given in~\eqref{eq:mut_info_early_time}.  At all times, we find 
 \begin{equation}\label{eq:Renyi_MI_charged}
    \mathcal{I}_{A_1:A_2}^{(n)}(t)=\begin{cases} 0 & t<r,\\
        2(t-r)s_{\rm imp}^{(n)}(\theta) & r\le t <\ell/2+r, \\
        4(t-\ell/2-r)(\bar{s}_{{\rm imp}}^{(n)}(\theta)-s^{(n)}(\theta))+2(\ell+r-t)s_{\rm imp}^{(n)}(\theta) & \ell/2+r\le t <\ell+r,\\
        2\ell(\bar s_{{\rm imp}}^{(n)}(\theta)-s^{(n)}(\theta)) & \ell+r\le t,
    \end{cases}
\end{equation}
where $\bar{s}_{\rm imp}^{(n)}(\theta)$ is defined in~\eqref{eq:S_imp_inf_swap}. Upon taking the replica limit, we get
 \begin{equation}
    \mathcal{I}_{A_1:A_2}(t)=\begin{cases} 0 & t<r,\\
        2(t-r)s_{\rm imp}(\theta) & r\le t <\ell/2+r,\\
        4(t-\ell/2-r)(\bar{s}_{{\rm imp}}(\theta)-s(\theta))+2(\ell+r-t)s_{\rm imp}(\theta) & \ell/2+r\le t <\ell+r,\\
        2\ell(\bar s_{{\rm imp}}(\theta)-s(\theta)) & \ell+r\le t.
    \end{cases}
\end{equation}
Compared to the result for an uncharged circuit in~\eqref{eq:mut_info_qpp_uncharged}, we see some striking differences.  Most notably, the long-time limit can be non-vanishing, with a density $\bar s_{{\rm imp}}(\theta)-s(\theta)\geq 0$. This is non-negative by definition of the mutual information, and implies the lower bound stated in~\eqref{eq:imp_bound}. When the impurity is symmetric, the long-time limit vanishes. Moreover, for $\theta=1/2$ we reproduce the result of the uncharged circuit, for which the mutual information also vanishes at long times. Aside from the non-vanishing long-time value, the presence of the symmetry-breaking impurity is signalled by the slower decrease of $\mathcal{I}_{A_1:A_2}(t)$ in the window $\ell/2+r\leq t\leq \ell+r$. 

\subsection{Entanglement membrane picture}
Let us now contrast the above behaviour with the one predicted by the entanglement membrane picture. To do this, we discuss the behaviour of the entanglement entropy for the disjoint subsystem $A=A_1\cup A_2$, from which we can determine $\mathcal{I}^{(n)}_{A_1:A_2}(t)$. At early times, $t<\ell/2$, the membrane picture can reproduce the exact calculation using dual unitarity and therefore agrees with the quasiparticle picture. 
At later times, however, they provide different predictions. 
Similar to the uncharged bulk circuit, two membrane configurations must be considered to compute $S^{(n)}[\rho_A(t)]$, with the final prediction being given by the minimum of the two expressions, as in~\eqref{eq:mem_minimum}. The first path (Path I) is the same as the one depicted in~\eqref{eq:early_time_two_interval_uncharged} for the uncharged case. Recalling that we must first decompose the diagram into charge sectors prior to using the membrane picture, we find
\begin{equation}
    S_{\rm I}^{(n)}(t)=\begin{cases}
        (2\ell+4r) s^{(n)}(\theta) + (4t-2\ell)s_{{\rm conf}} & \ell/2\le t<\ell/2+r,\\
        4t s^{(n)}(\theta) + 4r s_{{\rm conf}} &  \ell/2+r\le t <\ell+r.
    \end{cases}
\end{equation}
Similar to the uncharged circuit, this result does not depend on the impurity.  Due to the presence of the bulk charge, however, the same membrane configuration can exhibit a change in behaviour at $t=\ell/2+r$, triggered by the generation of long-range charge correlations between the membranes. 

The second configuration to consider (Path II) is the one depicted in~\eqref{eq:early_time_two_interval_uncharged_II}. 
The contribution from this diagram depends on the choice of impurity, as the two disjoint parts of the membrane interact through it. It can be generally expressed using two entropy
terms: one where the impurity gate mediates the interaction between a single point on each membrane, and one that involves two points on each membrane. Overall, we find that
\begin{equation}
    S_{\rm II}^{(n)}(t)=\begin{cases}
        2(2\ell+r-t) s^{(n)}(\theta) + 2(t-r)s^{(n)}(\theta_{\rm imp})-(t-r)J^{(n)}(\phi) & \ell/2\le t<\ell/2+r,\\
          2(2\ell+r-t) s^{(n)}(\theta) + 2(t-r)s^{(n)}(\theta_{\rm imp})&\\
          \hspace{60pt}-(\ell+r-t)J^{(n)}(\phi)+(\ell+2r-2t)I^{(n)}(\varphi) &  \ell/2+r\le t <\ell+r,\\
        2\ell [s^{(n)}(\theta)+s^{(n)}(\theta_{\rm imp})]-\ell I^{(n)}(\varphi) &  \ell+r\le t ,
    \end{cases}
\end{equation}
which contains the two new terms. Here, $I^{(n)}(\varphi)$ is the classical mutual information introduced in~\eqref{eq:class_mut_info}, evaluated for the distribution defined by 
\begin{eqnarray}
    \varphi_{j,k}=
\begin{tikzpicture}[baseline={([yshift=-0.6ex]current bounding box.center)},scale=0.5]
   \prop{0}{0}{colimp}{1}
   \tgridLine{0.5}{0.5}{0.75}{0.75}
     \tgridLine{-0.5}{-0.5}{-0.75}{-0.75}
     \tgridLine{0.5}{0.5}{0.75}{0.75}
     \tgridLine{-0.5}{0.5}{-0.75}{0.75}
     \tgridLine{0.5}{0.5}{0.75}{0.75}
     \tgridLine{0.5}{-0.5}{0.75}{-0.75}
    \circleb{-.45}{-.45}
     \circler{.45}{-.45}
     \circlelg{-.45}{.45}
     \circledg{.45}{.45}
     \circle{-0.75}{-0.75}
      \circle{0.75}{-0.75}
       \circle{-0.75}{0.75}
      \circle{0.75}{0.75};
 \end{tikzpicture}\, ,
\end{eqnarray}
where the top left and right solitons are projectors onto the $j$ and $k$ charge sectors respectively. This captures charge correlations between $A_1$ and $A_2$. The other new term, $J^{(n)}(\phi)$, also encodes charge correlations between the different parts of the subsystem.  We introduce the distribution $\phi_{ijkl}$, which is defined through the diagrams
\begin{eqnarray}
    \phi_{ijkl}=\begin{tikzpicture}[baseline={([yshift=-0.6ex]current bounding box.center)},scale=0.5]
   \prop{0}{0}{colimp}{1}
   \tgridLine{0.5}{0.5}{0.75}{0.75}
     \tgridLine{-0.5}{-0.5}{-1}{-1}
     \tgridLine{0.5}{0.5}{0.75}{0.75}
     \tgridLine{-0.5}{0.5}{-0.75}{0.75}
     \tgridLine{0.5}{0.5}{0.75}{0.75}
     \tgridLine{0.5}{-0.5}{1}{-1}
    \circledg{-.45}{-.45}
     \circlelg{-.45}{.45}
     \circledg{.45}{.45}
     \circlelg{.45}{-.45}
     \circleb{-0.75}{-0.75}
     \circler{0.75}{-0.75}
     \circle{-1.1}{-1.1}
      \circle{1.1}{-1.1}
       \circle{-0.75}{0.75}
      \circle{0.75}{0.75};
 \end{tikzpicture}\, .
\end{eqnarray}
The light and dark grey solitons are projectors onto the charge sectors $i,j,k,l$, starting from the top left and proceeding clockwise, while, as usual, the blue and red ones carry the initial state charge structure. The quantity $J^{(n)}(\phi)$ has the form of a mutual information over this distribution
\begin{eqnarray}\nonumber
    J^{(n)}(\phi)&=&\frac{1}{n-1}\Big[\log\!\sum_{i,j,k,l=0,1}\!\!\!\phi_{ijkl}^n-\log\sum_{i=0,1}\!\big(\!\sum_{j,k,l=0,1}\phi_{ijkl}\big)^n-\log\sum_{j=0,1}\!\big(\!\sum_{i,k,l=0,1}\phi_{ijkl}\big)^n\\
    &&\quad\quad-\log\sum_{k=0,1}\!\big(\!\sum_{i,j,l=0,1}\phi_{ijkl}\big)^n-\log\sum_{l=0,1}\!\big(\!\sum_{i,j,k=0,1}\phi_{ijkl}\big)^n\Big]~.
\end{eqnarray}
We can interpret it as measuring how close $\phi_{ijkl}$ is to being completely factorized. It captures correlations between two different parts of $A_1$ and $A_2$. Overall, we can interpret the contribution of Path II as consisting of terms that arise from the bulk and impurity gates: $s^{(n)}(\theta)$ and $s^{(n)}(\theta_{\rm imp})$, respectively. This is then reduced by the terms $J^{(n)}(\phi)$ and $I^{(n)}(\varphi)$, which account for the fact that the impurity creates non-trivial charge correlations within the subsystem $A$, thereby preventing entanglement with $\bar A$.

One may check that, for all possible impurity gates, Path I gives lower entropy than Path II for $\ell/2\le t<\ell-r$.  At long times, however, Path II becomes the dominant one, as $S_{\rm II}^{(n)}(t)$ converges to a time-independent value while $S_{\rm I}^{(n)}(t)$ continues to grow. From this, we infer that the mutual information between the intervals converges at $t=\ell+r$ to
\begin{equation}
    \mathcal{I}_{A_1:A_2}^{(n)} (t\ge\ell+r) = \ell I^{(n)}(\varphi).
\end{equation}
Thus, the membrane picture predicts that the mutual information can be non-vanishing at long times, and this coincides with the classical mutual information between the charge distributions in each of the intervals. In the absence of the impurity, it has been shown that charged chaotic circuits can exhibit a quasiparticle-like transient peak in the mutual information due to the classical charge correlations between the two intervals~\cite{foligno2024entanglement}. In the limit where the impurity is symmetric, we recover those results. 

\begin{figure}
    \centering
    \includegraphics[width=\linewidth]{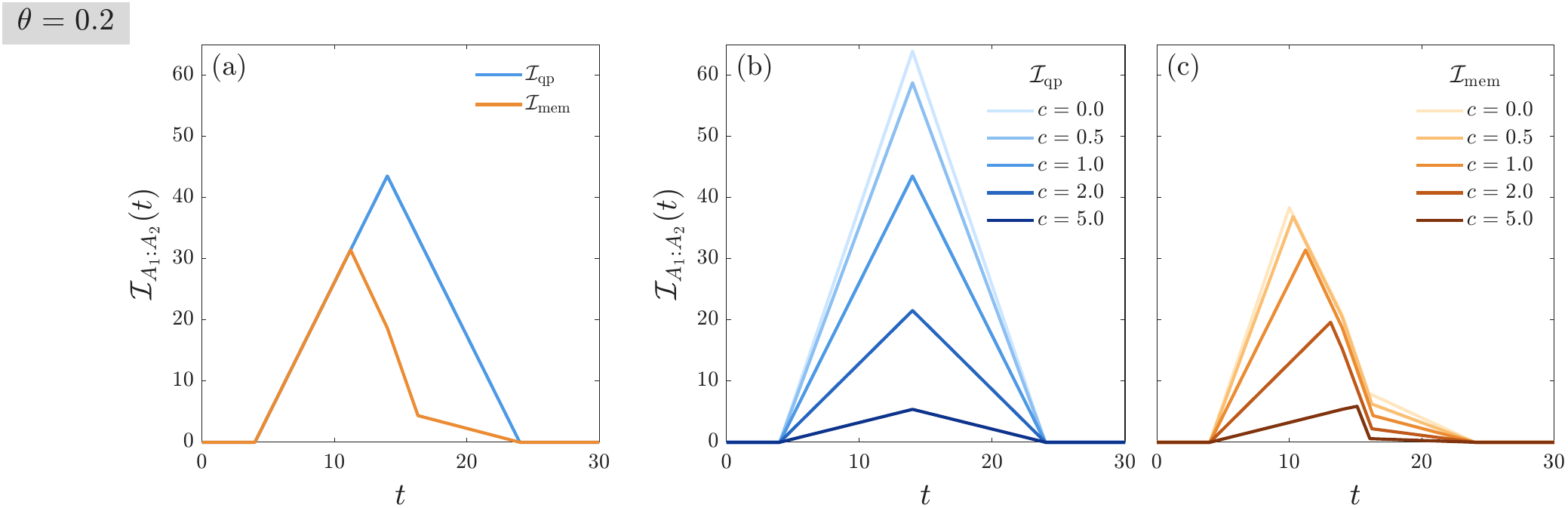}
    \includegraphics[width=\linewidth]{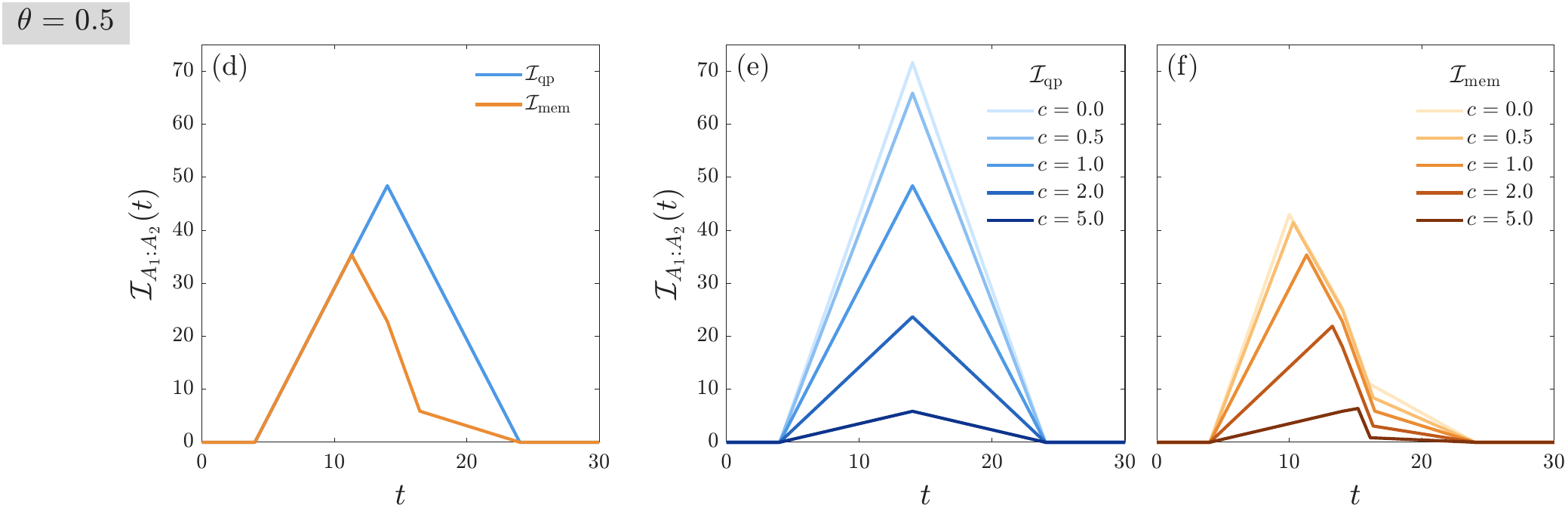}
    \caption{The evolution of the mutual information between two intervals in a charged dual-unitary circuit with an impurity $U_{\rm Id}(c)$, as defined in Eq.~\eqref{eq:ISWAP_impurity_def}. The local charge-sector dimension is $d=3$, and the parameters of the intervals $A_1$ and $A_2$ are fixed at $\ell=20$ and $r=4$. Top panels: Results for a symmetry-broken initial state with $\theta=0.2$; Panel (a) compares the predictions of the quasiparticle and membrane pictures for an impurity parameter $c=1$, while Panels (b) and (c) show the quasiparticle picture and membrane picture results (respectively) for various values of $c$. Bottom panels: Similar to top panels, but with $\theta=0.5$.}
    \label{fig:Two_interval_MI_ISWAP}
\end{figure}

\subsection{Comparing the quasiparticle and membrane pictures}
\begin{figure}
    \centering
    \includegraphics[width=\linewidth]{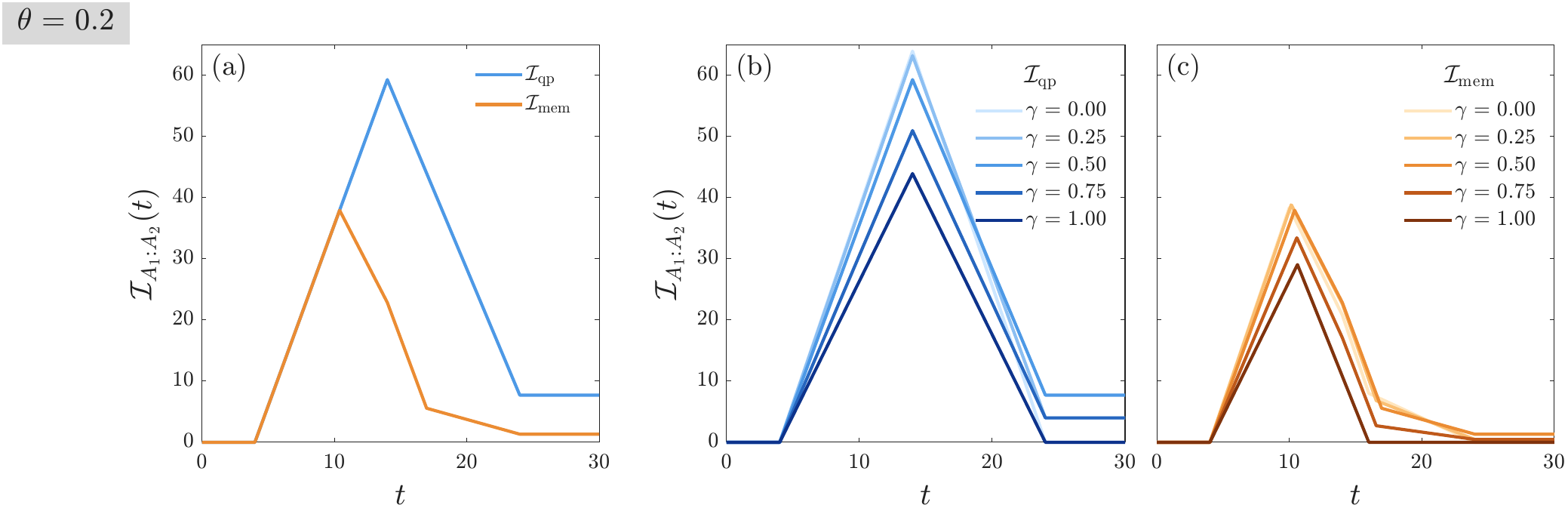}
    \includegraphics[width=\linewidth]{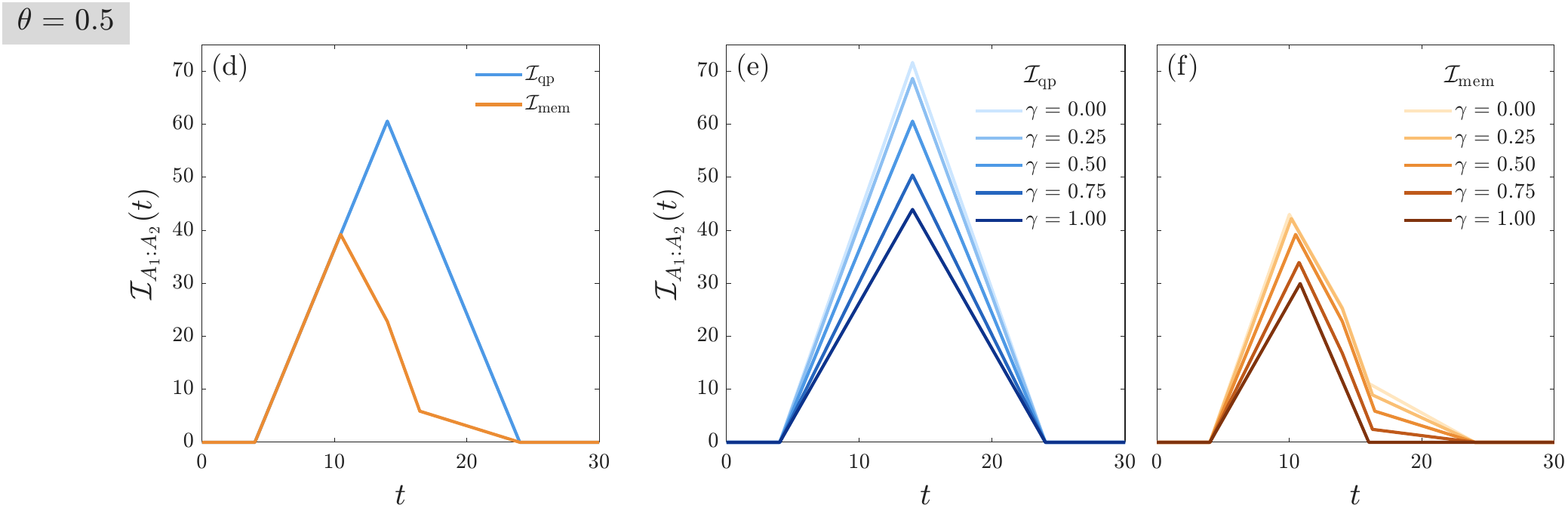}
    \caption{The evolution of the mutual information between two intervals in a charged dual-unitary circuit with an impurity $U_{\rm XY}(\gamma)$, as defined in Eq.~\eqref{eq:XY_impurity_def}. The local charge-sector dimension is $d=3$, and the parameters of the intervals $A_1$ and $A_2$ are fixed at $\ell=20$ and $r=4$. Top panels: Results for a symmetry-broken initial state with $\theta=0.2$; Panel (a) compares the predictions of the quasiparticle and membrane pictures for an impurity parameter $\gamma=0.5$, while Panels (b) and (c) show the quasiparticle picture and membrane picture results (respectively) for various values of $\gamma$. Bottom panels: Similar to top panels, but with a maximally symmetry-breaking initial state (i.e., $\theta=0.5$).}
    \label{fig:Two_interval_MI_XY}
\end{figure}

To compare the predictions of the two effective pictures for the mutual information dynamics, we plot them in Figs.~\ref{fig:Two_interval_MI_ISWAP} and \ref{fig:Two_interval_MI_XY} for the impurities defined in Eqs.~\eqref{eq:ISWAP_impurity_def} and \eqref{eq:XY_impurity_def}, respectively (the different values of the impurity and initial-state parameters are the same as those chosen for the single-interval entropy dynamics shown in Figs.~\ref{fig:Single_interval_entropy_ISWAP} and \ref{fig:Single_interval_entropy_XY}). Within the quasiparticle picture, the mutual information exhibits a linear rise followed by a linear fall before arriving at its final value, and its peak is always reached at $t=\ell/2+r$. In contrast, the membrane picture predicts a more intricate rise-and-fall profile with multiple changes of slope, and the time at which the mutual information peaks depends on the impurity.

As in the case of the single-interval entropy (cf.~Figs.~\ref{fig:Single_interval_entropy_ISWAP} and \ref{fig:Single_interval_entropy_XY}), the long-time values to which the mutual information converges reveal an important difference between the two impurity types: While for the impurity gate $U_{\rm Id}(c)$ the mutual information eventually vanishes, for the impurity gate $U_{\rm XY}(\gamma)$ it may converge to a positive value. This difference again arises because $U_{\rm Id}(c)$ respects the $U(1)$ symmetry while $U_{\rm XY}(\gamma)$ generically breaks it. The top panels of Fig.~\ref{fig:Two_interval_MI_XY} also indicate that the long-time mutual information predicted by the quasiparticle picture, when positive, is considerably larger than that predicted by the membrane picture for the same impurity and initial state. Unsurprisingly, the integrable bulk allows for stronger correlations between the two intervals in the long-time limit. This behaviour of the mutual information should be contrasted with that of the single-interval entropy, where we observed that the quasiparticle and membrane pictures both predict a very similar saturation entropy.

\begin{figure}
    \centering
    \includegraphics[width=\linewidth]{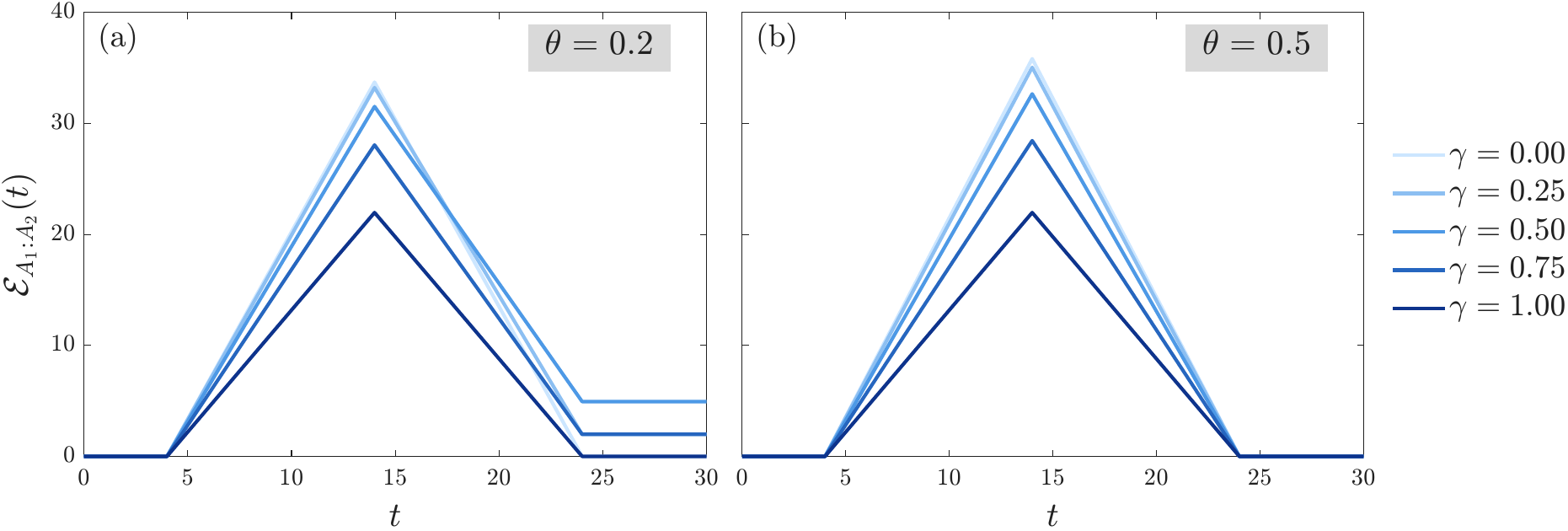}
    \caption{The evolution of the entanglement negativity between two intervals in a charged dual-unitary circuit with an impurity $U_{\rm XY}(\gamma)$, as defined in Eq.~\eqref{eq:XY_impurity_def}. The local charge-sector dimension is $d=3$, and the parameters of the intervals $A_1$ and $A_2$ are fixed at $\ell=20$ and $r=4$. Panels (a) and (b) show the prediction of the quasiparticle picture for various values of the impurity parameter $\gamma$. (a) Results for an initial state with $\theta=0.2$. (b) Same as (a), but with $\theta=0.5$.}
    \label{fig:Neg_Charged_XY}
\end{figure}

\subsection{Negativity dynamics}
The final quantity we wish to study is the entanglement negativity. We start from the diagram in Fig.~\ref{fig:TwoIntervalPartitionFuncNeg} and reduce it using dual unitarity and charge solvability. This can be accomplished for any choice of dual-unitary bulk, as long as $t<\ell/2$. Indeed, one finds that
\begin{equation}
    {\cal E}^{(n)}_{A_1:A_2}(t)=\begin{cases}
        8t(1-n)s^{(n)} (\theta)& t<r\,,\\
        4(t+r)(1-n)s^{(n)}(\theta) + (t-r)(2-n)s_{\rm imp}^{(n/2)}(\theta) & r\le t <\ell/2\,.
    \end{cases}
\end{equation}
We can obtain the quasiparticle prediction for all times by once again replacing all bulk gates by $P$. This reproduces the exact result at early times, and overall gives us
\begin{equation}\label{eq:Renyi_neg_charged}
    {\cal E}^{(n)}_{A_1:A_2}(t)=\begin{cases}
        4(\ell+r-t)(1-n)s^{(n)}(\theta) + (t-r)(2-n)s_{\rm imp}^{(n/2)}(\theta) & \ell/2\le t <\ell/2+r\,,\\
        2\ell(1-n)s^{(n)}(\theta)+ (\ell+r-t)(2-n)s_{\rm imp}^{(n/2)}(\theta)&\\
          \hspace{120pt}+(2t-\ell-2r)\bar{\varepsilon}_{{\rm imp}}^{(n)} (\theta)& \ell/2+r\le t <\ell+r\,,\\
        2\ell(1-n)s^{(n)}(\theta)+ \ell\bar{\varepsilon}_{{\rm imp}}^{(n)} (\theta)& t\ge \ell+r\,,
        
    \end{cases}
\end{equation}
where we defined the impurity contribution at long times as
\begin{equation}
     \bar{\varepsilon}_{\rm imp}^{(n)}(\theta)= \log\left(\, \begin{tikzpicture}[baseline={([yshift=-0.6ex]current bounding box.center)},scale=0.5]
   \prop{0}{0}{colimp}{}
   \tgridLine{0.5}{0.5}{0.75}{0.75}
     \tgridLine{-0.5}{-0.5}{-0.75}{-0.75}
     \tgridLine{0.5}{0.5}{0.75}{0.75}
     \tgridLine{-0.5}{0.5}{-0.75}{0.75}
     \tgridLine{0.5}{0.5}{0.75}{0.75}
     \tgridLine{0.5}{-0.5}{0.75}{-0.75}
    \circleb{-.45}{-.45}
     \circler{.45}{-.45}
     \circle{-0.75}{-0.75}
      \circle{0.75}{-0.75}
       \square{-0.75}{0.75}
      \triangleDown{0.75}{0.75};
 \end{tikzpicture}\,\right).
\end{equation}
This contribution can no longer be generally expressed in terms of a R\'enyi entropy as in previous calculations, which have all boiled down to the spectrum of a partially transposed \textit{pure-state} density matrix. Here, instead, $\bar{\varepsilon}_{\rm imp}^{(n)}(\theta)$ amounts to the R\'enyi negativity between two qudits within a mixed reduced state, which arises from the impurity gate acting on a pure four-qudit state. Note that for $\theta=1/2$ we get $\bar{\varepsilon}_{\rm imp}^{(n)}=2(1-n)\log(q)$.

Taking $n\to 1$ in Eq.~\eqref{eq:Renyi_neg_charged}, we see that only the impurity contribution can survive. Defining $\bar{\varepsilon}_{\rm imp}(\theta)=\lim_{n\to 1}\bar{\varepsilon}_{\rm imp}^{(n)}(\theta)$, we obtain 
\begin{equation}\label{eq:neg_charged}
\mathcal{E}_{A_1:A_2}(t) = \begin{cases}
        0 & t<r\,,\\
        (t-r)s_{\rm imp}^{(1/2)}(\theta) & r\le t <\ell/2+r\,,\\
        (\ell+r-t)s_{\rm imp}^{(1/2)}(\theta) + (2t-\ell-2r)\bar{\varepsilon}_{\rm imp}(\theta) & \ell/2+r\le t <\ell+r\,,\\
        \ell\bar{\varepsilon}_{\rm imp}(\theta) & \ell+r\le t\,.
    \end{cases}
\end{equation}
The symmetry-breaking impurity can therefore facilitate persistent, non-vanishing quantum correlations between $A_1$ and $A_2$, provided that $\bar \varepsilon_{\rm imp}(\theta)>0$. These quantum correlations can be attributed to the ability of the impurity to modify the charge fluctuations.  The results are illustrated in Fig.~\ref{fig:Neg_Charged_XY} for the impurity defined in Eq.~\eqref{eq:XY_impurity_def}. This impurity generally breaks the $U(1)$ symmetry and thus allows non-trivial entanglement at long times, as can be seen in Fig.~\ref{fig:Neg_Charged_XY}(a); in contrast, Fig.~\ref{fig:Neg_Charged_XY}(b) demonstrates that the negativity must vanish in the long-time limit for $\theta=1/2$.

\begin{figure}
    \centering
    \includegraphics[width=\linewidth]{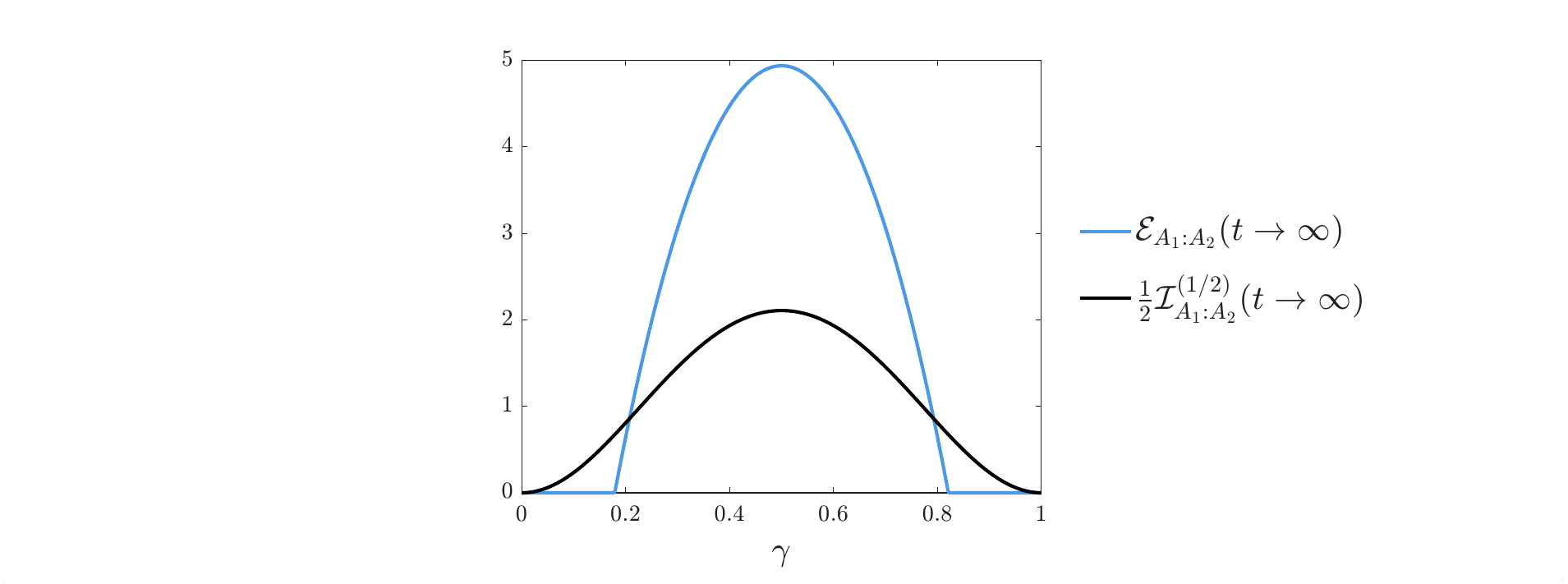}
    \caption{The long-time limits of the entanglement negativity and of the $n=1/2$ R\'enyi mutual information between two intervals in a charged dual-unitary circuit with an impurity $U_{\rm XY}(\gamma)$, as defined in Eq.~\eqref{eq:XY_impurity_def}. The local charge-sector dimension is $d=3$, and the half-lengths of the intervals $A_1$ and $A_2$ are fixed at $\ell=20$. The results are shown for an initial state with $\theta=0.2$ and as a function of the impurity parameter $\gamma$.}
    \label{fig:Neg_Charged_XY_long_time}
\end{figure}

The charge structure further gives rise to the breakdown of the familiar relation $\mathcal{E}_{A_1:A_2}(t)=\frac{1}{2}{\cal I}^{(1/2)}_{A_1:A_2}(t)$, which tends to hold within integrable dynamics obeying the quasiparticle picture~\cite{alba2019quantum}. This relation was noted in Sec.~\ref{sec:neg_uncharged}, as it holds in the uncharged case. Its breakdown in the charged case can be most easily seen by comparing the $t\to\infty$ limits of Eq.~\eqref{eq:neg_charged} and of $\frac{1}{2}{\cal I}^{(1/2)}_{A_1:A_2}$, where the latter is inferred from Eq.~\eqref{eq:Renyi_MI_charged}. Such a comparison is shown in Fig.~\ref{fig:Neg_Charged_XY_long_time} for the symmetry-breaking impurity defined in Eq.~\eqref{eq:XY_impurity_def}. The clear discrepancy between the two values can possibly be interpreted as a signature of the integrability being broken by the impurity.

The membrane picture prediction cannot be obtained for all times for the charged bulk. In addition, we were not able to determine the long-time limit using this effective theory. Although we have seen that the mutual information reduces to $\ell I(\varphi)$, which coincides with a classical mutual information, it is not certain that the correlations created are purely classical and therefore that $\mathcal{E}_{A_1:A_2}(t)\to 0$ at long times.  It would be very interesting to investigate whether it is possible within the membrane picture to sustain also persistent entanglement negativity for a symmetry-breaking impurity. Doing so would require an appropriate modification of the membrane picture, as we commented in Sec.~\ref{sec:neg_uncharged}.

\section{Conclusions}\label{sec:conclusions}
In this paper, we investigated the influence of symmetry-breaking impurities on the evolution of entanglement and mutual information in far-from-equilibrium quantum systems. We focused on dual-unitary brickwork quantum circuits with $U(1)$ charge conservation, which we perturb via a line of non-dual-unitary and non-charge-conserving gates. Using a combination of exact solvability afforded by dual unitarity and effective theories of entanglement dynamics, we derived analytical expressions for the entanglement entropy, mutual information, and logarithmic negativity. Our results show that the entanglement dynamics, and hence the local relaxation of the system, can be heavily modified by the presence of the impurity even if the impurity lies outside the subsystem of interest. 

We observed that an impurity can induce non-monotonic evolution of the entanglement entropy. This behaviour had previously been reported for uncharged circuits that obey the quasiparticle picture, but in the presence of an impurity that breaks the charge-conservation symmetry of the bulk, or even just the soliton structure of its charge dynamics, this is also true within the entanglement membrane picture. This non-monotonicity occurs because the symmetry-breaking impurity creates correlated charge fluctuations within the subsystem. Moreover, the long-time limit of the entanglement entropy is typically increased by the symmetry-breaking impurity, again due to the creation of charge fluctuations.       


We also studied the dynamics of the mutual information between two intervals straddling the impurity, in both charged and uncharged circuits. In the uncharged case, the impurity does not change the qualitative evolution of the mutual information, but only the height of its transient peak (and, within the membrane picture, its timing). Intriguingly, however, we found that for a strongly reflective impurity, the mutual information peak predicted by the membrane picture can be higher than the one reached when the bulk dynamics obeys the quasiparticle picture. This result motivates a future search for other examples where scrambling enhances the transmission of information across a weak bond, to shed further light on the mechanism behind this phenomenon.

In a charged circuit, a symmetry-breaking impurity can significantly modify the qualitative dynamics of the mutual information. For a bulk that obeys the membrane picture, we found that the mutual information can exhibit several stages of decline with different slopes, as opposed to its simple profile in the absence of the impurity, with a single stage of linear growth followed by a single stage of linear decrease. The cause of this more intricate pattern is the existence of charge correlations between different domain walls in the membrane picture description. Furthermore, for bulks described by either effective picture, soliton scattering off the impurity can result in a non-zero stationary value for the mutual information. In the membrane picture, this value coincides with the classical mutual information of the joint charge distribution defined by the impurity scattering properties. For the particular family of $U(1)$-symmetry-breaking impurities that we investigated in this context, we saw that the quasiparticle picture predicts a stationary value of the mutual information that is larger than the one predicted by the membrane picture. 

Finally, we calculated the entanglement negativity between the two intervals, where, apart from a universal early-time regime, we focused on a circuit with a non-interacting bulk. We showed that a symmetry-breaking impurity can lead to a non-zero stationary value also for the negativity. Moreover, this stationary value does not equal half of the R\'enyi-$1/2$ mutual information, breaking the relation between the two quantities that was previously reported in integrable models. We note that the existence of the stationary volume-law negativity and mutual information is in sharp contrast to the standard scenario of a quench in a translation-invariant system, which reaches a Gibbs or generalized Gibbs state featuring no extensive quantum correlations.

Characterizing the dynamics of the negativity for a chaotic-bulk circuit remains an open problem. The challenge of the microscopic calculation lies in the fact that the $n=2$ R\'enyi negativity is uninformative regarding the entanglement between the two intervals (as, by definition, it simply equals minus the $n=2$ R\'enyi entropy of their union). Thus, a calculation on two replicas, as we performed in the case of the entropies, provides no intuition into the negativity dynamics. Executing the calculation over a larger number of replicas could rely on a deeper analysis of relevant properties of dual-unitary gates with maximal entangling power, assisted by numerics. It might also draw inspiration from the line-tension description proposed for negativity dynamics in chaotic conformal field theories~\cite{Kudler-Flam2020quantum,Kudler-Flam2020correlation}. This would help us understand whether the non-zero stationary value of the mutual information, which the membrane picture predicts, is attributable to purely classical correlations. We leave the pursuit of this goal for future work.

Our work highlights the rich structure of quantum relaxation that can occur due to the presence of quantum impurities, and invites many questions for subsequent exploration. Here we considered only homogeneous initial states, but it will be illuminating to extend the analysis to non-homogeneous quenches, as they are known to give rise to distinct nonequilibrium entanglement structures in impurity systems~\cite{fraenkel2023extensive,rylands2023transport,gruber2020time}. Furthermore, while in this work we focused on impurities that break a continuous bulk symmetry, studying a counterpart problem with a discrete symmetry could reveal vastly different phenomenology. It will also be interesting to examine the fate of local symmetry restoration under the effect of the symmetry-breaking impurity~\cite{ares2023entanglement,bertini2024dynamics,foligno2024nonequilibrium}.

\section*{Acknowledgments}
S.F.~acknowledges funding from the BMW Endowment Fund. We acknowledge the use of generative AI to assist with constructing the code for data plotting and with proofreading the manuscript.

\section*{Data availability}
The code for generating the data plots is available upon reasonable request in Zenodo~\cite{fraenkel_2026_zenodo}.


\appendix

\section{Entanglement evolution in an interacting integrable bulk}\label{app:XXZ_single_interval}
In this appendix, we address the non-monotonic entanglement entropy evolution that is predicted by the quasiparticle picture for a single interval in the presence of an impurity. Specifically, we prove that this non-monotonic evolution arises in the case of an \textit{interacting} integrable dual-unitary bulk (Ref.~\cite{fraenkel2024entanglement} included an explicit computation only for the case of a non-interacting bulk). We focus on the dual-unitary XXZ Floquet model, where $q=2$. For this model, the bulk gates are uniform and given by
\begin{equation}\label{eq:XXZ_gate_decomp}
    U_{\rm XXZ}  = \exp\!\left[i\frac{\pi}{4}\left(X^{\otimes 2}+Y^{\otimes 2}\right)+iJZ^{\otimes2}\right]=e^{i\frac{\pi}{4}}\exp\!\left[i\left(J-\frac{\pi}{4}\right)Z^{\otimes2}\right]P,
\end{equation}
where $P$ is the swap gate and $J$ is a free real parameter. Ignoring the unimportant global phase, $J=\frac{\pi}{4}$ recovers the case of a swap-gate bulk, while $J=0$ leads to a free-fermionic evolution. Any other choice for $J$ leads to genuinely interacting dynamics. We originally used the swap-gate bulk case to compute the predictions of the quasiparticle picture, finding in particular that the entanglement entropy $S[\rho_A(t)]$ decreases linearly with time for $\ell/2\le t <\ell/2+r$~\cite{fraenkel2024entanglement}, as we recalled in Eq.~\eqref{eq:ent_swap_gates}. We now show that, for $\ell/2\le t <\ell/2+r$ and for any choice of $J$, the entropy $S[\rho_A(t)]$ agrees with the value obtained for the swap-gate bulk circuit up to a subleading correction, which therefore implies decreasing entanglement within this time regime.

To prove this statement, we employ a useful diagrammatic factorization that was introduced in Ref.~\cite{Rampp2024haydenpreskill} for XXZ dual-unitary circuits. Namely, we decompose each gate $U_{\rm XXZ}$ into a product of $P$ and the controlled-phase gate $V_J\equiv\exp\!\left[i\left(J-\frac{\pi}{4}\right)Z^{\otimes2}\right]$, as in Eq.~\eqref{eq:XXZ_gate_decomp}:
\begin{equation}
    U_{\rm XXZ} = \begin{tikzpicture}[baseline={([yshift=-0.6ex]current bounding box.center)},scale=0.35]
   \tgridLine{-0.8}{-0.8}{0.8}{0.8}
   \tgridLine{0.8}{-0.8}{0.1}{-0.1}
   \tgridLine{-0.8}{0.8}{-0.1}{0.1}
   \CPhaseLine{-0.5}{0.5}{0.5}{0.5}
 \end{tikzpicture}\, .
\end{equation}
The diagram encoding the $n$th R\'enyi entropy of the interval $A$ at a time $\ell/2\le t <\ell/2+r$ assumes then the form 
\begin{align}
\tr\left[\rho_A^n(t)\right] = q^{(1-n)\ell}\times\!\!\! \begin{tikzpicture}[baseline={([yshift=0.6ex]current bounding box.center)},scale=0.5]
{\swapprop{0}{4}{colSt}{}}
\foreach \x in {-1,1}{\swapprop{\x}{3}{colSt}{}}   
  \foreach \x in {-2,0,2}{\swapprop{\x}{2}{colSt}{}}   
  \foreach \x in {-3,-1,...,3}{\swapprop{\x}{1}{colSt}{}}   
    \foreach \x in {-4,-2,...,4}{\swapprop{\x}{0}{colSt}{}}
    \foreach \x in {-5,-3,...,5}{\swapprop{\x}{-1}{colSt}{}}
    \foreach \x in {-6,-4,...,6}{\swapprop{\x}{-2}{colSt}{}}
   \foreach \x in {-2,0} {\prop{4}{\x}{colimp}{}}
    \foreach \x in {-7.5,-5.5,...,6.5}{\istate{\x}{-2.5}{colSt}}
    \foreach \x in {-8,-7}{\tgridLine{\x+0.25}{\x+5.75}{\x+0.5}{\x+5.5}
    \circle{\x+0.25}{\x+5.75}}
    \foreach \x in {-6,...,-1}{\tgridLine{\x+0.25}{\x+5.75}{\x+0.5}{\x+5.5}
    \square{\x+0.25}{\x+5.75}}
    \foreach \x in {-4,...,3}{
    \tgridLine {4-\x-0.25}{\x+1.75} {4-\x-0.5}{\x+1.5}\circle
    {4-\x-0.25}{\x+1.75}}
    \foreach \x in {0,...,6}{\CPhaseLine{\x-6.4}{\x-1.6}{\x-5.6}{\x-1.6}}
    \foreach \x in {0,...,5}{\CPhaseLine{\x-4.4}{\x-1.6}{\x-3.6}{\x-1.6}}
    \foreach \x in {0,...,4}{\CPhaseLine{\x-2.4}{\x-1.6}{\x-1.6}{\x-1.6}}
    \foreach \x in {0,...,3}{\CPhaseLine{\x-0.4}{\x-1.6}{\x+0.4}{\x-1.6}}
    \foreach \x in {0,1}{\CPhaseLine{\x+1.6}{\x-1.6}{\x+2.4}{\x-1.6}}
    \CPhaseLine{4.6}{-0.6}{5.4}{-0.6}
    \CPhaseLine{5.6}{-1.6}{6.4}{-1.6}
  \end{tikzpicture}\!\!.
\end{align}

We now use the fact that the controlled-phase gates in the circuit commute with each other and with the swap gates, reducing the diagram to the following form:
\begin{align}
\tr\left[\rho_A^n(t)\right] = q^{2(1-n)(\ell+r-t)}\times\!\!\!\!\! \begin{tikzpicture}[baseline={([yshift=0.6ex]current bounding box.center)},scale=0.5]
   \foreach \x in {-2,1} {\prop{4}{\x}{colimp}{}}
    \foreach \x in {2.5,4.5}{\istate{\x}{-2.5}{colSt}}
    \foreach \x in {2.5,4.5}{\istate{\x}{0.5}{colSt}}
    \foreach \x in {-2,-1}{
    \tgridLine {4-\x-0.25}{\x+2.75} {4-\x-0.5}{\x+2.5}\circle
    {4-\x-0.25}{\x+2.75}}
    \foreach \x in {-2,-1}{
    \tgridLine {4-\x-0.25}{\x-0.25} {4-\x-0.5}{\x-0.5}\circle
    {4-\x-0.25}{\x-0.25}}
    \foreach \x in {2,3}{\tgridLine{\x-0.75}{\x-0.25}{\x+0.5}{\x-1.5}
    \square{\x-0.75}{\x-0.25}}
    \foreach \x in {2,3}{\tgridLine{\x-0.75}{\x-3.25}{\x+0.5}{\x-4.5}
    \square{\x-0.75}{\x-3.25}}
    \foreach \x in {-8,-6}{\tgridLine{\x+0.25}{-2.25}{\x+0.5}{-2.5}
    \circle{\x+0.25}{-2.25}\istate{\x+0.5}{-2.5}{colSt}
    \tgridLine{\x+1.5}{-2.5}{\x+6.75}{2.75}\circle{\x+6.75}{2.75}}
    \foreach \x in {0,1}{\CPhaseLine{\x-5.6}{\x-1.6}{\x+1.6}{\x-1.6}
    \CPhaseLine{\x-3.9}{\x-1.9}{\x+1.9}{\x-1.9}}
    \foreach \x in {0,1}{\CPhaseLine{\x-2.6}{\x+1.4}{\x+1.6}{\x+1.4}
    \CPhaseLine{\x-0.9}{\x+1.1}{\x+1.9}{\x+1.1}}
  \end{tikzpicture}\,\,\,.
\end{align}
We can equivalently write
\begin{equation}\label{eq:XXZ_interval_entropy_aux}
    S^{(n)}[\rho_A(t)] = 2(\ell+r-t)\log(q) + \frac{1}{1-n}\log\tr\left[\rho_{\tilde{B}}^n(t)\right],
\end{equation}
where $\rho_{\tilde{B}}(t)$ is the reduced density matrix of a subsystem $\tilde{B}$ within a pure state of $\tilde{A}\cup\tilde{B}\cup\tilde{C}$ (to be defined shortly), which gives rise to the remaining diagram that we need to contract:
\begin{align}
    \tr\left[\rho_{\tilde{B}}^n(t)\right] = \,\,\,\begin{tikzpicture}[baseline={([yshift=0.6ex]current bounding box.center)},scale=0.5]
   \foreach \x in {-2,1} {\prop{4}{\x}{colimp}{}}
    \foreach \x in {2.5,4.5}{\istate{\x}{-2.5}{colSt}}
    \foreach \x in {2.5,4.5}{\istate{\x}{0.5}{colSt}}
    \foreach \x in {-2,-1}{
    \tgridLine {4-\x-0.25}{\x+2.75} {4-\x-0.5}{\x+2.5}\circle
    {4-\x-0.25}{\x+2.75}}
    \foreach \x in {-2,-1}{
    \tgridLine {4-\x-0.25}{\x-0.25} {4-\x-0.5}{\x-0.5}\circle
    {4-\x-0.25}{\x-0.25}}
    \foreach \x in {2,3}{\tgridLine{\x-0.75}{\x-0.25}{\x+0.5}{\x-1.5}
    \square{\x-0.75}{\x-0.25}}
    \foreach \x in {2,3}{\tgridLine{\x-0.75}{\x-3.25}{\x+0.5}{\x-4.5}
    \square{\x-0.75}{\x-3.25}}
    \foreach \x in {-9,-7}{\tgridLine{\x+0.25}{-2.25}{\x+0.5}{-2.5}
    \circle{\x+0.25}{-2.25}\istate{\x+0.5}{-2.5}{colSt}
    \tgridLine{\x+1.5}{-2.5}{\x+6.75}{2.75}\circle{\x+6.75}{2.75}}
    \foreach \x in {0,1}{\CPhaseLine{\x-6.6}{\x-1.6}{\x+1.6}{\x-1.6}
    \CPhaseLine{\x-4.9}{\x-1.9}{\x+1.9}{\x-1.9}}
    \foreach \x in {0,1}{\CPhaseLine{\x-3.6}{\x+1.4}{\x+1.6}{\x+1.4}
    \CPhaseLine{\x-1.9}{\x+1.1}{\x+1.9}{\x+1.1}}
    \draw[semithick,decorate,decoration={brace}] (-9,3.25) -- (0,3.25) node[midway,xshift=0pt,yshift=15pt] {$\tilde{A}$};
    \draw[semithick,decorate,decoration={brace}] (1,3.25) -- (2.5,3.25) node[midway,xshift=0pt,yshift=15pt] {$\tilde{B}$};
    \draw[semithick,decorate,decoration={brace}] (4.5,3.25) -- (6,3.25) node[midway,xshift=0pt,yshift=15pt] {$\tilde{C}$};
  \end{tikzpicture}\,\,\,.
\end{align}
Here, subsystem $\tilde{A}$ is composed of $4t-2\ell$ qubits, and subsystems $\tilde{B}$ and $\tilde{C}$ are comprised of $2t-2r$ qubits each. $\tilde{C}$ is entangled to $\tilde{B}$ through $t-r$ impurity gates, so that its $n$th R\'enyi entanglement entropy is $S^{(n)}[\rho_{\tilde{C}}(t)]=(t-r)s_{\rm imp}^{(n)}$. On the other hand, $\tilde{A}$ can be seen as having been initially prepared as $2t-\ell$ Bell pairs, after which each Bell pair is independently entangled to $\tilde{B}$ via $2t-2r$ controlled-phase gates of the form of $V_J$. These gates are applied after all impurity gates have already been applied to $\tilde{B}\cup\tilde{C}$. 

Overall, we may write the global pure state of $\tilde{A}\cup\tilde{B}\cup\tilde{C}$ (before the reduction to $\rho_{\tilde{B}}$) as
\begin{equation}
    \ket{\Phi_{\tilde{A}\cup\tilde{B}\cup\tilde{C}}} = \prod_{j=1}^{2t-\ell}\prod_{m=1}^{2t-2r}V_J^{(j,m)}\left(\frac{1}{\sqrt{2}}\ket{00}+\frac{1}{\sqrt{2}}\ket{11}\right)^{\otimes(2t-\ell)}\ket{\phi_{\tilde{B}\cup\tilde{C}}},
\end{equation}
where $\ket{\phi_{\tilde{B}\cup\tilde{C}}}$ is the state of $\tilde{B}\cup\tilde{C}$ following the action of the impurity gates, and $V_J^{(j,m)}$ denotes the controlled-phase gate $V_J$ applied to one qubit out of the $j$th Bell pair in $\tilde{A}$ and to the $m$th qubit in $\tilde{B}$. The gates that entangle $\tilde{B}$ to $\tilde{A}$ can change $S^{(n)}[\rho_{\tilde{B}}(t)]$ compared to the simple case of a swap-gate bulk, where $J=\frac{\pi}{4}$ and $V_J$ becomes the trivial identity operator. Crucially, however, this change in entanglement is negligible, as we now explain.

Indeed, we first write
\begin{equation}
    \left(\frac{1}{\sqrt{2}}\ket{00}+\frac{1}{\sqrt{2}}\ket{11}\right)^{\otimes(2t-\ell)} = \frac{1}{2^{t-\ell/2}}\sum_{Q=0}^{2t-\ell}\sqrt{\binom{2t-\ell}{Q}}\ket{\phi_{\tilde{A}}(Q)},
\end{equation}
where each state $\ket{\phi_{\tilde{A}}(Q)}$ is defined as the equal superposition, with amplitudes given by $1/\sqrt{\binom{2t-\ell}{Q}}$, of all states in $\left\{\ket{00},\ket{11}\right\}^{\otimes(2t-\ell)}$ that have exactly $2Q$ qubits in the state $\ket{1}$. Next, we let $Z_m$ denote the $Z$ operator acting on the $m$th qubit in $\tilde{B}$, and we notice the following fact:
\begin{equation}
    \left(\prod_{j=1}^{2t-\ell}V_J^{(j,m)}\right)\ket{\phi_{\tilde{A}}(Q)}\ket{\phi_{\tilde{B}\cup\tilde{C}}}=\ket{\phi_{\tilde{A}}(Q)}\otimes\exp\!\left[i\left(J-\frac{\pi}{4}\right)\left(2t-\ell-2Q\right)Z_m\right]\ket{\phi_{\tilde{B}\cup\tilde{C}}}.
\end{equation}
This amounts to
\begin{equation}
    \left(\prod_{j=1}^{2t-\ell}\prod_{m=1}^{2t-2r}V_J^{(j,m)}\right)\ket{\phi_{\tilde{A}}(Q)}\ket{\phi_{\tilde{B}\cup\tilde{C}}} = \ket{\phi_{\tilde{A}}(Q)}\ket{\phi_{\tilde{B}\cup\tilde{C}}(Q)},
\end{equation}
where we defined
\begin{equation}
    \ket{\phi_{\tilde{B}\cup\tilde{C}}(Q)} = \exp\!\left[i\left(J-\frac{\pi}{4}\right)\left(2t-\ell-2Q\right)\sum_{m=1}^{2t-2r} Z_m\right]\ket{\phi_{\tilde{B}\cup\tilde{C}}}.
\end{equation}
In other words, if $\tilde{A}$ is in a state with a well-defined charge $Q$, then $Q$ simply determines the total phase gate that is applied to each qubit in $\tilde{B}$, and therefore the controlled phase gates do not generate any entanglement between $\tilde{A}$ and $\tilde{B}$. In all, this means that we can rewrite the global state as
\begin{equation}
    \ket{\Phi_{\tilde{A}\cup\tilde{B}\cup\tilde{C}}} = \frac{1}{2^{t-\ell/2}}\sum_{Q=0}^{2t-\ell}\sqrt{\binom{2t-\ell}{Q}}\ket{\phi_{\tilde{A}}(Q)}\ket{\phi_{\tilde{B}\cup\tilde{C}}(Q)}.
\end{equation}
The Schmidt decomposition of $\ket{\Phi_{\tilde{A}\cup\tilde{B}\cup\tilde{C}}}$ between $\tilde{A}$ and $\tilde{B}\cup\tilde{C}$ therefore involves at most $2t-\ell+1$ terms, and thus $S^{(n)}[\rho_{\tilde{A}}(t)] \le \log\left(2t-\ell+1\right)$.

Finally, by taking the limit $n\to 1$, we may exploit the sub-additivity of the von Neumann entropy, which implies that $\left|S[\rho_{\tilde{B}}]-S[\rho_{\tilde{C}}]\right|\le S[\rho_{\tilde{A}}]$, and so $S[\rho_{\tilde{B}}]$ differs from $S[\rho_{\tilde{C}}]=(t-r)s_{\rm imp}$ by a term that scales at most logarithmically with $2t-\ell$. Through Eq.~\eqref{eq:XXZ_interval_entropy_aux}, we infer that for $\ell/2\le t <\ell/2+r$, the entropy of the interval $A$ within the XXZ bulk circuit agrees with the result of the swap-gate bulk circuit (see Eq.~\eqref{eq:ent_swap_gates}), up to a subleading correction, which concludes the proof:
\begin{equation}
    \left|S[\rho_A(t)]-2(\ell+r-t)\log(q)-(t-r)s_{\rm imp}\right|\le\log\left(2t-\ell+1\right).
\end{equation}

\newpage
\section{Entropy of mirroring intervals in a bulk with maximal entangling power}\label{app:chaotic_2_interval_entropy}
Here we complement Sec.~\ref{sec:MI_uncharged_mem} and derive an exact result for the $n=2$ R\'enyi entropy of two mirroring intervals (as defined in Eq.~\eqref{eq:two_intervals_def}) in a circuit where the dual-unitary bulk has maximal entangling power $p=1$, in the large $q$ limit. 
While the entangling power (defined in Eq.~\eqref{eq:entangling_power_def}) of the bulk gates is assumed to be fixed, they are taken to be otherwise random. For this purpose, we employ the same technique used in Ref.~\cite{fraenkel2024entanglement}. 
Given that the dynamics of the R\'enyi entropy can be computed exactly for any index $n$ and any choice of bulk gates as long as $t<\ell/2$ or $t\ge\ell+r$, what remains is to analyze the regime $\ell/2\le t <\ell+r$. We will initially restrict the analysis to $\ell/2\le t <\ell/2+r$, and comment later on the straightforward extension to $\ell/2+r\le t < \ell+r$. 

First, we recall important properties that apply to random dual-unitary gates with $p=1$. Such gates can be generated by considering arbitrary dual-unitary gates with $p=1$ and inserting independent Haar-random single-qudit unitaries on all their legs. For two copies of the density matrix, it is then possible to write down an explicit form for the average (replicated) bulk gate. This is most conveniently expressed relative to the orthogonal basis $\left\{\begin{tikzpicture}[baseline={([yshift=-0.6ex]current bounding box.center)},scale=0.5]
    \circle{-0.5}{0.5}
      \gridLine{-0.38}{0.5}{0.2}{0.5}
 \end{tikzpicture},\begin{tikzpicture}[baseline={([yshift=-0.6ex]current bounding box.center)},scale=0.5]
    \circleblack{-0.5}{0.5}
      \gridLine{-0.38}{0.5}{0.2}{0.5}
 \end{tikzpicture}\right\}^{\otimes 2}$, where we defined  $\begin{tikzpicture}[baseline={([yshift=-0.6ex]current bounding box.center)},scale=0.5]
    \circleblack{-0.5}{0.5}
      \gridLine{-0.38}{0.5}{0.2}{0.5}
 \end{tikzpicture}=\begin{tikzpicture}[baseline={([yshift=-0.6ex]current bounding box.center)},scale=0.5]
    \square{-0.5}{0.5}
      \gridLine{-0.38}{0.5}{0.2}{0.5}
 \end{tikzpicture}-\begin{tikzpicture}[baseline={([yshift=-0.6ex]current bounding box.center)},scale=0.5]
    \circle{-0.5}{0.5}
      \gridLine{-0.38}{0.5}{0.2}{0.5}
 \end{tikzpicture}/q$. To leading order in $1/q$, the average gate is given by \cite{foligno2023growth}:
 \be\label{eq:AveragedGateMatrix}
\begin{aligned}
 &\begin{tikzpicture}[baseline={([yshift=-0.6ex]current bounding box.center)},scale=0.5]
   \prop{0}{0}{colSt}{}
 \end{tikzpicture}\approx\left(\begin{array}{cccc}
1 & 0 & 0 & 0\\
0 & 0 & 0 & 1/q\\
0 & 0 & 0 & 1/q\\
0 & 1/q & 1/q & 1
\end{array}\right).
\end{aligned}
\ee
In the derivation that follows, we will rely on two properties of this gate that were originally observed in Ref.~\cite{fraenkel2024entanglement}, namely
\begin{eqnarray}\label{eq:MaximalEP1stProperty}
\begin{tikzpicture}[baseline={([yshift=0]current bounding box.center)},scale=0.45] 
      \foreach \x in {0,...,5}{\prop{\x}{-\x}{colSt}{}}   
       \foreach \x in {-1,...,4}{\circle{\x+0.5}{-\x-1.5}}
 \circleblack
    {5.5}{-5.5}
      \draw[semithick,decorate,decoration={brace}] (4,-5.75) -- (-1.25,-0.5) node[midway,xshift=-20.5pt,yshift=0] {$m$};
  \end{tikzpicture}\approx \frac{1}{q^m} \begin{tikzpicture}[baseline={([yshift=-0.6ex]current bounding box.center)},scale=0.5]
  \tgridLine{0}{0}{-0.4}{0.4}
  \circleblack{0}{0}
  \end{tikzpicture}\otimes \Big[\begin{tikzpicture}[baseline={([yshift=-0.6ex]current bounding box.center)},scale=0.5]
  \tgridLine{0}{0}{0.4}{0.4}
  \circleblack{0}{0}
  \end{tikzpicture}\,\Big]^{\otimes m},
  \end{eqnarray}
  and
  \begin{eqnarray}\label{eq:MaximalEP2ndProperty}
\begin{tikzpicture}[baseline={([yshift=0]current bounding box.center)},scale=0.45] 
      \foreach \x in {0,...,5}{\prop{\x}{-\x}{colSt}{}}   
       \foreach \x in {-1,...,4}{\circleblack{\x+0.5}{-\x-1.5}}
       \square{-.5}{.5}
      \draw[semithick,decorate,decoration={brace}] (4,-5.75) -- (-1.25,-0.5) node[midway,xshift=-20.5pt,yshift=0] {$m$};
  \end{tikzpicture}\approx \ \Big[\begin{tikzpicture}[baseline={([yshift=-0.6ex]current bounding box.center)},scale=0.5]
  \tgridLine{0}{0}{0.4}{0.4}
  \square{0}{0}
  \end{tikzpicture}\,\Big]^{\otimes m}\otimes\begin{tikzpicture}[baseline={([yshift=-0.6ex]current bounding box.center)},scale=0.5]
  \tgridLine{0}{0}{0.4}{-0.4}
  \square{0}{0}
  \end{tikzpicture}-\frac{1}{q^2}\Big[\begin{tikzpicture}[baseline={([yshift=-0.6ex]current bounding box.center)},scale=0.5]
  \tgridLine{0}{0}{0.4}{0.4}
  \square{0}{0}
  \end{tikzpicture}\,\Big]^{\otimes \left(m-1\right)}\otimes \begin{tikzpicture}[baseline={([yshift=-0.6ex]current bounding box.center)},scale=0.5]
  \tgridLine{0}{0}{0.4}{0.4}
  \circle{0}{0}
  \end{tikzpicture} \otimes \begin{tikzpicture}[baseline={([yshift=-0.6ex]current bounding box.center)},scale=0.5]
  \tgridLine{0}{0}{0.4}{-0.4}
  \circle{0}{0}
  \end{tikzpicture}\,,
  \end{eqnarray}
where both approximations hold to leading order in $1/q$. Similarly useful are the resolution of the identity $\1=(\begin{tikzpicture}[baseline={([yshift=-0.6ex]current bounding box.center)},scale=0.65]
   \tgridLine{0}{0}{0.35}{0.35}
  \circle{0}{0}
   \tgridLine{-.5}{0}{-.8}{0.3}
  \circle{-0.5}{0}\end{tikzpicture}+\begin{tikzpicture}[baseline={([yshift=-0.6ex]current bounding box.center)},scale=0.65]
  \tgridLine{0}{0}{0.3}{0.3}
  \circleblack{0}{0}
  \tgridLine{-.5}{0}{-.8}{0.3}
  \circleblack{-0.5}{0}
  \end{tikzpicture})/q^2$ 
  and the resolution of the two-site initial state 
  $\begin{tikzpicture}[baseline={([yshift=-0.6ex]current bounding box.center)},scale=0.65]
  \istate{0}{0}{colSt}
\end{tikzpicture}=(\begin{tikzpicture}[baseline={([yshift=-0.6ex]current bounding box.center)},scale=0.65]
   \tgridLine{0}{0}{0.35}{0.35}
  \circle{0}{0}
   \tgridLine{-.5}{0}{-.8}{0.3}
  \circle{-0.5}{0}
\end{tikzpicture}+\begin{tikzpicture}[baseline={([yshift=-0.6ex]current bounding box.center)},scale=0.65]
  \tgridLine{0}{0}{0.3}{0.3}
  \circleblack{0}{0}
  \tgridLine{-.5}{0}{-.8}{0.3}
  \circleblack{-0.5}{0}
  \end{tikzpicture})/q^4$.

For $\ell/2\le t <\ell/2+r$, following Eq.~\eqref{eq:inter_two_inter}, the diagram we need to compute to obtain the entropy of $A=A_1\cup A_2$ is given by
\be
\tr \left[{\rho_A}^2\right]=q^{2\ell-8t}\times \!\!\!\!\! \begin{tikzpicture}[baseline={([yshift=0.6ex]current bounding box.center)},scale=0.5]
    \foreach \x in {-3,-1,...,5}{\prop{\x}{-3}{colSt}{}}
     \foreach \x in {5,7,...,15}{\prop{\x}{-3}{colSt}{}}
      \foreach \x in {-1,1,...,5}{\prop{\x}{-1}{colSt}{}}
           \foreach \x in {7,9,...,13}{\prop{\x}{-1}{colSt}{}}
        \foreach \x in {1,3}{\prop{\x}{1}{colSt}{}}
             \foreach \x in {9,11}{\prop{\x}{1}{colSt}{}}

    \foreach \x in {-2,0,...,6}{\prop{\x}{-4}{colSt}{}}
         \foreach \x in {8,10,...,14}{\prop{\x}{-4}{colSt}{}}
     \foreach \x in {-2,0,...,6}{\prop{\x}{-2}{colSt}{}}
          \foreach \x in {6,8,...,14}{\prop{\x}{-2}{colSt}{}}
    \foreach \x in {0,2,4}{\prop{\x}{0}{colSt}{}} 
         \foreach \x in {8,10,12}{\prop{\x}{0}{colSt}{}}
     \foreach \x in {2,10}{\prop{\x}{2}{colSt}{}}
      \foreach \x in {-4,-2} {\prop{6}{\x}{colimp}{}}
    \foreach \x in {-1.5,0.5,...,5.5}{\istate{\x}{-4.5}{colSt}}
        \foreach \x in {6.5,8.5,...,12.5}{\istate{\x}{-4.5}{colSt}}
    
    \foreach \x in {-4,...,1}{\tgridLine{\x+0.25}{\x+1.75}{\x+.5}{\x+1.5}\square{\x+0.25}{\x+1.75}}
    \foreach \x in {-4,...,-3}{\tgridLine{\x+0.5}{-\x-7.5}{\x+.75}{-\x-7.25}\circle{\x+0.5}{-\x-7.5}}
    \foreach \x in {6,...,9}{\tgridLine{\x+0.25}{\x+1.75-8}{\x+.5}{\x+1.5-8}\circle{\x+0.25}{\x+1.75-8}}

    \foreach \x in {10,...,15}{\tgridLine{\x+0.75}{-\x+3.75+9}{\x+0.5}{-\x+3.5+9}
    \square{\x+0.75}{-\x+3.75+9}}
    \foreach \x in {14,...,15}{\tgridLine{\x+0.5}{\x-18.5}{\x+0.25}{\x-18.25}
    \circle{\x+0.5}{\x-18.5}}
    \foreach \x in {2,...,5}{\tgridLine{\x+0.75}{-\x+3.75+1}{\x+0.5}{-\x+3.5+1}
    \circle{\x+0.75}{-\x+3.75+1}}
    
  \end{tikzpicture}.
\ee
By resolving the initial state on the leftmost pair of sites, and using the identities in Eqs.~\eqref{eq:MaximalEP1stProperty} and \eqref{eq:MaximalEP2ndProperty}, we obtain
\begin{align}\label{eq:Contributions_to_S2}
   \tr \left[{\rho_A}^2\right] & \approx q^{2\ell-8t-3}\times
   \begin{tikzpicture}[baseline={([yshift=0.6ex]current bounding box.center)},scale=0.5]
    \foreach \x in {-1,1,...,5}{\prop{\x}{-3}{colSt}{}}
     \foreach \x in {5,7,...,15}{\prop{\x}{-3}{colSt}{}}
      \foreach \x in {-1,1,...,5}{\prop{\x}{-1}{colSt}{}}
           \foreach \x in {7,9,...,13}{\prop{\x}{-1}{colSt}{}}
        \foreach \x in {1,3}{\prop{\x}{1}{colSt}{}}
             \foreach \x in {9,11}{\prop{\x}{1}{colSt}{}}
    \foreach \x in {0,2,...,6}{\prop{\x}{-4}{colSt}{}}
         \foreach \x in {8,10,...,14}{\prop{\x}{-4}{colSt}{}}
     \foreach \x in {-2,0,...,6}{\prop{\x}{-2}{colSt}{}}
          \foreach \x in {6,8,...,14}{\prop{\x}{-2}{colSt}{}}
    \foreach \x in {0,2,4}{\prop{\x}{0}{colSt}{}} 
         \foreach \x in {8,10,12}{\prop{\x}{0}{colSt}{}}
     \foreach \x in {2,10}{\prop{\x}{2}{colSt}{}}
      \foreach \x in {-4,-2} {\prop{6}{\x}{colimp}{}}
    \foreach \x in {0.5,2.5,...,5.5}{\istate{\x}{-4.5}{colSt}}
        \foreach \x in {6.5,8.5,...,12.5}{\istate{\x}{-4.5}{colSt}}    
    \foreach \x in {-3,...,1}{\tgridLine{\x+0.25}{\x+1.75}{\x+.5}{\x+1.5}\square{\x+0.25}{\x+1.75}}
    \foreach \x in {-3,...,-1}{\tgridLine{\x+0.5}{-\x-5.5}{\x+.75}{-\x-5.25}\circle{\x+0.5}{-\x-5.5}}
    \foreach \x in {6,...,9}{\tgridLine{\x+0.25}{\x+1.75-8}{\x+.5}{\x+1.5-8}\circle{\x+0.25}{\x+1.75-8}}
    \foreach \x in {10,...,15}{\tgridLine{\x+0.75}{-\x+3.75+9}{\x+0.5}{-\x+3.5+9}
    \square{\x+0.75}{-\x+3.75+9}}
    \foreach \x in {14,...,15}{\tgridLine{\x+0.5}{\x-18.5}{\x+0.25}{\x-18.25}
    \circle{\x+0.5}{\x-18.5}}
    \foreach \x in {2,...,5}{\tgridLine{\x+0.75}{-\x+3.75+1}{\x+0.5}{-\x+3.5+1}
    \circle{\x+0.75}{-\x+3.75+1}}   
  \end{tikzpicture} \nonumber \\
  & + q^{\ell-8t} \times \begin{tikzpicture}[baseline={([yshift=0.6ex]current bounding box.center)},scale=0.5]
     \foreach \x in {7,9,...,15}{\prop{\x}{-3}{colSt}{}}
           \foreach \x in {7,9,...,13}{\prop{\x}{-1}{colSt}{}}
             \foreach \x in {9,11}{\prop{\x}{1}{colSt}{}}
         \foreach \x in {8,10,...,14}{\prop{\x}{-4}{colSt}{}}
          \foreach \x in {6,8,...,14}{\prop{\x}{-2}{colSt}{}} 
         \foreach \x in {8,10,12}{\prop{\x}{0}{colSt}{}}
     \foreach \x in {10}{\prop{\x}{2}{colSt}{}}
      \foreach \x in {-4,-2} {\prop{6}{\x}{colimp}{}}
        \foreach \x in {6.5,8.5,...,12.5}{\istate{\x}{-4.5}{colSt}}    
    \foreach \x in {6,...,9}{\tgridLine{\x+0.25}{\x+1.75-8}{\x+.5}{\x+1.5-8}\circle{\x+0.25}{\x+1.75-8}}
    \foreach \x in {10,...,15}{\tgridLine{\x+0.75}{-\x+3.75+9}{\x+0.5}{-\x+3.5+9}
    \square{\x+0.75}{-\x+3.75+9}}
    \foreach \x in {14,...,15}{\tgridLine{\x+0.5}{\x-18.5}{\x+0.25}{\x-18.25}
    \circle{\x+0.5}{\x-18.5}}
    \foreach\y in {-4,-2}{\tgridLine{5.25}{\y-.75}{5.5}{\y-.5}
    \square{5.25}{\y-.75}}
    \foreach\y in {-4,-2}{\tgridLine{5.25}{\y+.75}{5.5}{\y+.5}
    \square{5.25}{\y+.75}}
  \end{tikzpicture}\nonumber\\
  & - q^{2\ell-8t-6}\times \begin{tikzpicture}[baseline={([yshift=0.6ex]current bounding box.center)},scale=0.5]
    \foreach \x in {1,3,...,5}{\prop{\x}{-3}{colSt}{}}
     \foreach \x in {5,7,...,15}{\prop{\x}{-3}{colSt}{}}
      \foreach \x in {3,5}{\prop{\x}{-1}{colSt}{}}
           \foreach \x in {7,9,...,13}{\prop{\x}{-1}{colSt}{}}
             \foreach \x in {9,11}{\prop{\x}{1}{colSt}{}}
    \foreach \x in {2,4,...,6}{\prop{\x}{-4}{colSt}{}}
         \foreach \x in {8,10,...,14}{\prop{\x}{-4}{colSt}{}}
     \foreach \x in {2,4,...,6}{\prop{\x}{-2}{colSt}{}}
          \foreach \x in {6,8,...,14}{\prop{\x}{-2}{colSt}{}}
    \foreach \x in {4}{\prop{\x}{0}{colSt}{}} 
         \foreach \x in {8,10,12}{\prop{\x}{0}{colSt}{}}
     \foreach \x in {10}{\prop{\x}{2}{colSt}{}}
      \foreach \x in {-4,-2} {\prop{6}{\x}{colimp}{}}
    \foreach \x in {2.5,4.5,...,5.5}{\istate{\x}{-4.5}{colSt}}
        \foreach \x in {6.5,8.5,...,12.5}{\istate{\x}{-4.5}{colSt}}    
    \foreach \x in {0,...,3}{\tgridLine{\x+0.25}{\x-2.25}{\x+.5}{\x-2.5}\square{\x+0.25}{\x-2.25}}
    \foreach \x in {0,...,1}{\tgridLine{\x+0.5}{-\x-3.5}{\x+.75}{-\x-3.25}\circle{\x+0.5}{-\x-3.5}}
    \foreach \x in {6,...,9}{\tgridLine{\x+0.25}{\x+1.75-8}{\x+.5}{\x+1.5-8}\circle{\x+0.25}{\x+1.75-8}}
    \foreach \x in {10,...,15}{\tgridLine{\x+0.75}{-\x+3.75+9}{\x+0.5}{-\x+3.5+9}
    \square{\x+0.75}{-\x+3.75+9}}
    \foreach \x in {14,...,15}{\tgridLine{\x+0.5}{\x-18.5}{\x+0.25}{\x-18.25}
    \circle{\x+0.5}{\x-18.5}}
    \foreach \x in {4,...,5}{\tgridLine{\x+0.75}{-\x+3.75+1}{\x+0.5}{-\x+3.5+1}
    \circle{\x+0.75}{-\x+3.75+1}}
    \draw[semithick,decorate,decoration={brace}] (-.25,-1.75) -- (3,1.25) node[midway,xshift=-10.5pt,yshift=7.5pt] {$\ell-2$};
             \draw[semithick,decorate,decoration={brace}] (1.25,-4.8) -- (0.,-3.55) node[midway,xshift=-12.5pt] {$2$};
  \end{tikzpicture}.
\end{align}
We may further simplify the first and third terms in the equation above by repeatedly resolving the initial state on the remaining leftmost pair of sites, and then keeping only the contribution from $\begin{tikzpicture}[baseline={([yshift=-0.6ex]current bounding box.center)},scale=0.65]
   \tgridLine{0}{0}{0.35}{0.35}
  \circle{0}{0}
   \tgridLine{-.5}{0}{-.8}{0.3}
  \circle{-0.5}{0}
\end{tikzpicture}$ and omitting the contribution from \begin{tikzpicture}[baseline={([yshift=-0.6ex]current bounding box.center)},scale=0.65]
   \tgridLine{0}{0}{0.35}{0.35}
  \circleblack{0}{0}
   \tgridLine{-.5}{0}{-.8}{0.3}
  \circleblack{-0.5}{0}
\end{tikzpicture}. This move is justified because, by using Eq.~\eqref{eq:MaximalEP1stProperty} and replacing the occurrences of $\begin{tikzpicture}[baseline={([yshift=-0.6ex]current bounding box.center)},scale=0.65]
   \tgridLine{0}{0}{0.35}{0.35}
  \circleblack{0}{0}
\end{tikzpicture}$ with $\begin{tikzpicture}[baseline={([yshift=-0.6ex]current bounding box.center)},scale=0.65]
   \tgridLine{0}{0}{0.35}{0.35}
  \square{0}{0}
\end{tikzpicture}$, the contribution from \begin{tikzpicture}[baseline={([yshift=-0.6ex]current bounding box.center)},scale=0.65]
   \tgridLine{0}{0}{0.35}{0.35}
  \circleblack{0}{0}
   \tgridLine{-.5}{0}{-.8}{0.3}
  \circleblack{-0.5}{0}
\end{tikzpicture} can be bounded from above\footnote{We emphasize that for all diagrams we treat here (when taken without the preceding coefficients), the resolution of the identity or the initial state always yields a sum of \textit{non-negative} contributions. This can be straightforwardly seen from the non-negative matrix elements of the averaged bulk gate in Eq.~\eqref{eq:AveragedGateMatrix}.} by a term that is negligible relative to the second term in Eq.~\eqref{eq:Contributions_to_S2}. This leads to the following result:
\begin{align}
   \tr \left[{\rho_A}^2\right] & \approx q^{2\ell-8t-3(\ell+r-t)}\times\!\!\!\!
   \begin{tikzpicture}[baseline={([yshift=0.6ex]current bounding box.center)},scale=0.5]
    \foreach \x in {5}{\prop{\x}{-3}{colSt}{}}
     \foreach \x in {5,7,...,15}{\prop{\x}{-3}{colSt}{}}
      \foreach \x in {3,5}{\prop{\x}{-1}{colSt}{}}
           \foreach \x in {7,9,...,13}{\prop{\x}{-1}{colSt}{}}
        \foreach \x in {1,3}{\prop{\x}{1}{colSt}{}}
             \foreach \x in {9,11}{\prop{\x}{1}{colSt}{}}
         \foreach \x in {8,10,...,14}{\prop{\x}{-4}{colSt}{}}
     \foreach \x in {4}{\prop{\x}{-2}{colSt}{}}
          \foreach \x in {6,8,...,14}{\prop{\x}{-2}{colSt}{}}
    \foreach \x in {2,4}{\prop{\x}{0}{colSt}{}} 
         \foreach \x in {8,10,12}{\prop{\x}{0}{colSt}{}}
     \foreach \x in {2,10}{\prop{\x}{2}{colSt}{}}
      \foreach \x in {-4,-2} {\prop{6}{\x}{colimp}{}}
        \foreach \x in {6.5,8.5,...,12.5}{\istate{\x}{-4.5}{colSt}}    
    \foreach \x in {0,...,1}{\tgridLine{\x+0.25}{\x+1.75}{\x+.5}{\x+1.5}\square{\x+0.25}{\x+1.75}}
    \foreach \x in {0,...,5}{\tgridLine{\x+0.5}{-\x+.5}{\x+.75}{-\x+0.75}\circle{\x+0.5}{-\x+.5}}
    \foreach \x in {6,...,9}{\tgridLine{\x+0.25}{\x+1.75-8}{\x+.5}{\x+1.5-8}\circle{\x+0.25}{\x+1.75-8}}
    \foreach \x in {10,...,15}{\tgridLine{\x+0.75}{-\x+3.75+9}{\x+0.5}{-\x+3.5+9}
    \square{\x+0.75}{-\x+3.75+9}}
    \foreach \x in {14,...,15}{\tgridLine{\x+0.5}{\x-18.5}{\x+0.25}{\x-18.25}
    \circle{\x+0.5}{\x-18.5}}
    \foreach \x in {2,...,5}{\tgridLine{\x+0.75}{-\x+3.75+1}{\x+0.5}{-\x+3.5+1}
    \circle{\x+0.75}{-\x+3.75+1}}  
    \draw[semithick,decorate,decoration={brace}] (-.25,2.25) -- (1,3.25) node[midway,xshift=-10.5pt,yshift=7.5pt] {$t-r$};
             \draw[semithick,decorate,decoration={brace}] (5.25,-4.8) -- (0,.45) node[midway,xshift=-12.5pt,yshift=-7.5pt] {$t+r$};
  \end{tikzpicture} \nonumber \\
  & + q^{\ell-8t} \times \begin{tikzpicture}[baseline={([yshift=0.6ex]current bounding box.center)},scale=0.5]
     \foreach \x in {7,9,...,15}{\prop{\x}{-3}{colSt}{}}
           \foreach \x in {7,9,...,13}{\prop{\x}{-1}{colSt}{}}
             \foreach \x in {9,11}{\prop{\x}{1}{colSt}{}}
         \foreach \x in {8,10,...,14}{\prop{\x}{-4}{colSt}{}}
          \foreach \x in {6,8,...,14}{\prop{\x}{-2}{colSt}{}} 
         \foreach \x in {8,10,12}{\prop{\x}{0}{colSt}{}}
     \foreach \x in {10}{\prop{\x}{2}{colSt}{}}
      \foreach \x in {-4,-2} {\prop{6}{\x}{colimp}{}}
        \foreach \x in {6.5,8.5,...,12.5}{\istate{\x}{-4.5}{colSt}}    
    \foreach \x in {6,...,9}{\tgridLine{\x+0.25}{\x+1.75-8}{\x+.5}{\x+1.5-8}\circle{\x+0.25}{\x+1.75-8}}
    \foreach \x in {10,...,15}{\tgridLine{\x+0.75}{-\x+3.75+9}{\x+0.5}{-\x+3.5+9}
    \square{\x+0.75}{-\x+3.75+9}}
    \foreach \x in {14,...,15}{\tgridLine{\x+0.5}{\x-18.5}{\x+0.25}{\x-18.25}
    \circle{\x+0.5}{\x-18.5}}
    \foreach\y in {-4,-2}{\tgridLine{5.25}{\y-.75}{5.5}{\y-.5}
    \square{5.25}{\y-.75}}
    \foreach\y in {-4,-2}{\tgridLine{5.25}{\y+.75}{5.5}{\y+.5}
    \square{5.25}{\y+.75}}
  \end{tikzpicture}\nonumber\\
  & - q^{2\ell-8t-3(\ell+r-t)}\times
   \begin{tikzpicture}[baseline={([yshift=0.6ex]current bounding box.center)},scale=0.5]
    \foreach \x in {5}{\prop{\x}{-3}{colSt}{}}
     \foreach \x in {5,7,...,15}{\prop{\x}{-3}{colSt}{}}
      \foreach \x in {3,5}{\prop{\x}{-1}{colSt}{}}
           \foreach \x in {7,9,...,13}{\prop{\x}{-1}{colSt}{}}
             \foreach \x in {9,11}{\prop{\x}{1}{colSt}{}}
         \foreach \x in {8,10,...,14}{\prop{\x}{-4}{colSt}{}}
     \foreach \x in {4}{\prop{\x}{-2}{colSt}{}}
          \foreach \x in {6,8,...,14}{\prop{\x}{-2}{colSt}{}}
    \foreach \x in {4}{\prop{\x}{0}{colSt}{}} 
         \foreach \x in {8,10,12}{\prop{\x}{0}{colSt}{}}
     \foreach \x in {10}{\prop{\x}{2}{colSt}{}}
      \foreach \x in {-4,-2} {\prop{6}{\x}{colimp}{}}
        \foreach \x in {6.5,8.5,...,12.5}{\istate{\x}{-4.5}{colSt}}    
    \foreach \x in {2,...,3}{\tgridLine{\x+0.25}{\x-4+1.75}{\x+.5}{\x-4+1.5}\square{\x+0.25}{\x-4+1.75}}
    \foreach \x in {2,...,5}{\tgridLine{\x+0.5}{-\x+.5}{\x+.75}{-\x+0.75}\circle{\x+0.5}{-\x+.5}}
    \foreach \x in {6,...,9}{\tgridLine{\x+0.25}{\x+1.75-8}{\x+.5}{\x+1.5-8}\circle{\x+0.25}{\x+1.75-8}}
    \foreach \x in {10,...,15}{\tgridLine{\x+0.75}{-\x+3.75+9}{\x+0.5}{-\x+3.5+9}
    \square{\x+0.75}{-\x+3.75+9}}
    \foreach \x in {14,...,15}{\tgridLine{\x+0.5}{\x-18.5}{\x+0.25}{\x-18.25}
    \circle{\x+0.5}{\x-18.5}}
    \foreach \x in {4,...,5}{\tgridLine{\x+0.75}{-\x+3.75+1}{\x+0.5}{-\x+3.5+1}
    \circle{\x+0.75}{-\x+3.75+1}}   
    \draw[semithick,decorate,decoration={brace}] (1.75,.25) -- (3,1.25) node[midway,xshift=-10.5pt,yshift=7.5pt] {$t-r$};
             \draw[semithick,decorate,decoration={brace}] (5.25,-4.8) -- (2,-1.55) node[midway,xshift=-18.5pt,yshift=-9.5pt] {$\ell+r-t$};
  \end{tikzpicture}.
\end{align}

Now, repeating the same steps from the right side of the impurity, we will end up with a sum of nine different terms, which become six once accounting for symmetry between left and right. Some of these terms are exactly reducible via the basic diagrammatic rules of Eqs.~\eqref{eq:local_gate_rels_1}--\eqref{eq:local_gate_rels_3}, and once we further omit subleading terms in $1/q$, we end up with
\begin{align}\label{eq:TwoIntervalEntropy_mid_result}
    \tr \left[{\rho_A}^2\right] & \approx q^{-4\ell} + q^{-4(t+r)} \nonumber\\
    & +2q^{-2\ell-5t-3r}\times\!\!\!\!\!\!\!\!
    \begin{tikzpicture}[baseline={([yshift=0.6ex]current bounding box.center)},scale=0.5]
    \foreach \x in {5}{\prop{\x}{-3}{colSt}{}}
      \foreach \x in {3,5}{\prop{\x}{-1}{colSt}{}}
        \foreach \x in {1,3}{\prop{\x}{1}{colSt}{}}
     \foreach \x in {4}{\prop{\x}{-2}{colSt}{}}
    \foreach \x in {2,4}{\prop{\x}{0}{colSt}{}} 
     \foreach \x in {2}{\prop{\x}{2}{colSt}{}}
      \foreach \x in {-4,-2} {\prop{6}{\x}{colimp}{}}
    \foreach \x in {0,...,1}{\tgridLine{\x+0.25}{\x+1.75}{\x+.5}{\x+1.5}\square{\x+0.25}{\x+1.75}}
    \foreach \x in {0,...,5}{\tgridLine{\x+0.5}{-\x+.5}{\x+.75}{-\x+0.75}\circle{\x+0.5}{-\x+.5}}
    \foreach \x in {2,...,5}{\tgridLine{\x+0.75}{-\x+3.75+1}{\x+0.5}{-\x+3.5+1}
    \circle{\x+0.75}{-\x+3.75+1}}
    \foreach\y in {-4,-2}{\tgridLine{6.5}{\y-.5}{6.25}{\y-.25}
    \square{6.5}{\y-.5}}
    \foreach\y in {-4,-2}{\tgridLine{6.5}{\y+.5}{6.25}{\y+.25}
    \square{6.5}{\y+.5}}
    \draw[semithick,decorate,decoration={brace}] (-.25,2.25) -- (1,3.25) node[midway,xshift=-10.5pt,yshift=7.5pt] {$t-r$};
             \draw[semithick,decorate,decoration={brace}] (5.25,-4.8) -- (0,.45) node[midway,xshift=-12.5pt,yshift=-7.5pt] {$t+r$};
  \end{tikzpicture}
  - 2q^{-2\ell-5t-3r} \times\!\!\!\!\!\!\!\!
    \begin{tikzpicture}[baseline={([yshift=0.6ex]current bounding box.center)},scale=0.5]
    \foreach \x in {5}{\prop{\x}{-3}{colSt}{}}
      \foreach \x in {3,5}{\prop{\x}{-1}{colSt}{}}
     \foreach \x in {4}{\prop{\x}{-2}{colSt}{}}
    \foreach \x in {4}{\prop{\x}{0}{colSt}{}} 
      \foreach \x in {-4,-2} {\prop{6}{\x}{colimp}{}}
    \foreach \x in {2,...,3}{\tgridLine{\x+0.25}{\x-4+1.75}{\x+.5}{\x-4+1.5}\square{\x+0.25}{\x-4+1.75}}
    \foreach \x in {2,...,5}{\tgridLine{\x+0.5}{-\x+.5}{\x+.75}{-\x+0.75}\circle{\x+0.5}{-\x+.5}}
    \foreach \x in {4,...,5}{\tgridLine{\x+0.75}{-\x+3.75+1}{\x+0.5}{-\x+3.5+1}
    \circle{\x+0.75}{-\x+3.75+1}}   
    \foreach\y in {-4,-2}{\tgridLine{6.5}{\y-.5}{6.25}{\y-.25}
    \square{6.5}{\y-.5}}
    \foreach\y in {-4,-2}{\tgridLine{6.5}{\y+.5}{6.25}{\y+.25}
    \square{6.5}{\y+.5}}
    \draw[semithick,decorate,decoration={brace}] (1.75,.25) -- (3,1.25) node[midway,xshift=-10.5pt,yshift=7.5pt] {$t-r$};
             \draw[semithick,decorate,decoration={brace}] (5.25,-4.8) -- (2,-1.55) node[midway,xshift=-18.5pt,yshift=-9.5pt] {$\ell+r-t$};
  \end{tikzpicture}\,\,.
\end{align}
Let us examine the first diagram appearing in this equation. Through the resolution of the identity and the use of the diagrammatic rules of Eqs.~\eqref{eq:local_gate_rels_1}--\eqref{eq:local_gate_rels_3} and~\eqref{eq:MaximalEP1stProperty}, we find that, to leading order in $1/q$,
\begin{align}\label{eq:TwoIntervalEntropy_mid_result_2}
    \begin{tikzpicture}[baseline={([yshift=0.6ex]current bounding box.center)},scale=0.5]
    \foreach \x in {5}{\prop{\x}{-3}{colSt}{}}
      \foreach \x in {3,5}{\prop{\x}{-1}{colSt}{}}
        \foreach \x in {1,3}{\prop{\x}{1}{colSt}{}}
     \foreach \x in {4}{\prop{\x}{-2}{colSt}{}}
    \foreach \x in {2,4}{\prop{\x}{0}{colSt}{}} 
     \foreach \x in {2}{\prop{\x}{2}{colSt}{}}
      \foreach \x in {-4,-2} {\prop{6}{\x}{colimp}{}}
    \foreach \x in {0,...,1}{\tgridLine{\x+0.25}{\x+1.75}{\x+.5}{\x+1.5}\square{\x+0.25}{\x+1.75}}
    \foreach \x in {0,...,5}{\tgridLine{\x+0.5}{-\x+.5}{\x+.75}{-\x+0.75}\circle{\x+0.5}{-\x+.5}}
    \foreach \x in {2,...,5}{\tgridLine{\x+0.75}{-\x+3.75+1}{\x+0.5}{-\x+3.5+1}
    \circle{\x+0.75}{-\x+3.75+1}}
    \foreach\y in {-4,-2}{\tgridLine{6.5}{\y-.5}{6.25}{\y-.25}
    \square{6.5}{\y-.5}}
    \foreach\y in {-4,-2}{\tgridLine{6.5}{\y+.5}{6.25}{\y+.25}
    \square{6.5}{\y+.5}}
    \draw[semithick,decorate,decoration={brace}] (-.25,2.25) -- (1,3.25) node[midway,xshift=-10.5pt,yshift=7.5pt] {$t-r$};
             \draw[semithick,decorate,decoration={brace}] (5.25,-4.8) -- (0,.45) node[midway,xshift=-12.5pt,yshift=-7.5pt] {$t+r$};
  \end{tikzpicture}
  & =
  \frac{1}{q^2}\,\begin{tikzpicture}[baseline={([yshift=-0.6ex]current bounding box.center)},scale=0.5]
   \prop{0}{0}{colimp}{}
   \square{.5}{.5}
   \square{.5}{-.5}
   \circle{-.5}{.5}
   \circle{-.5}{-.5}
 \end{tikzpicture}
 \,\begin{tikzpicture}[baseline={([yshift=0.6ex]current bounding box.center)},scale=0.5]
    \foreach \x in {5}{\prop{\x}{-3}{colSt}{}}
      \foreach \x in {3,5}{\prop{\x}{-1}{colSt}{}}
        \foreach \x in {1,3}{\prop{\x}{1}{colSt}{}}
     \foreach \x in {4}{\prop{\x}{-2}{colSt}{}}
    \foreach \x in {2,4}{\prop{\x}{0}{colSt}{}} 
     \foreach \x in {2}{\prop{\x}{2}{colSt}{}}
      \foreach \x in {-2} {\prop{6}{\x}{colimp}{}}
    \foreach \x in {0,...,1}{\tgridLine{\x+0.25}{\x+1.75}{\x+.5}{\x+1.5}\square{\x+0.25}{\x+1.75}}
    \foreach \x in {0,...,4}{\tgridLine{\x+0.5}{-\x+.5}{\x+.75}{-\x+0.75}\circle{\x+0.5}{-\x+.5}}
    \foreach \x in {2,...,5}{\tgridLine{\x+0.75}{-\x+3.75+1}{\x+0.5}{-\x+3.5+1}
    \circle{\x+0.75}{-\x+3.75+1}}
    \foreach\y in {-2}{\tgridLine{6.5}{\y-.5}{6.25}{\y-.25}
    \square{6.5}{\y-.5}}
    \foreach\y in {-2}{\tgridLine{6.5}{\y+.5}{6.25}{\y+.25}
    \square{6.5}{\y+.5}}
    \circle{5 .5}{-3.5}
    \end{tikzpicture}
    +
    \frac{1}{q^2}\,\begin{tikzpicture}[baseline={([yshift=-0.6ex]current bounding box.center)},scale=0.5]
   \prop{0}{0}{colimp}{}
   \square{.5}{.5}
   \square{.5}{-.5}
   \circleblack{-.5}{.5}
   \circle{-.5}{-.5}
 \end{tikzpicture}
 \,\begin{tikzpicture}[baseline={([yshift=0.6ex]current bounding box.center)},scale=0.5]
    \foreach \x in {5}{\prop{\x}{-3}{colSt}{}}
      \foreach \x in {3,5}{\prop{\x}{-1}{colSt}{}}
        \foreach \x in {1,3}{\prop{\x}{1}{colSt}{}}
     \foreach \x in {4}{\prop{\x}{-2}{colSt}{}}
    \foreach \x in {2,4}{\prop{\x}{0}{colSt}{}} 
     \foreach \x in {2}{\prop{\x}{2}{colSt}{}}
      \foreach \x in {-2} {\prop{6}{\x}{colimp}{}}
    \foreach \x in {0,...,1}{\tgridLine{\x+0.25}{\x+1.75}{\x+.5}{\x+1.5}\square{\x+0.25}{\x+1.75}}
    \foreach \x in {0,...,4}{\tgridLine{\x+0.5}{-\x+.5}{\x+.75}{-\x+0.75}\circle{\x+0.5}{-\x+.5}}
    \foreach \x in {2,...,5}{\tgridLine{\x+0.75}{-\x+3.75+1}{\x+0.5}{-\x+3.5+1}
    \circle{\x+0.75}{-\x+3.75+1}}
    \foreach\y in {-2}{\tgridLine{6.5}{\y-.5}{6.25}{\y-.25}
    \square{6.5}{\y-.5}}
    \foreach\y in {-2}{\tgridLine{6.5}{\y+.5}{6.25}{\y+.25}
    \square{6.5}{\y+.5}}
    \circleblack{5.5}{-3.5}
  \end{tikzpicture}\nonumber\\
  & \approx
  \frac{1}{q}\,\begin{tikzpicture}[baseline={([yshift=-0.6ex]current bounding box.center)},scale=0.5]
   \prop{0}{0}{colimp}{}
   \square{.5}{.5}
   \square{.5}{-.5}
   \circle{-.5}{.5}
   \circle{-.5}{-.5}
 \end{tikzpicture}
 \,\begin{tikzpicture}[baseline={([yshift=0.6ex]current bounding box.center)},scale=0.5]
      \foreach \x in {5}{\prop{\x}{-1}{colSt}{}}
        \foreach \x in {3}{\prop{\x}{1}{colSt}{}}
    \foreach \x in {4}{\prop{\x}{0}{colSt}{}} 
     \foreach \x in {2}{\prop{\x}{2}{colSt}{}}
      \foreach \x in {-2} {\prop{6}{\x}{colimp}{}}
    \foreach \x in {1}{\tgridLine{\x+0.25}{\x+1.75}{\x+.5}{\x+1.5}\square{\x+0.25}{\x+1.75}}
    \foreach \x in {1,...,5}{\tgridLine{\x+0.5}{-\x+2.5}{\x+.75}{-\x+2.75}\circle{\x+0.5}{-\x+2.5}}
    \foreach \x in {2,...,5}{\tgridLine{\x+0.75}{-\x+3.75+1}{\x+0.5}{-\x+3.5+1}
    \circle{\x+0.75}{-\x+3.75+1}}
    \foreach\y in {-2}{\tgridLine{6.5}{\y-.5}{6.25}{\y-.25}
    \square{6.5}{\y-.5}}
    \foreach\y in {-2}{\tgridLine{6.5}{\y+.5}{6.25}{\y+.25}
    \square{6.5}{\y+.5}}
    \end{tikzpicture}
    +
    \frac{1}{q^{t+r-1}}\,\begin{tikzpicture}[baseline={([yshift=-0.6ex]current bounding box.center)},scale=0.5]
   \prop{0}{0}{colimp}{}
   \square{.5}{.5}
   \square{.5}{-.5}
   \circleblack{-.5}{.5}
   \circle{-.5}{-.5}
 \end{tikzpicture}
 \,\begin{tikzpicture}[baseline={([yshift=0.6ex]current bounding box.center)},scale=0.5]
      \foreach \x in {5}{\prop{\x}{-1}{colSt}{}}
        \foreach \x in {3}{\prop{\x}{1}{colSt}{}}
    \foreach \x in {4}{\prop{\x}{0}{colSt}{}} 
     \foreach \x in {2}{\prop{\x}{2}{colSt}{}}
      \foreach \x in {-2} {\prop{6}{\x}{colimp}{}}
    \foreach \x in {1}{\tgridLine{\x+0.25}{\x+1.75}{\x+.5}{\x+1.5}\square{\x+0.25}{\x+1.75}}
    \foreach \x in {1,...,5}{\tgridLine{\x+0.5}{-\x+2.5}{\x+.75}{-\x+2.75}\circleblack{\x+0.5}{-\x+2.5}}
    \foreach \x in {2,...,5}{\tgridLine{\x+0.75}{-\x+3.75+1}{\x+0.5}{-\x+3.5+1}
    \circle{\x+0.75}{-\x+3.75+1}}
    \foreach\y in {-2}{\tgridLine{6.5}{\y-.5}{6.25}{\y-.25}
    \square{6.5}{\y-.5}}
    \foreach\y in {-2}{\tgridLine{6.5}{\y+.5}{6.25}{\y+.25}
    \square{6.5}{\y+.5}}
    \end{tikzpicture}\nonumber\\
    & \le q^3 \,\begin{tikzpicture}[baseline={([yshift=-0.6ex]current bounding box.center)},scale=0.5]
   \prop{0}{0}{colimp}{}
   \square{.5}{.5}
   \square{.5}{-.5}
   \circle{-.5}{.5}
   \circle{-.5}{-.5}
 \end{tikzpicture}
 \,\begin{tikzpicture}[baseline={([yshift=0.6ex]current bounding box.center)},scale=0.5]
      \foreach \x in {5}{\prop{\x}{-1}{colSt}{}}
        \foreach \x in {3}{\prop{\x}{1}{colSt}{}}
    \foreach \x in {4}{\prop{\x}{0}{colSt}{}} 
     \foreach \x in {2}{\prop{\x}{2}{colSt}{}}
      \foreach \x in {-2} {\prop{6}{\x}{colimp}{}}
    \foreach \x in {1}{\tgridLine{\x+0.25}{\x+1.75}{\x+.5}{\x+1.5}\square{\x+0.25}{\x+1.75}}
    \foreach \x in {1,...,5}{\tgridLine{\x+0.5}{-\x+2.5}{\x+.75}{-\x+2.75}\circle{\x+0.5}{-\x+2.5}}
    \foreach \x in {2,...,5}{\tgridLine{\x+0.75}{-\x+3.75+1}{\x+0.5}{-\x+3.5+1}
    \circle{\x+0.75}{-\x+3.75+1}}
    \foreach\y in {-2}{\tgridLine{6.5}{\y-.5}{6.25}{\y-.25}
    \square{6.5}{\y-.5}}
    \foreach\y in {-2}{\tgridLine{6.5}{\y+.5}{6.25}{\y+.25}
    \square{6.5}{\y+.5}}
    \end{tikzpicture}
    +q^{3(t-r)}\,.
\end{align}
The last inequality is obtained by replacing $\begin{tikzpicture}[baseline={([yshift=-0.6ex]current bounding box.center)},scale=0.65]
   \tgridLine{0}{0}{0.35}{0.35}
  \circleblack{0}{0}
\end{tikzpicture}$ with $\begin{tikzpicture}[baseline={([yshift=-0.6ex]current bounding box.center)},scale=0.65]
   \tgridLine{0}{0}{0.35}{0.35}
  \square{0}{0}
\end{tikzpicture}$ as before, and by using the bound
\be
\begin{tikzpicture}[baseline={([yshift=-0.6ex]current bounding box.center)},scale=0.5]
\prop{0}{0}{colimp}{}
   \square{.5}{.5}
   \square{.5}{-.5}
   \circle{-.5}{.5}
   \circle{-.5}{-.5}
 \end{tikzpicture} \le q^4,
\ee
which may be derived from the non-negativity of the R\'enyi entropy.
By repeatedly applying the inequality in Eq.~\eqref{eq:TwoIntervalEntropy_mid_result_2}, we may then deduce the bound
\be
2q^{-2\ell-5t-3r}\times\!\!\!\!\!\!\!\!
    \begin{tikzpicture}[baseline={([yshift=0.6ex]current bounding box.center)},scale=0.5]
    \foreach \x in {5}{\prop{\x}{-3}{colSt}{}}
      \foreach \x in {3,5}{\prop{\x}{-1}{colSt}{}}
        \foreach \x in {1,3}{\prop{\x}{1}{colSt}{}}
     \foreach \x in {4}{\prop{\x}{-2}{colSt}{}}
    \foreach \x in {2,4}{\prop{\x}{0}{colSt}{}} 
     \foreach \x in {2}{\prop{\x}{2}{colSt}{}}
      \foreach \x in {-4,-2} {\prop{6}{\x}{colimp}{}}
    \foreach \x in {0,...,1}{\tgridLine{\x+0.25}{\x+1.75}{\x+.5}{\x+1.5}\square{\x+0.25}{\x+1.75}}
    \foreach \x in {0,...,5}{\tgridLine{\x+0.5}{-\x+.5}{\x+.75}{-\x+0.75}\circle{\x+0.5}{-\x+.5}}
    \foreach \x in {2,...,5}{\tgridLine{\x+0.75}{-\x+3.75+1}{\x+0.5}{-\x+3.5+1}
    \circle{\x+0.75}{-\x+3.75+1}}
    \foreach\y in {-4,-2}{\tgridLine{6.5}{\y-.5}{6.25}{\y-.25}
    \square{6.5}{\y-.5}}
    \foreach\y in {-4,-2}{\tgridLine{6.5}{\y+.5}{6.25}{\y+.25}
    \square{6.5}{\y+.5}}
    \draw[semithick,decorate,decoration={brace}] (-.25,2.25) -- (1,3.25) node[midway,xshift=-10.5pt,yshift=7.5pt] {$t-r$};
             \draw[semithick,decorate,decoration={brace}] (5.25,-4.8) -- (0,.45) node[midway,xshift=-12.5pt,yshift=-7.5pt] {$t+r$};
  \end{tikzpicture} \le 2q^{-2\ell-2t-2r}.
\ee
Therefore, for $t<\ell-r$ this contribution is subleading relative to the term $q^{-4(t+r)}$ in Eq.~\eqref{eq:TwoIntervalEntropy_mid_result}, while for $t>\ell-r$ it is subleading relative to the term $q^{-4\ell}$. Through a similar process, the second diagram in Eq.~\eqref{eq:TwoIntervalEntropy_mid_result} can also be shown to contribute negligibly. We may finally write 
\be\label{eq:Purity_final_result_chaotic}
\tr \left[{\rho_A}^2\right] \approx q^{-4\ell} + q^{-4(t+r)}.
\ee

The same result for the average purity is also true for $\ell/2+r\le t <\ell+r$, as can be obtained by a similar derivation. Throughout this derivation, one encounters diagrams that are either fully reducible via the relations in Eqs.~\eqref{eq:local_gate_rels_1}--\eqref{eq:local_gate_rels_3}, or equivalent in their structure to the diagrams we already analyzed, and so we do not present these steps again. In all, we have from Eq.~\eqref{eq:Purity_final_result_chaotic} that the purity is composed of two additive terms, where the $q^{-4(t+r)}$ term dominates for $t<\ell-r$, whereas the $q^{-4\ell}$ term dominates for $t>\ell-r$. This leads to the final result for the second R\'enyi entropy, which agrees with Eq.~\eqref{eq:2_Renyi_ent_chaotic}.

\section{The membrane picture in the presence of a non-unitary impurity}\label{app:Non_unitary_imp}

In this appendix, we revisit the derivation from Ref.~\cite{fraenkel2024entanglement} of the exact single-interval purity $\tr \left[{\rho_A}^2\right]$ in a circuit where the dual-unitary bulk has maximal entangling power $p=1$ and is otherwise random, in the large $q$ limit. In Ref.~\cite{fraenkel2024entanglement} we assumed that the impurity gate was unitary, while here we generalize the result to accommodate a non-unitary impurity gate. This generalized result is used in formulating the membrane picture for a charged dual-unitary bulk in Sec.~\ref{sec:membrane_charged} (where $q$ needs to be replaced by $d=q/2$, as the gates are projected onto charge sectors).

In analyzing the case of a non-unitary impurity gate, we will encounter diagrams of the general form (with $n=2$ replicas)
\begin{equation}
{\cal D}\left(h,x\right) \equiv \begin{tikzpicture}[baseline={([yshift=0.6ex]current bounding box.center)},scale=0.45]
\foreach \x in {4}{\prop{\x}{4}{colSt}{}}
\foreach \x in {3,5}{\prop{\x}{3}{colSt}{}}
\foreach \x in {2,4}{\prop{\x}{2}{colSt}{}}
   \foreach \x in {1,3,5}{\prop{\x}{1}{colSt}{}}   
    \foreach \x in {2,4}{\prop{\x}{0}{colSt}{}}
    \foreach \x in {3,5}{\prop{\x}{-1}{colSt}{}}
    \foreach \x in {4,6}{\prop{\x}{-2}{colSt}{}}
    \foreach \x in {5}{\prop{\x}{-3}{colSt}{}}
    \foreach \x in {6}{\prop{\x}{-4}{colSt}{}}
   \foreach \x in {-4,-2,...,2} {\prop{6}{\x}{colimp}{}}
    \foreach \x in {0,...,3}{\square{\x+0.5}{\x+1.5}}
    \foreach \x in {0,...,5}{\circle{\x+0.5}{-\x+0.5}}
    \foreach \x in {-3,...,-1}{\circle
    {4-\x-0.5}{\x+5.5}}
     \foreach \x in {-6,...,0}{\circle
    {6.5}{\x+1.5}}
          \draw[semithick,decorate,decoration={brace}] (0,1.75) -- (3.25,5) node[midway,xshift=-10.5pt,yshift=7.5pt] {$x$};
          \draw[semithick,decorate,decoration={brace}] (5.25,-4.8) -- (0,0.45) node[midway,xshift=-20pt,yshift=-7pt] {$2h-x$};
  \end{tikzpicture}\,\,,
 \end{equation}
 and so we begin by analyzing the scaling of ${\cal D}\left(h,x\right)$ with $h$ and $x$.
 Note that $x$ is the number of impurity gates appearing in the diagram, while $2h$ is its overall height, and that we assume that $h>x$. If the impurity is unitary then the diagram can be easily reduced using Eq.~\eqref{eq:local_gate_rels_2}, yielding ${\cal D}\left(h,x\right)=q^{4h-x}$. To treat the more general case, we resolve the identity on the upper-left leg of the bottom impurity gate, and use Eqs.~\eqref{eq:MaximalEP1stProperty} and \eqref{eq:MaximalEP2ndProperty} to obtain
 \begin{eqnarray}\label{eq:D_diagram_initial_approx}
     {\cal D}\left(h,x\right) &\approx& \frac{1}{q}\times\!\!\!\!\!\!\!\! \begin{tikzpicture}[baseline={([yshift=0.6ex]current bounding box.center)},scale=0.45]
\foreach \x in {4}{\prop{\x}{4}{colSt}{}}
\foreach \x in {3,5}{\prop{\x}{3}{colSt}{}}
\foreach \x in {2,4}{\prop{\x}{2}{colSt}{}}
   \foreach \x in {3,5}{\prop{\x}{1}{colSt}{}}   
    \foreach \x in {4}{\prop{\x}{0}{colSt}{}}
    \foreach \x in {5}{\prop{\x}{-1}{colSt}{}}
    \foreach \x in {6}{\prop{\x}{-4}{colSt}{}}
   \foreach \x in {-4,-2,...,2} {\prop{6}{\x}{colimp}{}}
    \foreach \x in {1,...,3}{\square{\x+0.5}{\x+1.5}}
    \foreach \x in {1,...,5}{\circle{\x+0.5}{-\x+2.5}}
    \foreach \y in {0,1}{\circle{5.5}{\y-4.5}}
    \foreach \x in {-3,...,-1}{\circle
    {4-\x-0.5}{\x+5.5}}
     \foreach \x in {-6,...,0}{\circle
    {6.5}{\x+1.5}}
          \draw[semithick,decorate,decoration={brace}] (1,2.75) -- (3.25,5) node[midway,xshift=-12.5pt,yshift=7.5pt] {$x-1$};
          \draw[semithick,decorate,decoration={brace}] (5.25,-2.8) -- (1,1.45) node[midway,xshift=-25pt,yshift=-10pt] {$2h-x-1$};
  \end{tikzpicture} \,+\, \frac{1}{q^{2h-x-1}}\times\!\!\!\!\!\!\!\! \begin{tikzpicture}[baseline={([yshift=0.6ex]current bounding box.center)},scale=0.45]
\foreach \x in {4}{\prop{\x}{4}{colSt}{}}
\foreach \x in {3,5}{\prop{\x}{3}{colSt}{}}
\foreach \x in {2,4}{\prop{\x}{2}{colSt}{}}
   \foreach \x in {3,5}{\prop{\x}{1}{colSt}{}}   
    \foreach \x in {4}{\prop{\x}{0}{colSt}{}}
    \foreach \x in {5}{\prop{\x}{-1}{colSt}{}}
    \foreach \x in {6}{\prop{\x}{-4}{colSt}{}}
   \foreach \x in {-4,-2,...,2} {\prop{6}{\x}{colimp}{}}
    \foreach \x in {1,...,3}{\square{\x+0.5}{\x+1.5}}
    \foreach \x in {1,...,5}{\circleblack{\x+0.5}{-\x+2.5}}
    \foreach \y in {0}{\circle{5.5}{\y-4.5}}
    \foreach \y in {1}{\circleblack{5.5}{\y-4.5}}
    \foreach \x in {-3,...,-1}{\circle
    {4-\x-0.5}{\x+5.5}}
     \foreach \x in {-6,...,0}{\circle
    {6.5}{\x+1.5}}
          \draw[semithick,decorate,decoration={brace}] (1,2.75) -- (3.25,5) node[midway,xshift=-12.5pt,yshift=7.5pt] {$x-1$};
          \draw[semithick,decorate,decoration={brace}] (5.25,-2.8) -- (1,1.45) node[midway,xshift=-25pt,yshift=-10pt] {$2h-x-1$};
  \end{tikzpicture} \nonumber \\
  &\approx& \frac{1}{q} {\cal D}\left(h-1,x-1\right)\begin{tikzpicture}[baseline={([yshift=-0.6ex]current bounding box.center)},scale=0.5]
   \prop{1}{0}{colimp}{}
   \circle{1.5}{0.5}
    \circle{0.5}{0.5}
       \circle{1.5}{-0.5}
    \circle{0.5}{-0.5}
 \end{tikzpicture} + \frac{1}{q^{x-1}} \begin{tikzpicture}[baseline={([yshift=-0.6ex]current bounding box.center)},scale=0.5]
   \prop{1}{0}{colimp}{}
   \circle{1.5}{0.5}
    \circleblack{0.5}{0.5}
       \circle{1.5}{-0.5}
    \circle{0.5}{-0.5}
 \end{tikzpicture} \, \begin{tikzpicture}[baseline={([yshift=-0.6ex]current bounding box.center)},scale=0.5]
   \prop{1}{0}{colimp}{}
   \circle{1.5}{0.5}
    \square{0.5}{0.5}
       \circle{1.5}{-0.5}
    \circleblack{0.5}{-0.5}
 \end{tikzpicture} \left(\begin{tikzpicture}[baseline={([yshift=-0.6ex]current bounding box.center)},scale=0.5]
   \prop{1}{0}{colimp}{}
   \circle{1.5}{0.5}
    \square{0.5}{0.5}
       \circle{1.5}{-0.5}
    \square{0.5}{-0.5}
 \end{tikzpicture}\right)^{x-2} \nonumber \\
 && - \frac{1}{q^{x+1}} \left(\begin{tikzpicture}[baseline={([yshift=-0.6ex]current bounding box.center)},scale=0.5]
   \prop{1}{0}{colimp}{}
   \circle{1.5}{0.5}
    \circleblack{0.5}{0.5}
       \circle{1.5}{-0.5}
    \circle{0.5}{-0.5}
 \end{tikzpicture}\right)^2 \begin{tikzpicture}[baseline={([yshift=-0.6ex]current bounding box.center)},scale=0.5]
   \prop{1}{0}{colimp}{}
   \circle{1.5}{0.5}
    \square{0.5}{0.5}
       \circle{1.5}{-0.5}
    \circle{0.5}{-0.5}
 \end{tikzpicture} \left(\begin{tikzpicture}[baseline={([yshift=-0.6ex]current bounding box.center)},scale=0.5]
   \prop{1}{0}{colimp}{}
   \circle{1.5}{0.5}
    \square{0.5}{0.5}
       \circle{1.5}{-0.5}
    \square{0.5}{-0.5}
 \end{tikzpicture}\right)^{x-3}\,.
 \end{eqnarray}

To proceed, let us note that if $O_{\rm imp}$ denotes the impurity operator, then
\begin{equation}
    \begin{tikzpicture}[baseline={([yshift=-0.6ex]current bounding box.center)},scale=0.5]
   \prop{1}{0}{colimp}{}
   \circle{1.5}{0.5}
    \circle{0.5}{0.5}
       \circle{1.5}{-0.5}
    \circle{0.5}{-0.5}
 \end{tikzpicture} = \left( \tr [{O_{\rm imp}}^{\dagger}O_{\rm imp}] \right)^2,
\end{equation}
implying that this diagram vanishes if and only if $O_{\rm imp}=0$, in which case ${\cal D}\left(h,x\right)=0$. Assuming that $O_{\rm imp}\neq0$, we may observe that
\begin{eqnarray}\label{eq:non_uni_imp_diagram_bounds}
    \frac{1}{q} \le & \frac{1}{\begin{tikzpicture}[baseline={([yshift=-0.6ex]current bounding box.center)},scale=0.5]
   \prop{1}{0}{colimp}{}
   \circle{1.5}{0.5}
    \circle{0.5}{0.5}
       \circle{1.5}{-0.5}
    \circle{0.5}{-0.5}
 \end{tikzpicture}}\,\begin{tikzpicture}[baseline={([yshift=-0.6ex]current bounding box.center)},scale=0.5]
   \prop{1}{0}{colimp}{}
   \circle{1.5}{0.5}
    \square{0.5}{0.5}
       \circle{1.5}{-0.5}
    \circle{0.5}{-0.5}
 \end{tikzpicture} & \le 1, \nonumber \\
 \frac{1}{q^2} \le & \frac{1}{\begin{tikzpicture}[baseline={([yshift=-0.6ex]current bounding box.center)},scale=0.5]
   \prop{1}{0}{colimp}{}
   \circle{1.5}{0.5}
    \circle{0.5}{0.5}
       \circle{1.5}{-0.5}
    \circle{0.5}{-0.5}
 \end{tikzpicture}}\,\begin{tikzpicture}[baseline={([yshift=-0.6ex]current bounding box.center)},scale=0.5]
   \prop{1}{0}{colimp}{}
   \circle{1.5}{0.5}
    \square{0.5}{0.5}
       \circle{1.5}{-0.5}
    \square{0.5}{-0.5}
 \end{tikzpicture} & \le 1.
\end{eqnarray}
These relations follow from viewing the above ratios as values of purity for different subsystems within a (normalized) four-qudit state, and they indicate that, to leading order in $1/q$, we may write
\begin{equation}
    {\cal D}\left(h,x\right)\approx \frac{1}{q} {\cal D}\left(h-1,x-1\right),
\end{equation}
that is, we may ignore the remaining two terms appearing in Eq.~\eqref{eq:D_diagram_initial_approx}. Indeed, if we follow this rule (effectively keeping only the $\begin{tikzpicture}[baseline={([yshift=-0.6ex]current bounding box.center)},scale=0.5]
    \circle{-0.5}{0.5}
      \gridLine{-0.38}{0.5}{0.2}{0.5}
 \end{tikzpicture}$ contributions when resolving the legs of the impurity gates), then we will obtain
 \begin{equation}\label{eq:D_diagram_final_approx}
     {\cal D}\left(h,x\right)\approx \frac{1}{q^x}q^{4\left(h-x\right)}\left(\begin{tikzpicture}[baseline={([yshift=-0.6ex]current bounding box.center)},scale=0.5]
   \prop{1}{0}{colimp}{}
   \circle{1.5}{0.5}
    \circle{0.5}{0.5}
       \circle{1.5}{-0.5}
    \circle{0.5}{-0.5}
 \end{tikzpicture}\right)^x,
 \end{equation}
 which is parametrically larger than the last two terms originally appearing in Eq.~\eqref{eq:D_diagram_initial_approx}, justifying their omission.

Equipped with Eq.~\eqref{eq:D_diagram_final_approx}, we now turn to the computation of the averaged purity $\tr \left[{\rho_A}^2\right]$. For $t<\ell/2$, it is straightforward to compute the purity (or any moment of $\rho_A$, for any choice of dual-unitary bulk gates), regardless of the unitarity of the impurity gate. We therefore focus on the regime $t\ge\ell/2$, for which the purity is given by the diagram
\begin{equation}\label{eq:Purity_initial_diagram}
    \tr \left[{\rho_A}^2\right] = q^{\ell+2r-6t} \left(\frac{1}{q^4} \begin{tikzpicture}[baseline={([yshift=-0.6ex]current bounding box.center)},scale=0.5]
   \prop{1}{0}{colimp}{}
   \circle{1.5}{0.5}
    \circle{0.5}{0.5}
       \circle{1.5}{-0.5}
    \circle{0.5}{-0.5}
 \end{tikzpicture}\,\right)^{r-t} \times \begin{tikzpicture}[baseline={([yshift=0.6ex]current bounding box.center)},scale=0.45]
{\prop{2}{2}{colSt}{}}
   \foreach \x in {1,3}{\prop{\x}{1}{colSt}{}}   
    \foreach \x in {0,2,...,4}{\prop{\x}{0}{colSt}{}}
    \foreach \x in {-1,1,...,5}{\prop{\x}{-1}{colSt}{}}
    \foreach \x in {-2,0,...,6}{\prop{\x}{-2}{colSt}{}}
    \foreach \x in {-3,-1,...,5}{\prop{\x}{-3}{colSt}{}}
    \foreach \x in {-2,0,...,6}{\prop{\x}{-4}{colSt}{}}
   \foreach \x in {-4,-2} {\prop{6}{\x}{colimp}{}}
    \foreach \x in {-1.5,0.5,...,4.5}{\istate{\x}{-4.5}{colSt}}
    \foreach \x in {-4,...,1}{\square{\x+0.5}{\x+1.5}}
	\circle{-3.5}{-3.5}
	\circle{-2.5}{-4.5}
    \foreach \x in {-3,...,1}{\circle
    {4-\x-0.5}{\x+1.5}}
     \foreach \x in {-6,...,-4}{\circle
    {6.5}{\x+1.5}}
      \draw[semithick,decorate,decoration={brace}] (-4,-2.25) -- (1.25,3) node[midway,xshift=-7.5pt,yshift=7.5pt] {$\ell$};
      \draw[semithick,decorate,decoration={brace}] (-2.75,-4.8) -- (-4.,-3.55) node[midway,xshift=-21.5pt] {$2t-\ell$};
  \draw[semithick,decorate,decoration={brace}] (2.55,2.85) -- (5.8,-0.3) node[midway,xshift=7.5pt,yshift=7.5pt] {$2r$};
  \end{tikzpicture}\,,
\end{equation}
where we normalized the trace of the total density matrix and used the diagrammatic rules of Eqs.~\eqref{eq:local_gate_rels_1} and \eqref{eq:local_gate_rels_3} that apply to the bulk gates. The diagram drawn in Eq.~\eqref{eq:Purity_initial_diagram} implicitly assumed that $\ell/2\le t <\ell/2+r$, but the derivation steps for $t\ge\ell/2+r$ are equivalent.  By resolving the initial state at the leftmost pair of sites and using the properties in Eqs.~\eqref{eq:MaximalEP1stProperty}--\eqref{eq:MaximalEP2ndProperty}, we obtain
\begin{eqnarray}\label{eq:IntervalMembraneInitialDecomposition}
    \begin{tikzpicture}[baseline={([yshift=0.6ex]current bounding box.center)},scale=0.45]
{\prop{2}{2}{colSt}{}}
   \foreach \x in {1,3}{\prop{\x}{1}{colSt}{}}   
    \foreach \x in {0,2,...,4}{\prop{\x}{0}{colSt}{}}
    \foreach \x in {-1,1,...,5}{\prop{\x}{-1}{colSt}{}}
    \foreach \x in {-2,0,...,6}{\prop{\x}{-2}{colSt}{}}
    \foreach \x in {-3,-1,...,5}{\prop{\x}{-3}{colSt}{}}
    \foreach \x in {-2,0,...,6}{\prop{\x}{-4}{colSt}{}}
   \foreach \x in {-4,-2} {\prop{6}{\x}{colimp}{}}
    \foreach \x in {-1.5,0.5,...,4.5}{\istate{\x}{-4.5}{colSt}}
    \foreach \x in {-4,...,1}{\square{\x+0.5}{\x+1.5}}
	\circle{-3.5}{-3.5}
	\circle{-2.5}{-4.5}
    \foreach \x in {-3,...,1}{\circle
    {4-\x-0.5}{\x+1.5}}
     \foreach \x in {-6,...,-4}{\circle
    {6.5}{\x+1.5}}
      \draw[semithick,decorate,decoration={brace}] (-4,-2.25) -- (1.25,3) node[midway,xshift=-7.5pt,yshift=7.5pt] {$\ell$};
      \draw[semithick,decorate,decoration={brace}] (-2.75,-4.8) -- (-4.,-3.55) node[midway,xshift=-21.5pt] {$2t-\ell$};
  \draw[semithick,decorate,decoration={brace}] (2.55,2.85) -- (5.8,-0.3) node[midway,xshift=7.5pt,yshift=7.5pt] {$2r$};
  \end{tikzpicture} & \approx & \frac{1}{q^3} \begin{tikzpicture}[baseline={([yshift=0.6ex]current bounding box.center)},scale=0.45]
{\prop{2}{2}{colSt}{}}
   \foreach \x in {1,3}{\prop{\x}{1}{colSt}{}}   
    \foreach \x in {0,2,...,4}{\prop{\x}{0}{colSt}{}}
    \foreach \x in {-1,1,...,5}{\prop{\x}{-1}{colSt}{}}
    \foreach \x in {-2,0,...,6}{\prop{\x}{-2}{colSt}{}}
    \foreach \x in {-1,1,...,5}{\prop{\x}{-3}{colSt}{}}
    \foreach \x in {0,2,...,6}{\prop{\x}{-4}{colSt}{}}
   \foreach \x in {-4,-2} {\prop{6}{\x}{colimp}{}}
    \foreach \x in {0.5,2.5,4.5}{\istate{\x}{-4.5}{colSt}}
    \foreach \x in {-3,...,1}{\square{\x+0.5}{\x+1.5}}
\circle{-2.5}{-2.5}
	\circle{-1.5}{-3.5}
	\circle{-0.5}{-4.5}
    \foreach \x in {-3,...,1}{\circle
    {4-\x-0.5}{\x+1.5}}
     \foreach \x in {-6,...,-4}{\circle
    {6.5}{\x+1.5}}
          \draw[semithick,decorate,decoration={brace}] (-3,-1.25) -- (1.25,3) node[midway,xshift=-10.5pt,yshift=7.5pt] {$\ell-1$};
  \end{tikzpicture} + \frac{1}{q^\ell}\left(\,\begin{tikzpicture}[baseline={([yshift=-0.6ex]current bounding box.center)},scale=0.5]
   \prop{1}{0}{colimp}{}
   \circle{1.5}{0.5}
    \square{0.5}{0.5}
       \circle{1.5}{-0.5}
    \square{0.5}{-0.5}
 \end{tikzpicture}\,\right)^{t-r} \nonumber \\
 && - \frac{1}{q^6} \begin{tikzpicture}[baseline={([yshift=0.6ex]current bounding box.center)},scale=0.45]
    \foreach \x in {4}{\prop{\x}{0}{colSt}{}}
    \foreach \x in {3,5}{\prop{\x}{-1}{colSt}{}}
    \foreach \x in {2,4,6}{\prop{\x}{-2}{colSt}{}}
    \foreach \x in {1,3,5}{\prop{\x}{-3}{colSt}{}}
    \foreach \x in {2,4,6}{\prop{\x}{-4}{colSt}{}}
   \foreach \x in {-4,-2} {\prop{6}{\x}{colimp}{}}
    \foreach \x in {2.5,4.5}{\istate{\x}{-4.5}{colSt}}
     \foreach \x in {-6,...,-4}{\circle
    {6.5}{\x+1.5}}
     \foreach \x in {-2,...,1}{\square{\x+2.5}{\x-0.5}}
\foreach \x in {2,3}  {\circle{-1.5+\x}{-1.5-\x}}
\foreach \x in {2,3,4}  {\circle{2.5+\x}{2.5-\x}}
       \draw[semithick,decorate,decoration={brace}] (-0,-2) -- (3.25,1) node[midway,xshift=-10.5pt,yshift=7.5pt] {$\ell-2$};
             \draw[semithick,decorate,decoration={brace}] (1.25,-4.8) -- (0.,-3.55) node[midway,xshift=-12.5pt] {$2$};
  \end{tikzpicture}\,\,.
\end{eqnarray}
The last term on the right-hand side is, in fact, subleading in its absolute value relative to the first two. This can be seen by resolving the initial state and, after using Eq.~\eqref{eq:MaximalEP1stProperty}, replacing all occurrences of $\begin{tikzpicture}[baseline={([yshift=-0.6ex]current bounding box.center)},scale=0.5]
    \circleblack{-0.5}{0.5}
      \gridLine{-0.38}{0.5}{0.2}{0.5}
 \end{tikzpicture}$ with $\begin{tikzpicture}[baseline={([yshift=-0.6ex]current bounding box.center)},scale=0.5]
    \square{-0.5}{0.5}
      \gridLine{-0.38}{0.5}{0.2}{0.5}
 \end{tikzpicture}$, thus bounding the expression from above:
 \begin{eqnarray}
     \frac{1}{q^6} \begin{tikzpicture}[baseline={([yshift=0.6ex]current bounding box.center)},scale=0.45]
    \foreach \x in {4}{\prop{\x}{0}{colSt}{}}
    \foreach \x in {3,5}{\prop{\x}{-1}{colSt}{}}
    \foreach \x in {2,4,6}{\prop{\x}{-2}{colSt}{}}
    \foreach \x in {1,3,5}{\prop{\x}{-3}{colSt}{}}
    \foreach \x in {2,4,6}{\prop{\x}{-4}{colSt}{}}
   \foreach \x in {-4,-2} {\prop{6}{\x}{colimp}{}}
    \foreach \x in {2.5,4.5}{\istate{\x}{-4.5}{colSt}}
     \foreach \x in {-6,...,-4}{\circle
    {6.5}{\x+1.5}}
     \foreach \x in {-2,...,1}{\square{\x+2.5}{\x-0.5}}
\foreach \x in {2,3}  {\circle{-1.5+\x}{-1.5-\x}}
\foreach \x in {2,3,4}  {\circle{2.5+\x}{2.5-\x}}
       \draw[semithick,decorate,decoration={brace}] (-0,-2) -- (3.25,1) node[midway,xshift=-10.5pt,yshift=7.5pt] {$\ell-2$};
             \draw[semithick,decorate,decoration={brace}] (1.25,-4.8) -- (0.,-3.55) node[midway,xshift=-12.5pt] {$2$};
  \end{tikzpicture} & \le & \frac{1}{q^9} \, \begin{tikzpicture}[baseline={([yshift=0.6ex]current bounding box.center)},scale=0.45]
    \foreach \x in {4}{\prop{\x}{0}{colSt}{}}
    \foreach \x in {3,5}{\prop{\x}{-1}{colSt}{}}
    \foreach \x in {2,4,6}{\prop{\x}{-2}{colSt}{}}
    \foreach \x in {3,5}{\prop{\x}{-3}{colSt}{}}
    \foreach \x in {4,6}{\prop{\x}{-4}{colSt}{}}
   \foreach \x in {-4,-2} {\prop{6}{\x}{colimp}{}}
    \foreach \x in {4.5}{\istate{\x}{-4.5}{colSt}}
     \foreach \x in {-6,...,-4}{\circle
    {6.5}{\x+1.5}}
     \foreach \x in {-1,...,1}{\square{\x+2.5}{\x-0.5}}
\foreach \x in {3,4,5}  {\circle{-1.5+\x}{0.5-\x}}
\foreach \x in {2,3,4}  {\circle{2.5+\x}{2.5-\x}};
  \end{tikzpicture} + \frac{1}{q^{\ell+4}}\left(\,\begin{tikzpicture}[baseline={([yshift=-0.6ex]current bounding box.center)},scale=0.5]
   \prop{1}{0}{colimp}{}
   \circle{1.5}{0.5}
    \square{0.5}{0.5}
       \circle{1.5}{-0.5}
    \square{0.5}{-0.5}
 \end{tikzpicture}\,\right)^{t-r}.
 \end{eqnarray}
 Repeating this step while neglecting subleading terms in $1/q$, we may write
 \begin{equation}\label{eq:IntervalMembrane_SmallTerm_UpperBound}
     \frac{1}{q^6} \begin{tikzpicture}[baseline={([yshift=0.6ex]current bounding box.center)},scale=0.45]
    \foreach \x in {4}{\prop{\x}{0}{colSt}{}}
    \foreach \x in {3,5}{\prop{\x}{-1}{colSt}{}}
    \foreach \x in {2,4,6}{\prop{\x}{-2}{colSt}{}}
    \foreach \x in {1,3,5}{\prop{\x}{-3}{colSt}{}}
    \foreach \x in {2,4,6}{\prop{\x}{-4}{colSt}{}}
   \foreach \x in {-4,-2} {\prop{6}{\x}{colimp}{}}
    \foreach \x in {2.5,4.5}{\istate{\x}{-4.5}{colSt}}
     \foreach \x in {-6,...,-4}{\circle
    {6.5}{\x+1.5}}
     \foreach \x in {-2,...,1}{\square{\x+2.5}{\x-0.5}}
\foreach \x in {2,3}  {\circle{-1.5+\x}{-1.5-\x}}
\foreach \x in {2,3,4}  {\circle{2.5+\x}{2.5-\x}}
       \draw[semithick,decorate,decoration={brace}] (-0,-2) -- (3.25,1) node[midway,xshift=-10.5pt,yshift=7.5pt] {$\ell-2$};
             \draw[semithick,decorate,decoration={brace}] (1.25,-4.8) -- (0.,-3.55) node[midway,xshift=-12.5pt] {$2$};
  \end{tikzpicture}  \le \frac{1}{q^{3(\ell+r-t)}} {\cal D}\left(\frac{\ell}{2} , t-r \right) + \frac{1}{q^{\ell+4}}\left(\,\begin{tikzpicture}[baseline={([yshift=-0.6ex]current bounding box.center)},scale=0.5]
   \prop{1}{0}{colimp}{}
   \circle{1.5}{0.5}
    \square{0.5}{0.5}
       \circle{1.5}{-0.5}
    \square{0.5}{-0.5}
 \end{tikzpicture}\,\right)^{t-r}.
 \end{equation}
 On the other hand, the first term on the right-hand side of Eq.~\eqref{eq:IntervalMembraneInitialDecomposition} can be bounded from below by resolving the initial state and keeping only the $\begin{tikzpicture}[baseline={([yshift=-0.6ex]current bounding box.center)},scale=0.5]
    \circle{-0.5}{0.5}
      \gridLine{-0.38}{0.5}{0.2}{0.5}
 \end{tikzpicture}$ contributions, leading to
 \begin{equation}\label{eq:IntervalMembrane_LeadTerm_LowerBound}
     \frac{1}{q^3} \begin{tikzpicture}[baseline={([yshift=0.6ex]current bounding box.center)},scale=0.45]
{\prop{2}{2}{colSt}{}}
   \foreach \x in {1,3}{\prop{\x}{1}{colSt}{}}   
    \foreach \x in {0,2,...,4}{\prop{\x}{0}{colSt}{}}
    \foreach \x in {-1,1,...,5}{\prop{\x}{-1}{colSt}{}}
    \foreach \x in {-2,0,...,6}{\prop{\x}{-2}{colSt}{}}
    \foreach \x in {-1,1,...,5}{\prop{\x}{-3}{colSt}{}}
    \foreach \x in {0,2,...,6}{\prop{\x}{-4}{colSt}{}}
   \foreach \x in {-4,-2} {\prop{6}{\x}{colimp}{}}
    \foreach \x in {0.5,2.5,4.5}{\istate{\x}{-4.5}{colSt}}
    \foreach \x in {-3,...,1}{\square{\x+0.5}{\x+1.5}}
\circle{-2.5}{-2.5}
	\circle{-1.5}{-3.5}
	\circle{-0.5}{-4.5}
    \foreach \x in {-3,...,1}{\circle
    {4-\x-0.5}{\x+1.5}}
     \foreach \x in {-6,...,-4}{\circle
    {6.5}{\x+1.5}}
          \draw[semithick,decorate,decoration={brace}] (-3,-1.25) -- (1.25,3) node[midway,xshift=-10.5pt,yshift=7.5pt] {$\ell-1$};
  \end{tikzpicture} \ge  \frac{1}{q^{3(\ell+r-t)}} {\cal D}\left(t , t-r \right).
 \end{equation}
 Put together, Eqs.~\eqref{eq:D_diagram_final_approx}, \eqref{eq:IntervalMembrane_SmallTerm_UpperBound} and \eqref{eq:IntervalMembrane_LeadTerm_LowerBound} show that the last term in the right-hand side of Eq.~\eqref{eq:IntervalMembraneInitialDecomposition} may be neglected. If we then repeat the step of Eq.~\eqref{eq:IntervalMembraneInitialDecomposition} until we may use Eq.~\eqref{eq:D_diagram_final_approx}, we find that
\begin{equation}\label{eq:Purity_nonU_imp}
   \tr \left[{\rho_A}^2\right] \approx  \frac{1}{q^{2\ell}} + \frac{1}{q^{2(t+r)}}\left(\frac{1}{\begin{tikzpicture}[baseline={([yshift=-0.6ex]current bounding box.center)},scale=0.5]
   \prop{1}{0}{colimp}{}
   \circle{1.5}{0.5}
    \circle{0.5}{0.5}
       \circle{1.5}{-0.5}
    \circle{0.5}{-0.5}
 \end{tikzpicture}}\, \begin{tikzpicture}[baseline={([yshift=-0.6ex]current bounding box.center)},scale=0.5]
   \prop{1}{0}{colimp}{}
   \circle{1.5}{0.5}
    \square{0.5}{0.5}
       \circle{1.5}{-0.5}
    \square{0.5}{-0.5}
 \end{tikzpicture}\,\right)^{t-r}\, .
\end{equation} 

Other than the value of $t$ for which the two terms in Eq.~\eqref{eq:Purity_nonU_imp} are equal, at any other time one term renders the other negligible, so in fact the purity equals the maximum between them. Concretely, if $r<\ell/2$ then the purity equals the second term and decays with time until it reaches its minimal value $q^{-2\ell}$, while for $r\ge\ell/2$ the minimal value has already been reached by the time $t=\ell/2$.

\bibliographystyle{ytphys_new}
 \bibliography{Dual_Unitary_Imp}

@article{kane1992transport,
  title = {Transport in a one-channel {Luttinger} liquid},
  author = {Kane, C. L. and Fisher, Matthew P. A.},
  journal = {Phys. Rev. Lett.},
  volume = {68},
  issue = {8},
  pages = {1220--1223},
  numpages = {0},
  year = {1992},
  month = {Feb},
  publisher = {American Physical Society},
  doi = {10.1103/PhysRevLett.68.1220},
  url = {https://link.aps.org/doi/10.1103/PhysRevLett.68.1220}
}

@article{santos2014local,
  title = {Local quenches with global effects in interacting quantum systems},
  author = {Torres-Herrera, E. J. and Santos, Lea F.},
  journal = {Phys. Rev. E},
  volume = {89},
  issue = {6},
  pages = {062110},
  numpages = {10},
  year = {2014},
  month = {Jun},
  publisher = {American Physical Society},
  doi = {10.1103/PhysRevE.89.062110},
  url = {https://link.aps.org/doi/10.1103/PhysRevE.89.062110}
}

@article{fritzsch2022boundary,
  title = {Boundary chaos},
  author = {Fritzsch, Felix and Prosen, Toma{\v z}},
  journal = {Phys. Rev. E},
  volume = {106},
  issue = {1},
  pages = {014210},
  numpages = {17},
  year = {2022},
  month = {Jul},
  publisher = {American Physical Society},
  doi = {10.1103/PhysRevE.106.014210},
  url = {https://link.aps.org/doi/10.1103/PhysRevE.106.014210}
}

@Article{fritzsch2023boundary,
	title={{Boundary chaos: Exact entanglement dynamics}},
	author={Felix Fritzsch and Roopayan Ghosh and Toma{\v z} Prosen},
	journal={SciPost Phys.},
	volume={15},
	pages={092},
	year={2023},
	publisher={SciPost},
	doi={10.21468/SciPostPhys.15.3.092},
	url={https://scipost.org/10.21468/SciPostPhys.15.3.092},
}

@article{jonay2018coarse,
      title={Coarse-grained dynamics of operator and state entanglement}, 
      author={Cheryne Jonay and David A. Huse and Adam Nahum},
      year={2018},
      eprint={1803.00089},
      archivePrefix={arXiv},
      primaryClass={cond-mat.stat-mech}, 
}

@article{nahum2020entanglement,
  title = {Entanglement Membrane in Chaotic Many-Body Systems},
  author = {Zhou, Tianci and Nahum, Adam},
  journal = {Phys. Rev. X},
  volume = {10},
  issue = {3},
  pages = {031066},
  numpages = {37},
  year = {2020},
  month = {Sep},
  publisher = {American Physical Society},
  doi = {10.1103/PhysRevX.10.031066},
  url = {https://link.aps.org/doi/10.1103/PhysRevX.10.031066}
}

@article{nahum2017quantum,
  title = {Quantum Entanglement Growth under Random Unitary Dynamics},
  author = {Nahum, Adam and Ruhman, Jonathan and Vijay, Sagar and Haah, Jeongwan},
  journal = {Phys. Rev. X},
  volume = {7},
  issue = {3},
  pages = {031016},
  numpages = {30},
  year = {2017},
  month = {Jul},
  publisher = {American Physical Society},
  doi = {10.1103/PhysRevX.7.031016},
  url = {https://link.aps.org/doi/10.1103/PhysRevX.7.031016}
}

@article{nahum2018dynamics,
  title = {Dynamics of entanglement and transport in one-dimensional systems with quenched randomness},
  author = {Nahum, Adam and Ruhman, Jonathan and Huse, David A.},
  journal = {Phys. Rev. B},
  volume = {98},
  issue = {3},
  pages = {035118},
  numpages = {16},
  year = {2018},
  month = {Jul},
  publisher = {American Physical Society},
  doi = {10.1103/PhysRevB.98.035118},
  url = {https://link.aps.org/doi/10.1103/PhysRevB.98.035118}
}

@article{bertini2018exact,
  title = {Exact Spectral Form Factor in a Minimal Model of Many-Body Quantum Chaos},
  author = {Bertini, Bruno and Kos, Pavel and Prosen, Toma{\v z}},
  journal = {Phys. Rev. Lett.},
  volume = {121},
  issue = {26},
  pages = {264101},
  numpages = {6},
  year = {2018},
  month = {Dec},
  publisher = {American Physical Society},
  doi = {10.1103/PhysRevLett.121.264101},
  url = {https://link.aps.org/doi/10.1103/PhysRevLett.121.264101}
}

@article{bertini2019exact,
  title = {Exact Correlation Functions for Dual-Unitary Lattice Models in $1+1$ Dimensions},
  author = {Bertini, Bruno and Kos, Pavel and Prosen, Toma{\v z}},
  journal = {Phys. Rev. Lett.},
  volume = {123},
  issue = {21},
  pages = {210601},
  numpages = {6},
  year = {2019},
  month = {Nov},
  publisher = {American Physical Society},
  doi = {10.1103/PhysRevLett.123.210601},
  url = {https://link.aps.org/doi/10.1103/PhysRevLett.123.210601}
}

@article{gopalakrishnan2019unitary,
  title = {Unitary circuits of finite depth and infinite width from quantum channels},
  author = {Gopalakrishnan, Sarang and Lamacraft, Austen},
  journal = {Phys. Rev. B},
  volume = {100},
  issue = {6},
  pages = {064309},
  numpages = {15},
  year = {2019},
  month = {Aug},
  publisher = {American Physical Society},
  doi = {10.1103/PhysRevB.100.064309},
  url = {https://link.aps.org/doi/10.1103/PhysRevB.100.064309}
}

@article{prosen2021many,
   title={Many-body quantum chaos and dual-unitarity round-a-face},
   volume={31},
   ISSN={1089-7682},
   url={http://dx.doi.org/10.1063/5.0056970},
   DOI={10.1063/5.0056970},
   number={9},
   journal={Chaos: An Interdisciplinary Journal of Nonlinear Science},
   publisher={AIP Publishing},
   author={Prosen, Toma{\v z}},
   year={2021},
   month=sep, pages={093101} }

@article{kos2021correlations,
  title = {Correlations in Perturbed Dual-Unitary Circuits: Efficient Path-Integral Formula},
  author = {Kos, Pavel and Bertini, Bruno and Prosen, Toma{\v z}},
  journal = {Phys. Rev. X},
  volume = {11},
  issue = {1},
  pages = {011022},
  numpages = {30},
  year = {2021},
  month = {Feb},
  publisher = {American Physical Society},
  doi = {10.1103/PhysRevX.11.011022},
  url = {https://link.aps.org/doi/10.1103/PhysRevX.11.011022}
}

@article{kos2023circuits,
   title={Circuits of space and time quantum channels},
   volume={7},
   ISSN={2521-327X},
   url={http://dx.doi.org/10.22331/q-2023-05-24-1020},
   DOI={10.22331/q-2023-05-24-1020},
   journal={Quantum},
   publisher={Verein zur Forderung des Open Access Publizierens in den Quantenwissenschaften},
   author={Kos, Pavel and Styliaris, Georgios},
   year={2023},
   month=may, pages={1020} }

@article{claeys2021ergodic,
   title={Ergodic and Nonergodic Dual-Unitary Quantum Circuits with Arbitrary Local {Hilbert} Space Dimension},
   volume={126},
   ISSN={1079-7114},
   url={http://dx.doi.org/10.1103/PhysRevLett.126.100603},
   DOI={10.1103/physrevlett.126.100603},
   number={10},
   journal={Physical Review Letters},
   publisher={American Physical Society (APS)},
   author={Claeys, Pieter W. and Lamacraft, Austen},
   year={2021},
   month=mar,  pages = {100603} }

@article{rampp2024entanglement,
  title = {Entanglement membrane in exactly solvable lattice models},
  author = {Rampp, Michael A. and Rather, Suhail A. and Claeys, Pieter W.},
  journal = {Phys. Rev. Res.},
  volume = {6},
  issue = {3},
  pages = {033271},
  numpages = {26},
  year = {2024},
  month = {Sep},
  publisher = {American Physical Society},
  doi = {10.1103/PhysRevResearch.6.033271},
  url = {https://link.aps.org/doi/10.1103/PhysRevResearch.6.033271}
}

@article{Rampp2024haydenpreskill,
  doi = {10.22331/q-2024-08-08-1434},
  url = {https://doi.org/10.22331/q-2024-08-08-1434},
  title = {Hayden-{P}reskill recovery in chaotic and integrable unitary circuit dynamics},
  author = {Rampp, Michael A. and Claeys, Pieter W.},
  journal = {{Quantum}},
  issn = {2521-327X},
  publisher = {{Verein zur F{\"{o}}rderung des Open Access Publizierens in den Quantenwissenschaften}},
  volume = {8},
  pages = {1434},
  month = aug,
  year = {2024}
}

@article{foligno2024quantum,
  title = {Quantum Information Spreading in Generalized Dual-Unitary Circuits},
  author = {Foligno, Alessandro and Kos, Pavel and Bertini, Bruno},
  journal = {Phys. Rev. Lett.},
  volume = {132},
  issue = {25},
  pages = {250402},
  numpages = {7},
  year = {2024},
  month = {Jun},
  publisher = {American Physical Society},
  doi = {10.1103/PhysRevLett.132.250402},
  url = {https://link.aps.org/doi/10.1103/PhysRevLett.132.250402}
}

@article{foligno2024entanglement,
    author = "Foligno, Alessandro and Bertini, Bruno",
    title = {Entanglement of Disjoint Intervals in Dual-Unitary Circuits: Exact Results},
    doi = "10.22331/q-2025-03-26-1678",
    journal = "Quantum",
    volume = "9",
    pages = "1678",
    year = "2025"
}

@article{foligno2024nonequilibrium,
    author = "Foligno, Alessandro and Calabrese, Pasquale and Bertini, Bruno",
    title = {Nonequilibrium Dynamics of Charged Dual-Unitary Circuits},
    primaryClass = "cond-mat.stat-mech",
    doi = "10.1103/PRXQuantum.6.010324",
    journal = "PRX Quantum",
    volume = "6",
    number = "1",
    pages = "010324",
    year = "2025"
}

@article{foligno2023growth,
   title={Growth of entanglement of generic states under dual-unitary dynamics},
   volume={107},
   ISSN={2469-9969},
   url={http://dx.doi.org/10.1103/PhysRevB.107.174311},
   DOI={10.1103/physrevb.107.174311},
   number={17},
   journal={Physical Review B},
   publisher={American Physical Society (APS)},
   author={Foligno, Alessandro and Bertini, Bruno},
   year={2023},
   month=may, pages={174311} }

@article{piroli2020exact,
  title = {Exact dynamics in dual-unitary quantum circuits},
  author = {Piroli, Lorenzo and Bertini, Bruno and Cirac, J. Ignacio and Prosen, Toma\v{z}},
  journal = {Phys. Rev. B},
  volume = {101},
  issue = {9},
  pages = {094304},
  numpages = {16},
  year = {2020},
  month = {Mar},
  publisher = {American Physical Society},
  doi = {10.1103/PhysRevB.101.094304},
  url = {https://link.aps.org/doi/10.1103/PhysRevB.101.094304}
}

@article{piroli2020scrambling,
  title = {Scrambling in random unitary circuits: Exact results},
  author = {Bertini, Bruno and Piroli, Lorenzo},
  journal = {Phys. Rev. B},
  volume = {102},
  issue = {6},
  pages = {064305},
  numpages = {25},
  year = {2020},
  month = {Aug},
  publisher = {American Physical Society},
  doi = {10.1103/PhysRevB.102.064305},
  url = {https://link.aps.org/doi/10.1103/PhysRevB.102.064305}
}

@article{bertini2020operator,
   title={Operator Entanglement in Local Quantum Circuits {I}: Chaotic Dual-Unitary  Circuits},
   volume={8},
   ISSN={2542-4653},
   url={http://dx.doi.org/10.21468/SciPostPhys.8.4.067},
   DOI={10.21468/scipostphys.8.4.067},
   number={4},
   journal={SciPost Physics},
   publisher={Stichting SciPost},
   author={Bertini, Bruno and Kos, Pavel and Prosen, Tomaz},
   year={2020},
   month=apr, pages={067}}

@Article{bertini2020operator2,
	title={Operator Entanglement in Local Quantum Circuits {II}: Solitons in Chains of Qubits},
	author={Bruno Bertini and Pavel Kos and Tomaz Prosen},
	journal={SciPost Phys.},
	volume={8},
	pages={068},
	year={2020},
	publisher={SciPost},
	doi={10.21468/SciPostPhys.8.4.068},
	url={https://scipost.org/10.21468/SciPostPhys.8.4.068},
}

@article{holdendye2023fundamental,
  doi = {10.22331/q-2025-01-30-1615},
  url = {https://doi.org/10.22331/q-2025-01-30-1615},
  title = {Fundamental charges for dual-unitary circuits},
  author = {Holden-Dye, Tom and Masanes, Lluis and Pal, Arijeet},
  journal = {{Quantum}},
  issn = {2521-327X},
  publisher = {{Verein zur F{\"{o}}rderung des Open Access Publizierens in den Quantenwissenschaften}},
  volume = {9},
  pages = {1615},
  month = jan,
  year = {2025}
}

@article{borsi2022construction,
  title = {Construction and the ergodicity properties of dual unitary quantum circuits},
  author = {Borsi, M\'arton and Pozsgay, Bal\'azs},
  journal = {Phys. Rev. B},
  volume = {106},
  issue = {1},
  pages = {014302},
  numpages = {16},
  year = {2022},
  month = {Jul},
  publisher = {American Physical Society},
  doi = {10.1103/PhysRevB.106.014302},
  url = {https://link.aps.org/doi/10.1103/PhysRevB.106.014302}
}

@article{giudice2022temporal,
   title={Temporal Entanglement, Quasiparticles, and the Role of Interactions},
   volume={128},
   ISSN={1079-7114},
   url={http://dx.doi.org/10.1103/PhysRevLett.128.220401},
   DOI={10.1103/physrevlett.128.220401},
   number={22},
   journal={Physical Review Letters},
   publisher={American Physical Society (APS)},
   author={Giudice, Giacomo and Giudici, Giuliano and Sonner, Michael and Thoenniss, Julian and Lerose, Alessio and Abanin, Dmitry A. and Piroli, Lorenzo},
   year={2022},
   month=jun, pages={220401} }

@article{foligno2023temporal,
  title = {Temporal Entanglement in Chaotic Quantum Circuits},
  author = {Foligno, Alessandro and Zhou, Tianci and Bertini, Bruno},
  journal = {Phys. Rev. X},
  volume = {13},
  issue = {4},
  pages = {041008},
  numpages = {37},
  year = {2023},
  month = {Oct},
  publisher = {American Physical Society},
  doi = {10.1103/PhysRevX.13.041008},
  url = {https://link.aps.org/doi/10.1103/PhysRevX.13.041008}
}

@article{vernier2023integrable,
   title={Integrable Digital Quantum Simulation: Generalized {Gibbs} Ensembles and {Trotter} Transitions},
   volume={130},
   ISSN={1079-7114},
   url={http://dx.doi.org/10.1103/PhysRevLett.130.260401},
   DOI={10.1103/physrevlett.130.260401},
   number={26},
   journal={Physical Review Letters},
   publisher={American Physical Society (APS)},
   author={Vernier, Eric and Bertini, Bruno and Giudici, Giuliano and Piroli, Lorenzo},
   year={2023},
   month=jun, pages={260401} }

@article{rather2020creating,
   title={Creating Ensembles of Dual Unitary and Maximally Entangling Quantum Evolutions},
   volume={125},
   ISSN={1079-7114},
   url={http://dx.doi.org/10.1103/PhysRevLett.125.070501},
   DOI={10.1103/physrevlett.125.070501},
   number={7},
   journal={Physical Review Letters},
   publisher={American Physical Society (APS)},
   author={Rather, Suhail Ahmad and Aravinda, S. and Lakshminarayan, Arul},
   year={2020},
   month=aug, pages={070501} }

@article{jonnadula2020entangling,
   title={Entanglement measures of bipartite quantum gates and their thermalization under arbitrary interaction strength},
   volume={2},
   ISSN={2643-1564},
   url={http://dx.doi.org/10.1103/PhysRevResearch.2.043126},
   DOI={10.1103/physrevresearch.2.043126},
   number={4},
   journal={Physical Review Research},
   publisher={American Physical Society (APS)},
   author={Jonnadula, Bhargavi and Mandayam, Prabha and Życzkowski, Karol and Lakshminarayan, Arul},
   year={2020},
   month=oct, pages={043126} }

@article{calabrese2005evolution,
   title={Evolution of entanglement entropy in one-dimensional systems},
   volume={2005},
   ISSN={1742-5468},
   url={http://dx.doi.org/10.1088/1742-5468/2005/04/P04010},
   DOI={10.1088/1742-5468/2005/04/p04010},
   number={04},
   journal={Journal of Statistical Mechanics: Theory and Experiment},
   publisher={IOP Publishing},
   author={Calabrese, Pasquale and Cardy, John},
   year={2005},
   month=apr, pages={P04010} }

@article{alba2017entanglement,
   title={Entanglement and thermodynamics after a quantum quench in integrable systems},
   volume={114},
   ISSN={1091-6490},
   url={http://dx.doi.org/10.1073/pnas.1703516114},
   DOI={10.1073/pnas.1703516114},
   number={30},
   journal={Proceedings of the National Academy of Sciences},
   publisher={Proceedings of the National Academy of Sciences},
   author={Alba, Vincenzo and Calabrese, Pasquale},
   year={2017},
   month=jul, pages={7947--7951} }

@article{alba2019quantum,
   title={Quantum information dynamics in multipartite integrable systems},
   volume={126},
   ISSN={1286-4854},
   url={http://dx.doi.org/10.1209/0295-5075/126/60001},
   DOI={10.1209/0295-5075/126/60001},
   number={6},
   journal={EPL (Europhysics Letters)},
   publisher={IOP Publishing},
   author={Alba, Vincenzo and Calabrese, Pasquale},
   year={2019},
   month=jul, pages={60001} }

@article{ares2023entanglement,
   title={Entanglement asymmetry as a probe of symmetry breaking},
   volume={14},
   ISSN={2041-1723},
   url={http://dx.doi.org/10.1038/s41467-023-37747-8},
   DOI={10.1038/s41467-023-37747-8},
   number={1},
   journal={Nature Communications},
   publisher={Springer Science and Business Media LLC},
   author={Ares, Filiberto and Murciano, Sara and Calabrese, Pasquale},
   year={2023},
   month=apr, pages={2036} }

@article{bertini2024dynamics,
  title = {Dynamics of charge fluctuations from asymmetric initial states},
  author = {Bertini, Bruno and Klobas, Katja and Collura, Mario and Calabrese, Pasquale and Rylands, Colin},
  journal = {Phys. Rev. B},
  volume = {109},
  issue = {18},
  pages = {184312},
  numpages = {28},
  year = {2024},
  month = {May},
  publisher = {American Physical Society},
  doi = {10.1103/PhysRevB.109.184312},
  url = {https://link.aps.org/doi/10.1103/PhysRevB.109.184312}
}

@article{alba2019scrambling,
  title = {Quantum information scrambling after a quantum quench},
  author = {Alba, Vincenzo and Calabrese, Pasquale},
  journal = {Phys. Rev. B},
  volume = {100},
  issue = {11},
  pages = {115150},
  numpages = {7},
  year = {2019},
  month = {Sep},
  publisher = {American Physical Society},
  doi = {10.1103/PhysRevB.100.115150},
  url = {https://link.aps.org/doi/10.1103/PhysRevB.100.115150}
}

@article{asplund2015entanglement,
author={Asplund, Curtis T.
and Bernamonti, Alice
and Galli, Federico
and Hartman, Thomas},
title={Entanglement scrambling in 2d conformal field theory},
journal={Journal of High Energy Physics},
year={2015},
month={Sep},
day={17},
volume={2015},
number={9},
pages={110},
issn={1029-8479},
doi={10.1007/JHEP09(2015)110},
url={https://doi.org/10.1007/JHEP09(2015)110}
}

@article{znidaric2025intergability,
  title = {Integrability is generic in homogeneous {$U(1)$}-invariant nearest-neighbor qubit circuits},
  author = {\ifmmode \check{Z}\else \v{Z}\fi{}nidari\ifmmode \check{c}\else \v{c}\fi{}, Marko and Duh, Urban and Zadnik, Lenart},
  journal = {Phys. Rev. B},
  volume = {112},
  issue = {2},
  pages = {L020302},
  numpages = {6},
  year = {2025},
  month = {Jul},
  publisher = {American Physical Society},
  doi = {10.1103/tqy8-ynpd},
  url = {https://link.aps.org/doi/10.1103/tqy8-ynpd}
}

@article{VidalPRA2002,
  title = {Computable measure of entanglement},
  author = {Vidal, G. and Werner, R. F.},
  journal = {Phys. Rev. A},
  volume = {65},
  issue = {3},
  pages = {032314},
  numpages = {11},
  year = {2002},
  publisher = {American Physical Society},
  doi = {10.1103/PhysRevA.65.032314},
  url = {https://link.aps.org/doi/10.1103/PhysRevA.65.032314}
}

@article{Plenio2005,
  title = {Logarithmic Negativity: A Full Entanglement Monotone That is not Convex},
  author = {Plenio, M. B.},
  journal = {Phys. Rev. Lett.},
  volume = {95},
  issue = {9},
  pages = {090503},
  numpages = {4},
  year = {2005},
  publisher = {American Physical Society},
  doi = {10.1103/PhysRevLett.95.090503},
  url = {https://link.aps.org/doi/10.1103/PhysRevLett.95.090503}
}

@article{fagotti2008evolution,
   title={Evolution of entanglement entropy following a quantum quench: Analytic results for the {$XY$} chain in a transverse magnetic field},
   author = {Fagotti, Maurizio and Calabrese, Pasquale},
  journal = {Phys. Rev. A},
  volume = {78},
  issue = {1},
  pages = {010306(R)},
  numpages = {4},
  year = {2008},
  month = {Jul},
  publisher = {American Physical Society},
  doi = {10.1103/PhysRevA.78.010306},
  url = {https://link.aps.org/doi/10.1103/PhysRevA.78.010306} }

@article{goldstein2018symmetry,
  title = {Symmetry-Resolved Entanglement in Many-Body Systems},
  author = {Goldstein, Moshe and Sela, Eran},
  journal = {Phys. Rev. Lett.},
  volume = {120},
  issue = {20},
  pages = {200602},
  numpages = {6},
  year = {2018},
  month = {May},
  publisher = {American Physical Society},
  doi = {10.1103/PhysRevLett.120.200602},
  url = {https://link.aps.org/doi/10.1103/PhysRevLett.120.200602}
}

@article{xavier2018equipartition,
   title={Equipartition of the entanglement entropy},
   volume={98},
   ISSN={2469-9969},
   url={http://dx.doi.org/10.1103/PhysRevB.98.041106},
   DOI={10.1103/physrevb.98.041106},
   number={4},
   journal={Physical Review B},
   publisher={American Physical Society (APS)},
   author={Xavier, J. C. and Alcaraz, F. C. and Sierra, G.},
   year={2018},
   month=jul }

@article{bertini2016transport,
  title = {Transport in Out-of-Equilibrium {XXZ} Chains: Exact Profiles of Charges and Currents},
  author = {Bertini, Bruno and Collura, Mario and De Nardis, Jacopo and Fagotti, Maurizio},
  journal = {Phys. Rev. Lett.},
  volume = {117},
  issue = {20},
  pages = {207201},
  numpages = {8},
  year = {2016},
  month = {Nov},
  publisher = {American Physical Society},
  doi = {10.1103/PhysRevLett.117.207201},
  url = {https://link.aps.org/doi/10.1103/PhysRevLett.117.207201}
}

@article{castro2016emergent,
  title = {Emergent Hydrodynamics in Integrable Quantum Systems Out of Equilibrium},
  author = {Castro-Alvaredo, Olalla A. and Doyon, Benjamin and Yoshimura, Takato},
  journal = {Phys. Rev. X},
  volume = {6},
  issue = {4},
  pages = {041065},
  numpages = {17},
  year = {2016},
  month = {Dec},
  publisher = {American Physical Society},
  doi = {10.1103/PhysRevX.6.041065},
  url = {https://link.aps.org/doi/10.1103/PhysRevX.6.041065}
}

@article{andrei1983solution,
  title = {Solution of the {K}ondo problem},
  author = {Andrei, N. and Furuya, K. and Lowenstein, J. H.},
  journal = {Rev. Mod. Phys.},
  volume = {55},
  issue = {2},
  pages = {331--402},
  numpages = {0},
  year = {1983},
  month = {Apr},
  publisher = {American Physical Society},
  doi = {10.1103/RevModPhys.55.331},
  url = {https://link.aps.org/doi/10.1103/RevModPhys.55.331}
}

@article{bertini2025exactly,
  title = {Exactly solvable quantum many-body dynamics from space-time duality},
  author = {Bertini, Bruno and Claeys, Pieter W. and Prosen, Toma{\ifmmode \check{z}\else \v{z}\fi{}}},
  journal = {Rev. Mod. Phys.},
  volume = {98},
  issue = {2},
  pages = {025001},
  numpages = {63},
  year = {2026},
  month = {Apr},
  publisher = {American Physical Society},
  doi = {10.1103/yx73-dk86},
  url = {https://link.aps.org/doi/10.1103/yx73-dk86}
}

@ARTICLE{affleck1991kondo,
       author = {{Affleck}, Ian and {Ludwig}, Andreas W.~W.},
        title = "{The Kondo effect, conformal field theory and fusion rules}",
      journal = {Nuclear Physics B},
         year = 1991,
       volume = {352},
       number = {3},
        pages = {849-862},
          doi = {10.1016/0550-3213(91)90109-B},
       adsurl = {https://ui.adsabs.harvard.edu/abs/1991NuPhB.352..849A}
}

@ARTICLE{affleck1993exact,
       author = {{Affleck}, Ian and {Ludwig}, Andreas W.~W.},
        title = "{Exact conformal-field-theory results on the multichannel Kondo effect: Single-fermion Green's function, self-energy, and resistivity}",
      journal = {Phys. Rev. B},
         year = 1993,
       volume = {48},
       number = {10},
        pages = {7297-7321},
          doi = {10.1103/PhysRevB.48.7297},
       adsurl = {https://ui.adsabs.harvard.edu/abs/1993PhRvB..48.7297A}
}

@ARTICLE{kondo1964resistance,
       author = {{Kondo}, J.},
        title = {Resistance Minimum in Dilute Magnetic Alloys},
      journal = {Progress of Theoretical Physics},
         year = 1964,
       volume = {32},
       number = {1},
        pages = {37-49},
          doi = {10.1143/PTP.32.37},
       adsurl = {https://ui.adsabs.harvard.edu/abs/1964PThPh..32...37K}
}

@ARTICLE{rylands2016quantum,
       author = {{Rylands}, Colin and {Andrei}, Natan},
        title = "{Quantum impurity in a Luttinger liquid: Exact solution of the Kane-Fisher model}",
      journal = {Phys. Rev. B},
         year = 2016,
       volume = {94},
       number = {11},
          eid = {115142},
        pages = {115142},
          doi = {10.1103/PhysRevB.94.115142},
       adsurl = {https://ui.adsabs.harvard.edu/abs/2016PhRvB..94k5142R}
}

@article{li2025dynamics,
  title = {Dynamics in the presence of local symmetry-breaking impurities},
  author = {Li, Yahui and Sala, Pablo and Pollmann, Frank and Moudgalya, Sanjay and Motrunich, Olexei},
  journal = {Phys. Rev. B},
  volume = {112},
  issue = {15},
  pages = {155108},
  numpages = {44},
  year = {2025},
  month = {Oct},
  publisher = {American Physical Society},
  doi = {10.1103/p1cm-9z8n},
  url = {https://link.aps.org/doi/10.1103/p1cm-9z8n}
}

@article{capizzi2023entanglement,
	title={{Entanglement evolution after a global quench across a conformal defect}},
	author={Luca Capizzi and Viktor Eisler},
	journal={SciPost Phys.},
	volume={14},
	pages={070},
	year={2023},
	publisher={SciPost},
	doi={10.21468/SciPostPhys.14.4.070},
	url={https://scipost.org/10.21468/SciPostPhys.14.4.070},
}

@article{bertini2022entanglement,
  title = {Entanglement Negativity and Mutual Information after a Quantum Quench: Exact Link from Space-Time Duality},
  author = {Bertini, Bruno and Klobas, Katja and Lu, Tsung-Cheng},
  journal = {Phys. Rev. Lett.},
  volume = {129},
  issue = {14},
  pages = {140503},
  numpages = {8},
  year = {2022},
  month = {Sep},
  publisher = {American Physical Society},
  doi = {10.1103/PhysRevLett.129.140503},
  url = {https://link.aps.org/doi/10.1103/PhysRevLett.129.140503}
}

@article{fraenkel2024entanglement,
  title = {Entanglement in dual unitary quantum circuits with impurities},
  author = {Fraenkel, Shachar and Rylands, Colin},
  journal = {Phys. Rev. B},
  volume = {112},
  issue = {6},
  pages = {064305},
  numpages = {7},
  year = {2025},
  month = {Aug},
  publisher = {American Physical Society},
  doi = {10.1103/wfcg-39f4},
  url = {https://link.aps.org/doi/10.1103/wfcg-39f4}
}

@article{rylands2023transport,
  title = {Transport and Entanglement across Integrable Impurities from Generalized Hydrodynamics},
  author = {Rylands, Colin and Calabrese, Pasquale},
  journal = {Phys. Rev. Lett.},
  volume = {131},
  issue = {15},
  pages = {156303},
  numpages = {7},
  year = {2023},
  month = {Oct},
  publisher = {American Physical Society},
  doi = {10.1103/PhysRevLett.131.156303},
  url = {https://link.aps.org/doi/10.1103/PhysRevLett.131.156303}
}

@Article{fraenkel2021entanglement,
	title={{Entanglement measures in a nonequilibrium steady state: Exact results in one dimension}},
	author={Shachar Fraenkel and Moshe Goldstein},
	journal={SciPost Phys.},
	volume={11},
	pages={085},
	year={2021},
	publisher={SciPost},
	doi={10.21468/SciPostPhys.11.4.085},
	url={https://scipost.org/10.21468/SciPostPhys.11.4.085},
}

@Article{fraenkel2023extensive,
	title={{Extensive long-range entanglement in a nonequilibrium steady state}},
	author={Shachar Fraenkel and Moshe Goldstein},
	journal={SciPost Phys.},
	volume={15},
	pages={134},
	year={2023},
	publisher={SciPost},
	doi={10.21468/SciPostPhys.15.4.134},
	url={https://scipost.org/10.21468/SciPostPhys.15.4.134},
}

@article{capizzi2023domain,
   title={Domain wall melting across a defect},
   volume={141},
   ISSN={1286-4854},
   url={http://dx.doi.org/10.1209/0295-5075/acb50a},
   DOI={10.1209/0295-5075/acb50a},
   number={3},
   journal={Europhysics Letters},
   publisher={IOP Publishing},
   author={Capizzi, Luca and Scopa, Stefano and Rottoli, Federico and Calabrese, Pasquale},
   year={2023},
   month=jan, pages={31002} }

@article{arvinda2021bernoulli,
  title = {From dual-unitary to quantum {Bernoulli} circuits: Role of the entangling power in constructing a quantum ergodic hierarchy},
  author = {Aravinda, S. and Rather, Suhail Ahmad and Lakshminarayan, Arul},
  journal = {Phys. Rev. Res.},
  volume = {3},
  issue = {4},
  pages = {043034},
  numpages = {30},
  year = {2021},
  month = {Oct},
  publisher = {American Physical Society},
  doi = {10.1103/PhysRevResearch.3.043034},
  url = {https://link.aps.org/doi/10.1103/PhysRevResearch.3.043034}
}

@article{zanardi2000entangling,
  title = {Entangling power of quantum evolutions},
  author = {Zanardi, Paolo and Zalka, Christof and Faoro, Lara},
  journal = {Phys. Rev. A},
  volume = {62},
  issue = {3},
  pages = {030301(R)},
  numpages = {4},
  year = {2000},
  month = {Aug},
  publisher = {American Physical Society},
  doi = {10.1103/PhysRevA.62.030301},
  url = {https://link.aps.org/doi/10.1103/PhysRevA.62.030301}
}

@article{Gogolin_2016,
doi = {10.1088/0034-4885/79/5/056001},
url = {https://dx.doi.org/10.1088/0034-4885/79/5/056001},
year = {2016},
publisher = {IOP Publishing},
volume = {79},
number = {5},
pages = {056001},
author = {Christian Gogolin and Jens Eisert},
title = {Equilibration, thermalisation, and the emergence of statistical mechanics in closed quantum systems},
journal = {Rep. Prog. Phys.},
}

@article{Calabrese_2016,
doi = {10.1088/1742-5468/2016/06/064001},
url = {https://dx.doi.org/10.1088/1742-5468/2016/06/064001},
year = {2016},
publisher = {IOP Publishing and SISSA},
number = {6},
pages = {064001},
author = {Calabrese, Pasquale and Essler, Fabian H L and Mussardo, Giuseppe},
title = {Introduction to Quantum Integrability in Out of Equilibrium Systems},
journal = {J. Stat. Mech.}
}

@article{Essler_quench,
doi = {10.1088/1742-5468/2016/06/064002},
url = {https://dx.doi.org/10.1088/1742-5468/2016/06/064002},
year = {2016},
publisher = {IOP Publishing and SISSA},
number = {6},
pages = {064002},
author = {Fabian H L Essler and Maurizio Fagotti},
title = {Quench Dynamics and Relaxation in Isolated Integrable Quantum Spin Chains},
journal = {J. Stat. Mech.}
}

@article{Deutsch1991,
  title = {Quantum statistical mechanics in a closed system},
  author = {Deutsch, J. M.},
  journal = {Phys. Rev. A},
  volume = {43},
  issue = {4},
  pages = {2046},
  numpages = {0},
  year = {1991},
  publisher = {American Physical Society},
  doi = {10.1103/PhysRevA.43.2046},
  url = {https://link.aps.org/doi/10.1103/PhysRevA.43.2046}
}

@article{srednicki1,
  title = {Chaos and quantum thermalization},
  author = {Srednicki, Mark},
  journal = {Phys. Rev. E},
  volume = {50},
  issue = {2},
  pages = {888},
  numpages = {0},
  year = {1994},
  publisher = {American Physical Society},
  doi = {10.1103/PhysRevE.50.888},
  url = {https://link.aps.org/doi/10.1103/PhysRevE.50.888}
}

@article{Rigol:2007juv,
    author = "Rigol, Marcos and Dunjko, Vanja and Olshanii, Maxim",
    title = "{Thermalization and its mechanism for generic isolated quantum systems}",
    doi = "10.1038/nature06838",
    journal = "Nature",
    volume = "452",
    number = "7189",
    pages = "854",
    year = "2008"
}

@article{polkovnikov,
  title = {Colloquium: Nonequilibrium dynamics of closed interacting quantum systems},
  author = {Polkovnikov, Anatoli and Sengupta, Krishnendu and Silva, Alessandro and Vengalattore, Mukund},
  journal = {Rev. Mod. Phys.},
  volume = {83},
  issue = {3},
  pages = {863},
  numpages = {0},
  year = {2011},
  publisher = {American Physical Society},
  doi = {10.1103/RevModPhys.83.863},
  url = {https://link.aps.org/doi/10.1103/RevModPhys.83.863}
}

@article{DAlessio:2015qtq,
    author = "D'Alessio, Luca and Kafri, Yariv and Polkovnikov, Anatoli and Rigol, Marcos",
    title = "{From quantum chaos and eigenstate thermalization to statistical mechanics and thermodynamics}",
    doi = "10.1080/00018732.2016.1198134",
    journal = "Adv. Phys.",
    volume = "65",
    number = "3",
    pages = "239",
    year = "2016"
}

@ARTICLE{rylands2020nonequilibrium,
       author = {{Rylands}, Colin and {Andrei}, Natan},
        title = {Nonequilibrium Aspects of Integrable Models},
      journal = {Annual Review of Condensed Matter Physics},
         year = 2020,
        month = mar,
       volume = {11},
        pages = {147-168},
          doi = {10.1146/annurev-conmatphys-031119-050630},
       adsurl = {https://ui.adsabs.harvard.edu/abs/2020ARCMP..11..147R}
}

@article{gruber2020time,
   title={Time evolution of entanglement negativity across a defect},
   volume={53},
   ISSN={1751-8121},
   url={http://dx.doi.org/10.1088/1751-8121/ab831c},
   DOI={10.1088/1751-8121/ab831c},
   number={20},
   journal={Journal of Physics A: Mathematical and Theoretical},
   publisher={IOP Publishing},
   author={Gruber, Matthias and Eisler, Viktor},
   year={2020},
   month=Apr, pages={205301} }

@article{horodecki2009entanglement,
  title = {Quantum entanglement},
  author = {Horodecki, Ryszard and Horodecki, Pawe{\l{}} and Horodecki, Micha{\l{}} and Horodecki, Karol},
  journal = {Rev. Mod. Phys.},
  volume = {81},
  issue = {2},
  pages = {865--942},
  numpages = {0},
  year = {2009},
  month = {Jun},
  publisher = {American Physical Society},
  doi = {10.1103/RevModPhys.81.865},
  url = {https://link.aps.org/doi/10.1103/RevModPhys.81.865}
}

@article{groisman2005correlations,
  title = {Quantum, classical, and total amount of correlations in a quantum state},
  author = {Groisman, Berry and Popescu, Sandu and Winter, Andreas},
  journal = {Phys. Rev. A},
  volume = {72},
  issue = {3},
  pages = {032317},
  numpages = {11},
  year = {2005},
  month = {Sep},
  publisher = {American Physical Society},
  doi = {10.1103/PhysRevA.72.032317},
  url = {https://link.aps.org/doi/10.1103/PhysRevA.72.032317}
}

@article{Helwig2012AME,
  title = {Absolute maximal entanglement and quantum secret sharing},
  author = {Helwig, Wolfram and Cui, Wei and Latorre, Jos{\'e} Ignacio and Riera, Arnau and Lo, Hoi-Kwong},
  journal = {Phys. Rev. A},
  volume = {86},
  issue = {5},
  pages = {052335},
  numpages = {5},
  year = {2012},
  month = {Nov},
  publisher = {American Physical Society},
  doi = {10.1103/PhysRevA.86.052335},
  url = {https://link.aps.org/doi/10.1103/PhysRevA.86.052335}
}

@article{Pastawski2015holographic,
author={Pastawski, Fernando
and Yoshida, Beni
and Harlow, Daniel
and Preskill, John},
title={Holographic quantum error-correcting codes: toy models for the bulk/boundary correspondence},
journal={Journal of High Energy Physics},
year={2015},
month={Jun},
day={23},
volume={2015},
number={6},
pages={149},
issn={1029-8479},
doi={10.1007/JHEP06(2015)149},
url={https://doi.org/10.1007/JHEP06(2015)149}
}

@article{Kudler-Flam2020correlation,
author={Kudler-Flam, Jonah
and Kusuki, Yuya
and Ryu, Shinsei},
title={Correlation measures and the entanglement wedge cross-section after quantum quenches in two-dimensional conformal field theories},
journal={Journal of High Energy Physics},
year={2020},
month={Apr},
day={14},
volume={2020},
number={4},
pages={74},
issn={1029-8479},
doi={10.1007/JHEP04(2020)074},
url={https://doi.org/10.1007/JHEP04(2020)074}
}

@article{Kudler-Flam2020quantum,
author={Kudler-Flam, Jonah
and Nozaki, Masahiro
and Ryu, Shinsei
and Tan, Mao Tian},
title={Quantum vs. classical information: operator negativity as a probe of scrambling},
journal={Journal of High Energy Physics},
year={2020},
month={Jan},
day={07},
volume={2020},
number={1},
pages={31},
doi={10.1007/JHEP01(2020)031},
url={https://doi.org/10.1007/JHEP01(2020)031}
}

@software{fraenkel_2026_zenodo,
  author       = {Fraenkel, Shachar and
                  Rylands, Colin},
  title        = {Entanglement and information dynamics in charged
                   dual-unitary circuits with impurities
                  },
  year         = 2026,
  publisher    = {Zenodo},
  doi          = {10.5281/zenodo.24164251},
  url          = {https://doi.org/10.5281/zenodo.24164251},
}

\end{document}